\documentclass[12pt,letterpaper]{article}
\usepackage[a4paper, total={7in, 10in}]{geometry}
\usepackage{graphicx}
\usepackage{authblk}
\usepackage{hyperref}
\usepackage{amsmath} 
\usepackage{amssymb} 
\usepackage{orcidlink} 
\usepackage[super,comma,sort&compress]  
   {natbib}
\usepackage{booktabs}

\title{Low-regret Strategies for Energy Systems Planning in a
Highly Uncertain Future}
\date{May 19, 2025}

\author[1,\orcidlink{0000-0001-7508-872X}]{Gabriel Wiest}
\author[1]{Niklas Nolzen}
\author[1]{Florian Baader}
\author[1]{André Bardow}
\author[1,*]{Stefano Moret}

\affil[1]{Energy and Process Systems Engineering, Department of Mechanical and Process
Engineering, ETH Zürich, Tannenstrasse 3, Zürich, 8092, Switzerland}
\affil[*]{Correspondence: morets@ethz.ch}

\begin{document}

\maketitle

\hrule
\section*{Abstract}

Large uncertainties in the energy transition urge decision-makers to develop \emph{low-regret} strategies, i.e., strategies that perform well regardless of how the future unfolds.
To address this challenge, we introduce a decision-support framework that identifies low-regret strategies in energy system planning under uncertainty. Our framework (i) automatically identifies strategies, (ii) evaluates their performance in terms of regret, (iii) assesses the key drivers of regret, and (iv) supports the decision process with intuitive decision trees, regret curves and decision maps.
    
We apply the framework to evaluate the optimal use of biomass in the transition to net-zero energy systems, considering all major biomass utilization options: biofuels, biomethane, chemicals, hydrogen, biochar, electricity, and heat.
Producing fuels and chemicals from biomass performs best across various decision-making criteria. In contrast, the current use of biomass, mainly for low-temperature heat supply, results in high regret, making it a \emph{must-avoid} in the energy transition.

\vspace{10pt}

\noindent \emph{Keywords:} Energy system, Strategic planning, Uncertainty, Decision-making, Low-regret,\\ Biomass
\vspace{10pt}
\hrule

\section{Introduction}

To mitigate climate change, energy systems need to transition rapidly towards net-zero. The urgency for the transition is in conflict with long life spans of energy infrastructure calling for careful planning. However, the planning of the energy transition faces large uncertainties \cite{panos_deep_2023} along various dimensions: social (e.g., technology acceptance\cite{rogelj_probabilistic_2013}), financial (e.g., discount rates \cite{garcia-gusano_role_2016}), technological (e.g., future deployment of carbon capture and storage \cite{lane_uncertain_2021}), and geopolitical (e.g., isolationism \cite{eser_trade-offs_2018}). 
These uncertainties impair a fast transition as they often delay the decision-making process \cite{pindyck_irreversibility_1990}, resulting in higher total costs and not achieving climate goals \cite{luderer_economic_2013}\cite{heuberger_impact_2018}\cite{victoria_early_2020}.
Therefore, uncertainties must be included in planning the energy transition to identify strategies that are robust against the unknown future. 

Strategic energy planning often relies on energy system optimization models.
These models consider uncertainties by either intrusive or non-intrusive methods:
Intrusive methods, like stochastic \cite{birge_introduction_2011} and robust optimization \cite{bertsimas_theory_2011}\cite{moret_decision_2020}, include parametric uncertainties in the mathematical formulation of the optimization problem. 
However, computational tractability (stochastic optimization) and difficult-to-interpret parameters (e.g., defining decision-makers' risk-aversion in $\Gamma$-robustness \cite{brown_satisficing_2009}) limit the applicability of these intrusive approaches.
Non-intrusive methods vary the input parameters according to their probability distribution and independently compute the model's response for each realization of input parameters (i.e., a \emph{scenario}). Optimizing the system for each sampled scenario often results in a large space of possible designs that can overwhelm decision-makers \cite{baader_streamlining_2023}. Because of these challenges in both methods, uncertainty is to date seldom included in energy system studies \cite{yue_review_2018}, leaving decision-makers with the question: 
\newline \textit{``What is the best strategy to choose today in the face of a highly uncertain future?''}
\newline This paper solves this problem and identifies strategies for a highly uncertain future by using the concept of \emph{regret}
\cite{savage_theory_1951}. Regret assesses the additional cost when making a wrong choice and has been observed to be a naturally used criterion in decision-making \cite{bell_regret_1982}\cite{loomes_regret_1982}. Regret has found application in multiple disciplines, e.g., finance \cite{arisoy_investor_2024}, climate policy \cite{rezai_climate_2017}, and medicine \cite{connolly_regret_2005}.
Despite the importance of regret for decision-making, in particular for renewable energy systems \cite{boeri_importance_2017}, the concept has not yet been widely adopted in the field of energy system planning. While some studies \cite{mersch_impact_2023}\cite{klemm_potential-risk_2024}, use the terms ``no/low-regret'' to describe must-do choices in energy system planning, they identify these choices without explicitly quantifying the regret. In other cases, minimax regret has been applied as a decision criterion, e.g., for greenhouse gas abatement strategies \cite{loulou_minimax_1999}\cite{li_interval-valued_2011}, for the energy system of Accra in Ghana \cite{yazdanie_strengthening_2024}, and for European hydrogen supply chains \cite{ganter_minimum-regret_2024}. However, these studies consider only a handful of different input scenarios and thus do not cover the full range of uncertainties faced in the energy transition.

This work introduces a quantitative framework to identify low-regret strategies in energy systems planning for a large number of future uncertainties. The method first explores the design space for the energy system, then automatically identifies potential strategies, and finally, quantifies the performance of these strategies in terms of regret and analyzes their sensitivity. Visualizing the strategies in decision trees, cumulative regret curves, and decision maps makes the otherwise often overwhelming outcomes of uncertainty studies accessible to decision-makers. 

We apply our framework to investigate the optimal biomass allocation under uncertainty for a net-zero Switzerland in 2050. We focus on biomass as it takes on a central role in the transition towards net-zero energy systems due to its removal of CO$_{2}$ from the atmosphere during growth. Today, the largest fraction of biomass is used for low-temperature heat applications such as residential heating and cooking \cite{iea_net_2023}. However, these applications can efficiently be electrified, freeing biomass for other uses. The favorable properties of biomass - low cost, regional availability, and potential for negative emissions \cite{millinger_diversity_2025} - together with its versatility \cite{colla_optimal_2022} make it an attractive resource across various energy sectors, including aviation \cite{bergero_pathways_2023}, the chemical industry \cite{meys_achieving_2021}, combined heat and power (CHP) \cite{ozolina_can_2022}, hydrogen \cite{pal_review_2022}, biomethane \cite{hamelin_harnessing_2021}, biochar \cite{cha_production_2016}, and process heat \cite{lenz_status_2020}. Sector-specific studies often identify biomass as the optimal choice. 
Moreover, biomass can provide system services, such as reducing import dependency \cite{gokcol_importance_2009}, securing the power supply amidst rising intermittent renewables \cite{hahn_review_2014}, and storing energy seasonally \cite{converse_seasonal_2012}. However, biomass cannot fulfil all these roles simultaneously due to its limited sustainable supply \cite{burg_analyzing_2018}. This limitation leads to strong competition for biomass use, necessitating a holistic approach to identify the optimal allocation of biomass, especially when factoring in the large uncertainties associated with long-term energy systems planning.

For the case study, we apply the energy system model EnergyScope \cite{limpens_energyscope_2019} to Switzerland, extending it by including additional biomass conversion and circular plastic pathways. Dedicated energy crops are excluded due to their controversial climate mitigation potential \cite{lark_environmental_2022} and negative impacts on biodiversity \cite{nunezregueiro_effects_2021} and food security \cite{muscat_battle_2020}. Moreover, biomass imports are prohibited to avoid burden shifting to other countries. Instead, we focus on the four domestic sustainable biomass types: woody biomass residues, wet biomass, animal manure, and sewage sludge.
We consider uncertainties in the demands, discount rate, technologies (investment and maintenance costs), and resources (availability and cost). We sample a set of 1000 scenarios to which we apply our method of strategy identification and regret quantification. Our findings reveal low-regret and must-avoid pathways for biomass use in the energy transition.

\section{Results}

\subsection{Large variations in biomass allocation across scenarios}

The energy system fully utilizes the available biomass potential (27.5 - 31.3 TWh/y) in all scenarios, underscoring biomass's important role in achieving a net-zero system. However, the biomass allocation varies widely across scenarios.

\begin{figure}[h]
    \includegraphics[width=\textwidth]{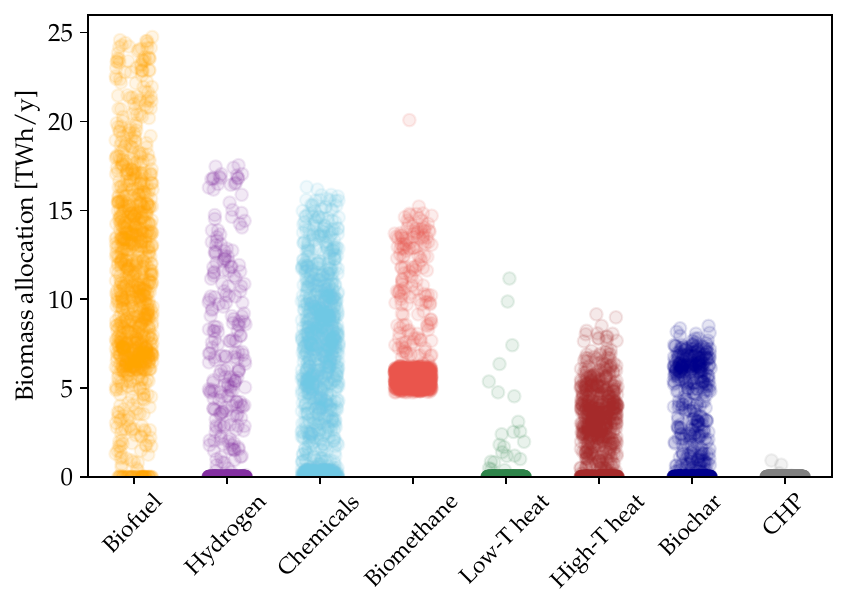}
    \caption{Scatterplot showing the distribution of biomass flow across all 1000 scenarios into different usage options (i.e., \emph{outputs of interest}) of the energy system. The scatterpoints of each output of interest are randomly spread along the x-axis to enhance visualization. The outputs of interest are ordered from left to right based on their range of observed values across the scenarios.}
    \label{fig: scatterplot}
\end{figure}

We assess the allocation of biomass across the 1000 scenarios in eight \emph{outputs of interest}: residential heat (Low-T heat), process heat (High-T heat), combined heat and power (CHP), biomethane, biochar, biofuels, chemicals, and hydrogen (Fig.~\ref{fig: scatterplot}). These outputs of interest represent different biomass usage options, measured in terms of allocated energy per year.

The conversion of biomass into biofuels shows the largest variation of allocation across scenarios, ranging from 0 to 24.8 TWh per year. The liquid fuel demand almost exclusively stems from the aviation sector (over 99.7\%), to which biofuels contribute, on average, 26\%. 

Hydrogen production exhibits the second largest range (0 – 17.5 TWh/y) of biomass allocation. We find that the endogenous demand for hydrogen in the energy system varies widely (3.1 - 59.3 TWh/y) with an average of 20.1 TWh/y. However, biomass provides, on average, only 4.2\% of the hydrogen demand.

For the production of chemicals, up to 16.3 TWh/y of biomass serve as carbon feedstock. The largest share of chemicals is subsequently converted into plastics. Across scenarios, recycling dominates the carbon feedstock for plastics with a share of 38.6\%, followed by biomass (33.4\%), fossil imports (26.4\%), and carbon capture and utilization (1.7\%).

Biomass allocated for biomethane production ranges from 4.8 to 15.2 TWh/y, with one outlier above 20 TWh/y. The minimum allocation of 4.8 TWh/y is due to a mandatory anaerobic digestion step for wet biomass, animal manure, and sewage sludge. On average, 45\% of the methane used in the energy system is biogenic.

Allocations for process heat (up to 9.2 TWh/y) and biochar (up to 8.5 TWh/y) remain relatively low. Furthermore, biomass is scarcely used for low-temperature heat, exceeding 1 TWh/y in less than 2\% of the scenarios. Combined heat and power receives negligible biomass allocation, even when integrating carbon capture.

While these results provide some insight into potential biomass allocations, developing a clear strategy for biomass use remains challenging due to the diversity of available options.

\subsection{Interpretable biomass strategies derived from decision tree}
\label{subsec:2A}

\begin{figure}
    \includegraphics[width=0.92\textwidth]{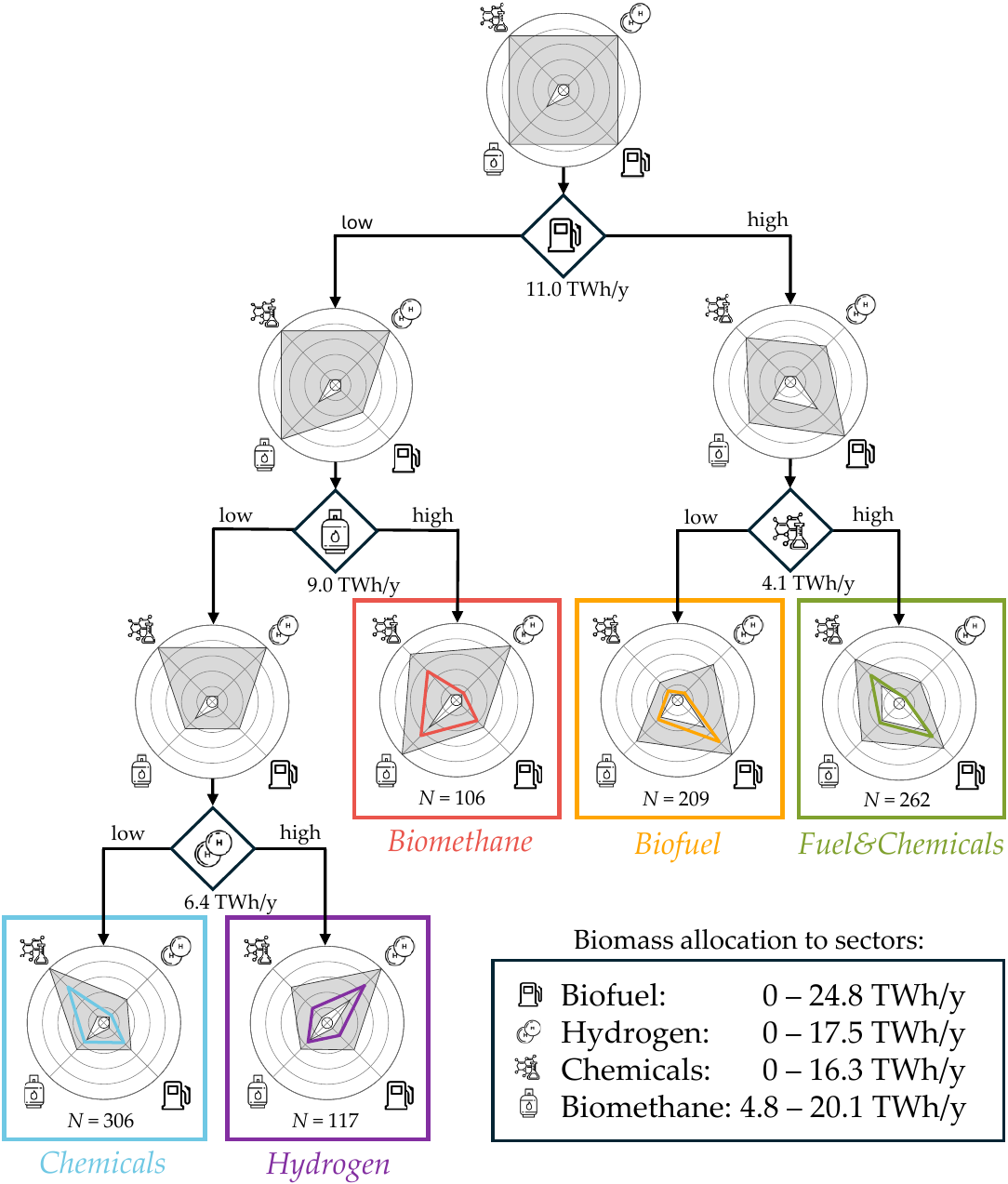}
    \caption{Decision tree applied to the space of the outputs of interests. Decisions are indicated by the symbols in the tilted squares, with the corresponding split given below each box. The axes of the radar plots show the biomass allocation to the specific sectors. Each axis is normalized to the maximum observed value of the respective output of interest.
    The grey-shaded area in each node indicates the range between the minimum and maximum observed value of each output of interest. Each $N$ represents the number of scenarios contained in a leaf node.
    The leaf nodes of the decision tree return five strategies for the use of biomass. The solid colored line indicates the average value of the outputs of interests in the leaf node. We name the strategies by the biomass allocation that distinguishes them from the other leaves. Hence, from now on, we refer to the five strategies as \emph{Chemicals}, \emph{Hydrogen}, \emph{Biomethane}, \emph{Biofuel}, and \emph{Fuel\&Chemicals}  strategy.}
    \label{fig: strategy_tree}
\end{figure}

To address this challenge, we cluster the eight outputs of interest across the scenarios and train a decision tree (Fig.~\ref{fig: strategy_tree}), following the approach proposed by Baader et al. \cite{baader_streamlining_2023}. 

The visualization focuses on the four outputs of interest with the widest range of biomass allocation across scenarios: biofuel, hydrogen, chemicals, and biomethane.
The radar plot at the top of the tree shows the full decision space representing the various possibilities for biomass utilization (as in Fig.~\ref{fig: scatterplot}). Starting from this point, the decision tree sequentially narrows the decision space by making decisions on the outputs of interest, unveiling the trade-offs in the allocation of biomass to competing sectors.
For instance, the first decision determines whether to allocate more or less than 11.0 TWh/y of biomass for biofuel production. Allocating more than 11.0 TWh/y restricts the maximum biomass allocation for other outputs of interest, whereas allocating less than 11.0 TWh/y leaves greater flexibility for the allocation to the production of hydrogen, biomethane, and chemicals.

The tree identifies five leaf nodes, corresponding to five main strategies for biomass allocation: \emph{Chemicals}, \emph{Hydrogen}, \emph{Biomethane}, \emph{Biofuel}, and \emph{Fuels\&Chemicals}. We name the strategies based on their specific emphasis on one or more outputs of interest that distinguish them from the other strategies.
\newpage

\subsection{Cumulative regret curves enable risk-aware choice of strategy}
\label{subsec:2B}

To assess the performance of biomass allocation strategies under varying future conditions, we apply the concept of regret. Regret is the difference in cost between selecting a suboptimal strategy and the optimal strategy for a given scenario (Eq.~\ref{eq: regret}). 

\begin{figure}[h]
    \includegraphics[width=\textwidth]{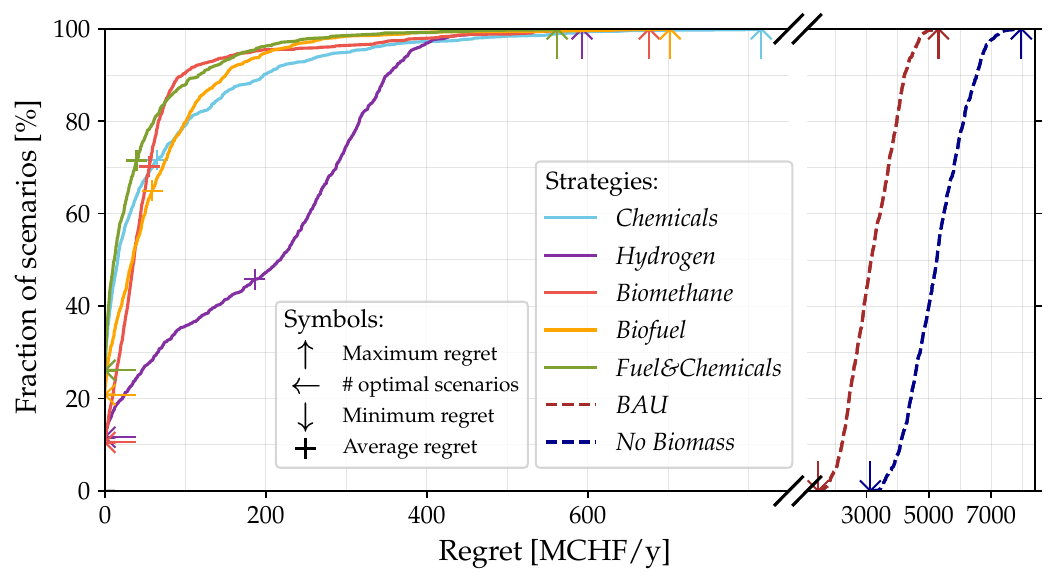}
    \caption{The cumulative regret curves display the regret of the biomass strategies across all 1000 scenarios. Leftward arrows highlight the y-intercept, indicating the number of scenarios in which a strategy is the optimal choice. Downward arrows mark the minimal regret for strategies that are never optimal, while upward arrows show the maximum regret for each strategy. Crosses represent the average regret across all scenarios. The value-at-risk \emph{VaR}$_{\alpha}$ for a strategy is determined by selecting a confidence level $\alpha\in[0,100]$ on the y-axis and identifying the corresponding regret value on the x-axis. This regret value represents the level that will be exceeded in (100-$\alpha$)\% of scenarios. Solid lines represent the cumulative regret of strategies derived from the decision tree (Fig.~\ref{fig: strategy_tree}), while dashed lines show the regret curves for the additionally evaluated \emph{Business-as-Usual} (\emph{BAU}) and \emph{No Biomass} strategies.}
    \label{fig: regret_curve}
\end{figure}

To compare strategies, we construct cumulative regret curves, which show the regret of each strategy across all scenarios (Fig.~\ref{fig: regret_curve}). The presented curves enable a direct comparison of strategies based on various decision-making criteria. Depending on the attitude to risk, stakeholders may apply different criteria \cite{mavromatidis_comparison_2018}, e.g., the number of optimal scenarios, minimum regret, average regret, maximum regret, and $\alpha$-th percentile of regret given by the value-at-risk \emph{VaR$_{\alpha}$} (Tab.~\ref{tab: regret table}). Notably, using expected regret as a decision criterion is mathematically equivalent to the commonly used expected costs (Eq.~\ref{avg_regret_equivalent_to_avg_cost}).

\begin{table}[h]
    \centering
    \caption{Summary of key results from the regret evaluation of biomass strategies. The best-performing strategy with respect to each criterion is highlighted by an asterisk(*).}
    \begin{tabular}{l c c c c c c}
        \toprule
        Strategy &  Optimal & Minimum & \emph{VaR}$_{50}$ & Average & \emph{VaR}$_{90}$ & Maximum\\
         &  scenarios & regret & & regret & & regret\\
        & [\%] & [MCHF/y] & [MCHF/y] & [MCHF/y] & [MCHF/y] & [MCHF/y]\\
        \midrule
        \emph{Fuel\&Chemicals} & 26.2* & 0*   & 11*  & 39*  & 115  & 562* \\
        \emph{Chemicals}       & 26.2* & 0*   & 16   & 65   & 197  & 816 \\
        \emph{Biofuel}         & 20.9  & 0*   & 34   & 59   & 150  & 702 \\
        \emph{Hydrogen}        & 11.7  & 0*   & 217  & 186  & 351  & 593 \\
        \emph{Biomethane}      & 10.6  & 0*   & 36   & 55   & 96*  & 677 \\
        \emph{BAU}             & 0     & 1454 & 3155 & 3206 & 4215 & 5310 \\
        \emph{No Biomass}      & 0     & 3127 & 5254 & 5268 & 6532 & 7963 \\
        \bottomrule
    \end{tabular}
    \label{tab: regret table}

\end{table}

In the regret analysis, the \emph{Fuel\&Chemicals} strategy stands out, performing best across all decision criteria except for the \emph{VaR$_{90}$}, where it ranks second. The \emph{Biofuel}, \emph{Chemicals}, and \emph{Biomethane} strategies also perform competitively, with only marginally higher average regret. Among the three, \emph{Chemicals} has a greater number of scenarios with no or low regret but also exhibits a longer tail towards higher regrets, resulting in the highest maximum regret. 
Although the \emph{Hydrogen} strategy has the second-lowest maximum regret, its average regret is more than four times higher than that of \emph{Fuel\&Chemicals}, making it a less favorable option.
Notably, the ranking order of strategies substantially depends on the choice of the decision criterion.

Additionally, we evaluate two baseline strategies: First, \emph{Business-as-Usual} (\emph{BAU}) representing the current utilization of biomass in Switzerland \cite{prognos_ag_energieperspektiven_2021}, where 19 TWh/y of biomass are primarily used for low-temperature heat and CHP. Second, \emph{No Biomass}, where biomass is excluded entirely. Both reference strategies result in substantially higher regret compared to the strategies derived from the decision tree. 
 
In terms of total annual system costs, \emph{BAU} leads to cost increases up to 13\%, while \emph{No Biomass} results in even higher increases up to 20\%, both corresponding to additional costs of several billion CHF per year for the energy system.

\subsection{Policymakers' decisions influence the regret of biomass strategies}
\label{subsec:2C}

\begin{figure}[h]
    \centering
    \includegraphics[width=0.95\textwidth]{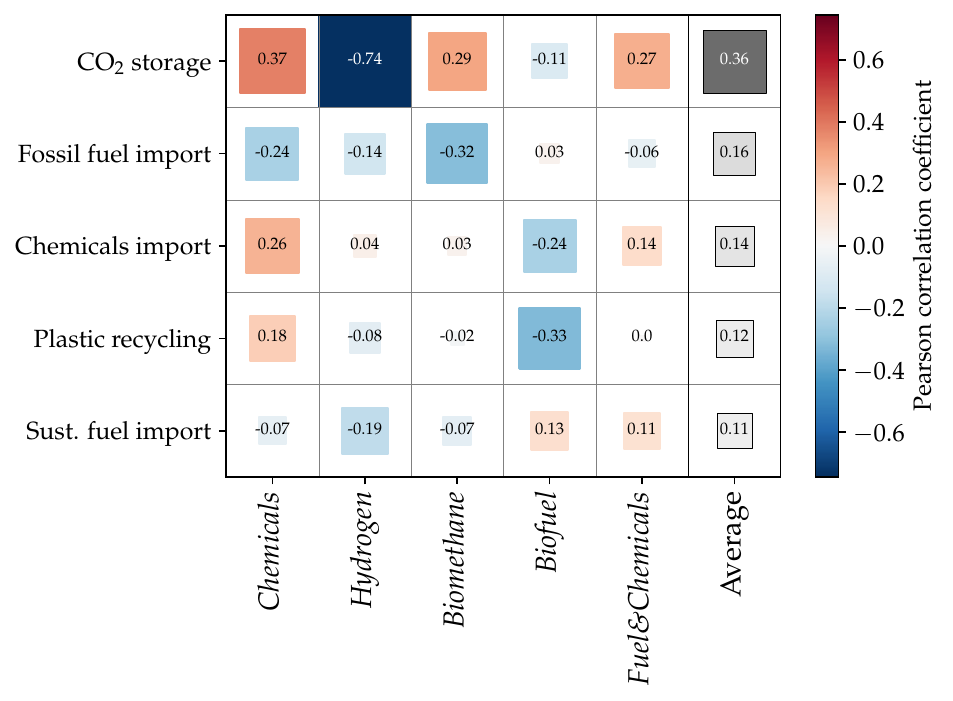}
    \caption{The Pearson correlation coefficients indicate the sensitivity of different strategies' regret to variations in input parameter values. Positive correlations (red) suggest that higher input parameter values increase regret, whereas negative correlations (blue) imply that higher values reduce regret. The five input parameters with the highest correlations (ordered by the average of absolute values across all strategies) are displayed for the five strategies derived from the decision tree.} 
    \label{fig: correlation}
\end{figure}

To understand the drivers of regret for each biomass strategy, we analyze its sensitivity with respect to the uncertain input parameters. Specifically, we measure the Pearson correlation between the regret values and the input parameters (Fig.~\ref{fig: correlation}).
By averaging the absolute values of the correlations over the different strategies, we identify the five parameters that are influencing regret most strongly. 
Overall, the correlation values remain low, indicating that regret emerges from interactions between multiple factors rather than being driven by individual parameters.

The availability of permanent CO$_2$ storage emerges as the most influential factor across strategies, highlighting the critical role of biogenic carbon. The CO$_2$ storage availability particularly affects the \emph{Hydrogen} strategy since converting biomass to hydrogen releases the biogenic carbon as CO$_2$ instead of utilizing it. When CO$_2$ storage availability is low, the excess CO$_2$ must either be re-combined with the produced hydrogen to synthesize methane or fuel, which is less efficient than producing them directly from biomass or be offset at a high price outside of Switzerland. Both options lead to high regrets for the \emph{Hydrogen} strategy.

Beyond CO$_2$ storage, limited availability of fossil and sustainable liquid fuel imports drive biofuel demand, thereby increasing regret for strategies with restricted biofuel production. Similarly, low recycling rates and limited chemical feedstock imports increase reliance on biomass as a carbon source for the chemical sector.

Among all strategies, \emph{Fuel\&Chemicals} exhibits the lowest sensitivity to any single input parameter, thus indicating robustness. 
Interestingly, the most influential parameters are not directly related to biomass but are instead factors that can be actively shaped by policymakers. For example, increasing the plastic recycling rate could be achieved through public awareness campaigns \cite{grodzinska-jurczak_effects_2006} and improvements in waste collection management \cite{andrews_comparison_2013}.

\subsection{Decision maps help to navigate decision-making under uncertainty}

\begin{figure}[h]
\centering
    \includegraphics[width=0.81\textwidth]{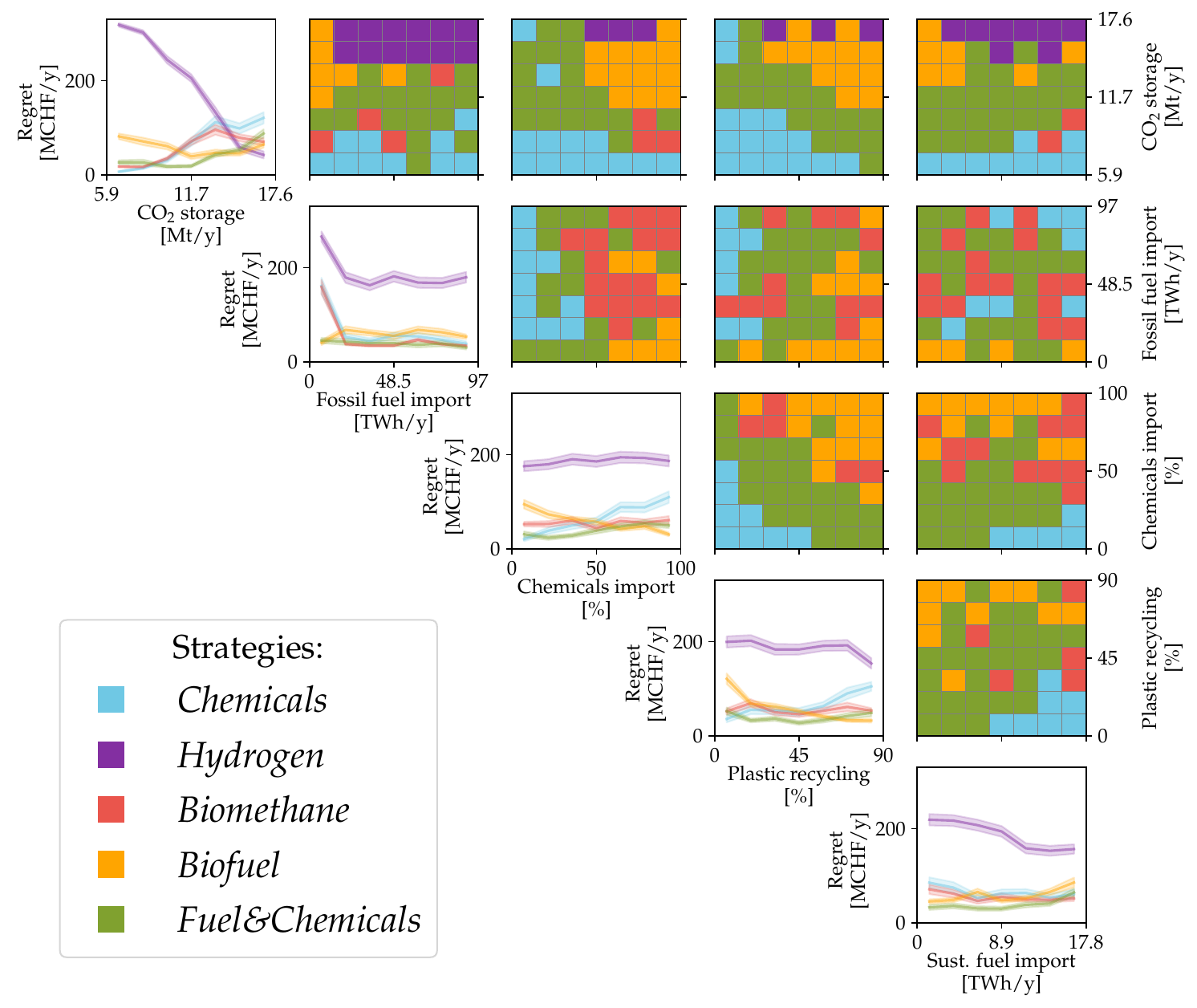}
    \caption{The figure illustrates the dependence of regret for the biomass strategies on the five most influential parameters. The diagonal panels display the dependence on individual parameters, while the upper triangular panels present decision maps for pairwise parameter combinations. These decision maps highlight the conditions under which one strategy outperforms the others, with the color of a pixel representing the strategy with the lowest mean regret under the given conditions. Each parameter range is divided into $n=7$ equally sized bins, resulting in an average of $N/n = 1000/7 \approx 143$ scenarios per bin and $N/(n\times n) = 1000/49 \approx 20.4$ scenarios per pixel. In the diagonal panels, the mean regret (solid line) and the standard error of the mean (shaded area) are shown for each bin.}
    \label{fig: regret_map}
\end{figure}

To provide decision-makers with actionable insights, we further analyze the relation between inputs and the regret of strategies, focusing on the five most influential input parameters previously identified (Fig.~\ref{fig: correlation}). We visualize these relations in decision maps, which highlight the strategy with the lowest regret under given conditions (Fig.~\ref{fig: regret_map}). These maps serve as a practical tool for navigating uncertainty. For instance, if decision-makers estimate, based on their expert judgment, that a parameter is unlikely to exceed a given threshold, e.g., recycling rates $>$ 70\%, they can exclude the corresponding area from consideration and focus their assessment on the remaining part of the parameter space.

We find that the \emph{Fuel\&Chemicals} strategy achieves the lowest regret most frequently, particularly when parameters remain close to the center of the uncertainty ranges. In contrast, the \emph{Hydrogen} strategy only emerges as the lowest-regret option when availability for permanent CO$_2$ storage is high. This finding underscores the necessity of establishing large-scale CO$_2$ storage infrastructure if hydrogen production from biomass is pursued. Conversely, when CO$_2$ storage is limited, the \emph{Chemicals} strategy becomes the dominant choice, as it effectively utilizes the carbon from biomass feedstock rather than producing CO$_2$.

High recycling rates and abundant chemical imports favor the \emph{Biofuels} strategy, as reduced demand for carbon feedstock in the chemical sector allows for the use of biomass for biofuel production instead, mitigating aviation emissions. 
\emph{Biomethane} emerges as the lowest-regret strategy when imports of chemicals and liquid fuels are available.

Moreover, the regret associated with \emph{Biomethane}, \emph{Chemicals}, and \emph{Hydrogen} increases sharply when the availability of imported liquid fossil fuels falls below 10 TWh/y. This sensitivity highlights a key risk for these strategies, which, however, policymakers can mitigate by diversifying trade partners and ensuring stable access to energy imports.\cite{mersch_energy_2024}

\section{Discussion}

This study introduces a regret-based decision-making framework to identify strategies for the energy system and quantify their performance under uncertainty.
By using decision trees, the framework simplifies the decision space and provides a structured and transparent approach to finding strategies. 
Evaluating the strategies with regret-based decision criteria is intuitive and allows for a risk-aware assessment. 
The sensitivity analysis reveals the most influential drivers of regret, providing insights into both opportunities and vulnerabilities of strategies.
Decision maps further 
offer policymakers a practical tool to navigate uncertainty by indicating the lowest-regret strategy under given circumstances.

Applying this framework to biomass allocation in Switzerland reveals that producing \emph{Fuel\&} \emph{Chemicals} is the most robust strategy across multiple decision-making criteria. We find that parameters which can be shaped by policymakers, such as CO$_2$ storage and fuel import availability, influence biomass allocation decisions more strongly than biomass technology costs or biomass resource prices. Given the framework's generalizability, it can be applied to investigate strategies for the energy system beyond biomass allocation, providing a tool for uncertainty-aware decision-making. 

\subsection{Limitations of the study}
Despite its advantages, the framework has limitations:  Evaluating several strategies across $N$ scenarios requires a large number of model runs, which can be challenging for computationally intensive models; however, as each scenario is optimized independently, computational time can be reduced substantially by parallelization. 

Another limitation lies in the representation of uncertainties. The study assumes uniform probability distributions and independent parameters, which does not fully reflect real-world conditions where correlations between uncertainties exist. Such correlations can significantly impact optimal decarbonization pathways \cite{rodriguez-matas_how_2025}. Future work could address this limitation by incorporating correlation structures, e.g., with copula-based methods \cite{lu_multi-objective_2024}. Nevertheless, the decision maps (Fig.~\ref{fig: regret_map}) already provide policymakers with the flexibility to weigh scenarios based on their own probability assessments. 

Defining strategies based on the decision thresholds from the decision tree, while intuitive, allows for substantial adaptation of a strategy during regret evaluation. Alternatively, a more rigid strategy definition could be used, such as fixing the outputs of interest to the exact values of a strategy representative (see SI Fig.~S1-S3, Tab.~S1-S2). 

\subsection{Biomass allocation in the context of previous work}
To put our findings into perspective, we compare them with existing research on biomass use in energy systems.
Previous studies have shown conflicting results for the production of biofuels. 
While some works on the European energy system suggest that biofuel mandates increase system costs \cite{millinger_are_2022}, others observe biofuel production in cost-optimal solutions \cite{millinger_diversity_2025}\cite{wu_strategic_2023}. 
Our case study finds biofuel production to be the dominant use of biomass across scenarios, primarily due to the demand for sustainable aviation fuels. 
This result aligns with prior research on Switzerland’s energy system, which found that a substantial amount of biomass is dedicated to producing jet fuel \cite{li_decarbonization_2020}. 

Energy system models still often overlook non-energy demands despite their growing importance in achieving net-zero systems. 
Previous studies that include non-energy demands found at the European level that biomass is used for methanol and methane production only for scenarios with high biomass availability \cite{wu_strategic_2023}. Research on Belgium’s energy system observed that for more ambitious emission reduction targets, biomass is increasingly used for non-energy applications \cite{rixhon_integration_2022}.
We find that using biomass as a carbon feedstock for the chemical industry proves particularly effective when the availability for permanent CO$_2$ storage is low, plastic recycling rates are low, and chemical imports are limited.

Contrary to previous work that suggests a role for bioelectricity combined with carbon capture and storage in the European power system
\cite{millinger_diversity_2025}, our results show negligible biomass allocation for electricity generation in Switzerland. Switzerland’s high share of hydropower already provides sufficient flexibility to the power system, diminishing the need for dispatchable biomass-based electricity generation.

The continued use of biomass for low-temperature heat, as represented by the \emph{BAU} strategy, leads to consistently higher regret compared to alternative strategies. Previous studies on Switzerland’s energy system support this finding, suggesting that low-temperature heat is not an optimal use of biomass in the long-term \cite{codina_girones_optimal_2017}\cite{codina_girones_assessment_2018}. Thus, the continued use of biomass for low-T heat is a must-avoid option, as transitioning toward more efficient biomass use is inevitable to mitigate regret.

Our study quantifies the cost increases incurred when excluding biomass from the system, ranging from 8\% to 20\%, with an average increase of 13\%. Similar cost penalties have been reported for European-scale studies, with estimates ranging from 10\% to 20\% depending on the emission targets \cite{millinger_diversity_2025} and the reference biomass availability \cite{wu_strategic_2023}.

\subsection{Conclusion}

Overall, this study contributes to energy system planning by introducing a regret-based decision-making framework which identifies robust strategies and quantifies their performance under uncertainty. The framework enables a comprehensive assessment of regret drivers, helping policymakers evaluate trade-offs between strategies.

Applying this framework to Switzerland’s energy system identified five potential biomass allocation strategies and assessed their performance across diverse future scenarios. Our findings show the conditions under which different strategies become favorable, offering policymakers insights for long-term decision-making. 
While our results largely align with previous studies on biomass allocation in the energy system, our framework deepens the understanding of risks for regret and conditions for optimality for biomass strategies.

\section{Methods}

The energy system designs in our study are obtained by solving a linear optimization problem. The objective of the problem is to minimize a linear cost function $f$:
\begin{align}
    \begin{split}
        \min_{\boldsymbol{x}} \,& f(\boldsymbol{x}, \boldsymbol{\theta}) \\
        \mathrm{s.t.} \,\,\,&\boldsymbol{c}(\boldsymbol{x}, \boldsymbol{\theta}) \leq 0 \\
    \end{split}
    \label{eq: optimization problem}
\end{align} over the vector of decision variables \( \boldsymbol{x} \) (e.g., investment and operation decisions for technologies), given the parameters $\boldsymbol{\theta} $ (e.g., technology cost and resource prices/availabilities), subject to a set of linear constraints $\boldsymbol{c}$ (e.g., energy balance, operational limits of technologies, etc.). The solution to the problem is a vector of decision variables \( \boldsymbol{x}^* \) for which $f$ returns the minimal cost: 

\begin{equation}
    C^{\mathrm{opt}} = f(\boldsymbol{x}^*, \boldsymbol{\theta}).
\end{equation}

\subsection*{Generation of scenarios energy system designs}
Many parameters, such as prices and demands, are uncertain. We model the uncertain parameters $\boldsymbol{\theta}$ as uniformly distributed between a lower bound $\underline{\boldsymbol{\theta}}$ and an upper bound $\overline{\boldsymbol{\theta}}$, i.e., 

\begin{equation}
    \underline{\boldsymbol{\theta}}\leq \boldsymbol{\theta} \leq\overline{\boldsymbol{\theta}}.
\end{equation}

To account for this uncertainty in the parameters, we employ an \emph{exploratory modeling} \cite{kwakkel_exploratory_2013} approach. Specifically, we use Latin Hypercube Sampling \cite{mckay_comparison_1979} to generate $N$ different realizations of $\boldsymbol{\theta}$. In the sampling, we assume that the input parameters are independent of each other. Thereby, we obtain a set of scenarios $\Omega$

\begin{equation}
    i \in \Omega = \{\boldsymbol{\theta}_1,\, ...\,, \,\boldsymbol{\theta}_N \}.
\end{equation}

By solving the optimization problem (Eq.~\ref{eq: optimization problem}) for each scenario $i$, a set of optimal decision variables $\mathcal{X}^*$

\begin{equation}
    \boldsymbol{x}^*_i \in \mathcal{X}^* = \{\boldsymbol{x}_1^*,\, ...\,, \,\boldsymbol{x}_N^* \}
\end{equation}

with their respective optimal costs $\mathcal{C}^{\mathrm{opt}}$

\begin{equation}
    C^{\mathrm{opt}}_i \in \mathcal{C}^{\mathrm{opt}} = \{ C^{\mathrm{opt}}_1,\, ...\,, \, C^{\mathrm{opt}}_N \}
\end{equation}
is computed.

\subsection{Identification of strategies}

An energy system optimization model typically has of the order of $|\boldsymbol{x}|\approx10^6$ decision variables, where $|\cdot|$ represents the cardinality of the vector.
Therefore, a vector of \emph{outputs of interest} $\boldsymbol{y}$ must be chosen such that it captures the system of interest, here: biomass allocation, but at the same time reduces dimensionality significantly, i.e., $|\boldsymbol{y}| \ll |\boldsymbol{x}|$. The outputs of interest are calculated as function $h$ of the decision variables $\boldsymbol{x}$:
\begin{align}
    \begin{split}
        &\boldsymbol{y} = h(\boldsymbol{x})\\
    \end{split}
\end{align} 
To keep the optimization problem for the regret evaluation (Eq.~\ref{eq: optimization problem strategy}) linear, $h$ must also be linear. Evaluating the set of optimal decision variables $\mathcal{X}^*$ under $h$, gives the set of outputs of interests $\mathcal{Y}^*$:

\begin{equation}
    \boldsymbol{y}^*_i \in \mathcal{Y}^* = \{\boldsymbol{y}^*_1,\, ...\,, \,\boldsymbol{y}^*_N \}.
\end{equation}

Based on this set, $k$ different strategies can be defined:

\begin{equation}
    s \in \mathcal{S} = \{S_1,\, ...\,, \,S_k \}.
\end{equation}

We recommend choosing $k< 10$ strategies, to avoid overwhelming decision-makers with too many options \cite{cowan_magical_2010}.
We define a strategy by restricting the output of interests to bounds $\underline{\boldsymbol{y}}^s$, $\overline{\boldsymbol{y}}^s$ that represent the strategy $s$. We denote the space of outputs of interests that fulfills a given strategy $s$ by $\mathcal{Y}^s$:

\begin{equation}
    \mathcal{Y}^s = \{\boldsymbol{y} \,|\,\underline{\boldsymbol{y}}^s \leq \boldsymbol{y} \leq \overline{\boldsymbol{y}}^s\}.
\end{equation}
Note that in this work, the strategy defining bounds are obtained by training a decision tree on the set of outputs of interests $\mathcal{Y}^*$.

Another, more restrictive, strategy definition is to fix the outputs of interests to a vector $\boldsymbol{y}^s$ that is representative of the strategy $s$, i.e., $\underline{\boldsymbol{y}}^s = \boldsymbol{y}^s = \overline{\boldsymbol{y}}^s$.
The strategy representative $\boldsymbol{y}^s$ can be obtained, for example, by clustering the set of outputs of interests $\mathcal{Y}^*$ and then selecting the average/centroid of each cluster.

\subsection{Regret calculation}

Our methodology aims at evaluating each strategy's performance over the whole set of scenarios $\Omega$.
Therefore, we modify the optimization problem (Eq.~\ref{eq: optimization problem}) to enforce a strategy $s$ by adding its defining constraints $\mathcal{Y}^s$:

\begin{align}
    \begin{split}
        \min_{\boldsymbol{x}} \,& f(\boldsymbol{x}, \boldsymbol{\theta}) \\
        \mathrm{s.t.} \,\,\,&\boldsymbol{c}(\boldsymbol{x}, \boldsymbol{\theta}) \leq 0 \\
        \,\,\,\,\,\,\,\,\,\,\,\, &\boldsymbol{y}\in \mathcal{Y}^s.
    \end{split}
    \label{eq: optimization problem strategy}
\end{align}

Solving the new problem gives the decision variables $\boldsymbol{x}^{s,*}$ that minimize the system cost $C^s$ while fulfilling the strategy defining constraints 

\begin{equation}
    C^s = f(\boldsymbol{x}^{s,*}, \boldsymbol{\theta}).
\end{equation}

Optimizing the new problem for all scenarios $i \in \Omega$, returns the set of system costs with strategy $s$ enforced:

\begin{equation}
    C^s_i \in \mathcal{C}^s = \{ C^s_{1},\, ...\,, \, C^s_N \}.
\end{equation}

We use regret as a measure to compare the different strategies, which is defined as the additional costs encountered when not choosing the optimal strategy, i.e., the regret $R^{s}_{i}$ for strategy $s$ in scenario $i$ is given by:

\begin{align}
    \label{eq: regret}
    \begin{split}
    &R^{s}_{i} = C^{s}_{i} - C^{\mathrm{opt}}_{i}\\ 
    &\mathrm{with}\,\,\, R^{s}_{i} \geq 0.
    \end{split}
\end{align}

By construction, $R^{s}_{i}$ is always non-negative, as the optimization problem without the strategy constraints (Eq.~\ref{eq: optimization problem}) performs equally or better than the problem with strategy constraints (Eq.~\ref{eq: optimization problem strategy}).

Notably, using average regret $\left<R^s\right>$ as a decision-making criterion to compare strategies is equivalent to using the average cost $\left<C^s\right>$ criterion as the average of the optimal costs $\left<C^{\mathrm{opt}}\right>$ is independent of the chosen strategy:
\begin{align}
    \begin{split}
    \left<R^s\right> &= \frac{1}{N} \sum^{N}_{i=1} R^s_{i} = \frac{1}{N} \sum^{N}_{i=1} \left( C^s_{i} - C^{\mathrm{opt}}_{i} \right) = \\ &=\frac{1}{N} \sum^{N}_{i=1} C^s_{i} - \frac{1}{N} \sum^{N}_{i=1} C^{\mathrm{opt}}_{i} = \\ &=\left<C^s\right> - \left<C^{\mathrm{opt}}\right>.
    \label{avg_regret_equivalent_to_avg_cost}
    \end{split}
\end{align}

\subsection{Model and case study}
We use EnergyScope TD \cite{limpens_energyscope_2019}, a comprehensive energy system model that integrates multiple sectors, including electricity, heat (low-, medium-, and high-temperature), mobility (public, private, and aviation), and non-energy demands (methanol, ammonia, high-value chemicals (HVC), plastic, and cement). We extended the model to include further biomass conversion pathways, as well as circular plastic pathways. We optimize the energy system with 12 typical days, each with an hourly resolution. The spatial resolution is a single-node system, representing Switzerland as an energy entity, but allowing for imports from outside.

Our case study focuses on a net-zero Switzerland in 2050, assuming a complete phase-out of nuclear power. To represent our system of interest, we measure the amount of biomass allocated to different technologies in units of energy per year. Furthermore, we group these technologies into one of eight potential biomass usage sectors: Low-T heat, High-T heat, CHP, biomethane, biochar, biofuels, chemicals, and hydrogen (Tab.~\ref{tab:outputs_of_interest}). Our \emph{outputs of interest} are defined as the total allocated biomass within each sector over one year.
\begin{table}[h]
\centering
\caption{Grouping of biomass conversion technologies to sectors}

\begin{tabular}{ l l }
     \toprule
     Outputs of interest  &  Conversion technologies \\
     \toprule
     Biofuel & Biomass-to-liquid \\
            &  Hydrothermal liquefaction\\
     \midrule
     Biomethane &  Wood gasification to methane \\
          &      Hydrothermal gasification \\
          &      Anaerobic digestion of manure \\
           &     Anaerobic digestion of wet biomass \\
          &      Waste water treatment plant\\
     \midrule
     Hydrogen &  Wood gasification to hydrogen \\
     \midrule
     Low-T heat &    Wood-fired boiler\\
     \midrule
     High-T heat &   Industrial wood furnace \\
         &   Industrial digestate furnace\\
     \midrule
     CHP &      Wood-fired combined heat and power \\
         &      Digestate-fired combined heat and power \\
         &      Manure-based biogas plant\\    

     \midrule
     Chemicals &  Biomass-to-methanol\\
                & Biomass-to-HVC \\
     \midrule
     Biochar &  Slow pyrolysis of wood\\
             &  Hydrothermal carbonization \\
     \bottomrule
     \label{tab:outputs_of_interest}
\end{tabular}
\end{table}

Based on the methodology proposed in Baader et al. \cite{baader_streamlining_2023}, we train a decision tree (Fig.~\ref{fig: strategy_tree}) using these outputs of interest. Since all outputs share the same unit, no normalization is required. Strategies are then defined by the set of decisions leading to each leaf node in the decision tree. 

By construction, the leftmost strategy of the tree is defined solely by upper bounds on biomass usage, without explicitly directing biomass to any specific sector. However, we observe that this strategy predominantly focuses on the production of chemicals from biomass. To ensure a consistent comparison with the other strategies, we enforce a lower bound on the allocation of biomass to produce chemicals in the \emph{Chemicals} strategy. 
Specifically, we set the threshold at the mean minus one standard deviation of the biomass use for chemicals in the leaf node. Thereby, we retain 86\% of the designs previously included in the leaf.

In addition to the strategies identified by the decision tree, we analyze a \emph{Business-as-Usual}\cite{prognos_ag_energieperspektiven_2021} (\emph{BAU}) strategy and a scenario in which biomass usage is entirely excluded (\emph{No Biomass}). Tab.~\ref{tab: strategy constraints} summarizes the defining constraints for all seven strategies.

\begin{table}[h]
    \centering
    \caption{Defining constraints for the strategies}
    \begin{tabular}{ l r }
        
         \toprule
         
        Strategy name & Defining constraints  \\
         \toprule
         \emph{Biofuel} & Biofuel $\geq$ 11.0 TWh/y \\
                 & Chemicals \textless \,\,\,\ 4.1  TWh/y \\
         \midrule
         \emph{Fuel\&Chemicals} &  Biofuel $\geq$ 11.0 TWh/y \\
                    &    Chemicals $\geq$ \, 4.1 TWh/y\\
         \midrule
         \emph{Biomethane} &   Biofuel \textless \, 11.0 TWh/y \\
                 &      Biomethane $\geq$ 9.0 TWh/y\\
         \midrule
         \emph{Hydrogen} &   Biofuel \textless \, 11.0 TWh/y \\
                  &      Biomethane \textless \, 9.0 TWh/y\\
                  &      Hydrogen $\geq$ \, 6.4 TWh/y\\
         \midrule
         \emph{Chemicals} &   Biofuel \textless \, 11.0 TWh/y \\
                   &     Biomethane \textless \, 9.0 TWh/y\\
                   &     Hydrogen \textless \,\,\,\, 6.7 TWh/y\\
                   &     Chemicals\, $\geq$ \, 6.5 TWh/y\\
         \midrule
         \emph{Business-as-Usual}  &  Biofuel \,=\,\,\,  0.3 TWh/y \\
          (\emph{BAU})      &    Biomethane \,=\,\,\,  1.4 TWh/y\\
                    &    Hydrogen \,=\,\,\,\,\,\,\,\,  0 TWh/y\\
                    &    Chemicals \,=\,\,\,\,\,\,\,\,  0 TWh/y\\
                    &    Low-T heat \,=\,\,\,  7.7 TWh/y\\
                    &    Biochar \,=\,\,\,\,\,\,\,\,  0 TWh/y\\
                    &    High-T heat \,=\,\,\,  3.1 TWh/y\\
                    &    CHP \,=\,\,\,  6.5 TWh/y\\
         \midrule
         \emph{No Biomass} &  Biofuel \,=\,\,\,\,\,\,\,\,  0 TWh/y \\
                    &    Biomethane \,=\,\,\,\,\,\,\,\,  0 TWh/y\\
                    &    Hydrogen \,=\,\,\,\,\,\,\,\,  0 TWh/y\\
                    &    Chemicals \,=\,\,\,\,\,\,\,\,  0 TWh/y\\
                    &    Low-T heat \,=\,\,\,\,\,\,\,\,  0 TWh/y\\
                    &    Biochar \,=\,\,\,\,\,\,\,\,  0 TWh/y\\
                    &    High-T heat \,=\,\,\,\,\,\,\,\,  0 TWh/y\\
                    &    CHP \,=\,\,\,\,\,\,\,\,  0 TWh/y\\
         \bottomrule
    \end{tabular}
    \label{tab: strategy constraints}
\end{table}

\clearpage

\section*{Resource availabilty}

All data is documented in the Supplemental Information in Tables S3-S24 Figures S4-S100. 
Additionally, all original code and data associated with this study are publicly available on GitHub:\\ \url{https://gitlab.ethz.ch/epse/systems-design-public/low-regret-strategies}.

\section*{Acknowledgements}

S.M. and G.W. acknowledge support from the Swiss National Science Foundation (SNSF) under Grant No. PZ00P2 202117. A.B. acknowledges the support of the Swiss Federal Office of Energy through the SWEET consortium PATHFNDR under Grant No. SI/502259.

\section*{Author contributions}

Conceptualization, G.W., A.B., and S.M..; Methodology, G.W., N.N., F.B., A.B., and S.M.; Investigation, G.W.; Writing – Original Draft, G.W.; Writing – Review \& Editing, G.W., N.N., A.B., and S.M.; Funding Acquisition, A.B. and S.M.; Resources, A.B. and S.M.; Supervision,  S.M.

\section*{Declaration of interests}

A.B. served on review committees for research and development at ExxonMobil and TotalEnergies. A.B. and S.M. have ownership interests in firms that render services to industry, some of which may provide energy planning services. All other authors have no competing financial interests.

\newpage

\bibliography{references}

\bigskip

\end{document}


\maketitle

\tableofcontents

\newpage
\section{Supplementary Results}

To evaluate the robustness of our findings, we conduct an additional regret analysis using an alternative definition of strategies (Table S\ref{tab: alternative strategy definition}). In this formulation, each strategy is defined by the average biomass allocation within a given leaf node of the decision tree, rather than by the decision thresholds that lead to that node. This definition is more restrictive, as it removes the flexibility to adapt allocations during the regret evaluation process.

Specifying outputs of interest as absolute allocation values could lead to infeasibility, particularly when biomass availability is low. To avoid infeasibility, we instead define the outputs of interest as relative shares of the total used biomass. This approach guarantees feasibility across all scenarios. 

However, the new formulation based on relative shares may result in underutilization of the available biomass potential, as the predefined shares may not fully align with the supplied biomass types, i.e., a strategy defined by a low share of biomethane production cannot fully utilize all biomass in a scenario with large availability of biomass that must undergo an anaerobic digestion step (i.e.,  manure, sludge, green waste) and low availability of wood.

\begin{table}[h!]
\centering
\label{tab: alternative strategy definition}
\caption{Strategy definition based on the average of the leaf nodes. The values in the table give the share (in \%) of consumed biomass for an output of interest. The sum over each column returns 100\%, i.e., the total consumed biomass. Related to Table~3 in the main paper.}
\resizebox{\textwidth}{!}{%
\begin{tabular}{lccccc}
\toprule
& \emph{Fuel\&Chemicals\_avg} & \emph{Biomethane\_avg} & \emph{Hydrogen\_avg} & \emph{Chemicals\_avg} & \emph{Biofuel\_avg}  \\
\midrule
Biofuel     & 47.87 & 26.01 & 12.20 & 25.11 & 62.21 \\
Hydrogen    & 0.28  & 1.97  & 38.58 & 2.49  & 1.88  \\
Biomethane  & 19.27 & 42.04 & 19.00 & 19.48 & 19.59 \\
Chemicals   & 25.95 & 26.54 & 10.40 & 34.72 & 4.44  \\
HTH         & 4.62  & 2.70  & 12.13 & 7.75  & 6.41  \\
Biochar     & 1.99  & 0.73  & 7.24  & 9.89  & 5.35  \\
CHP         & 0.00  & 0.00  & 0.05  & 0.00  & 0.00  \\
LTH         & 0.02  & 0.00  & 0.41  & 0.55  & 0.14  \\
\bottomrule
\end{tabular}}
\end{table}

\begin{figure}[h]
    \includegraphics[width=\textwidth]{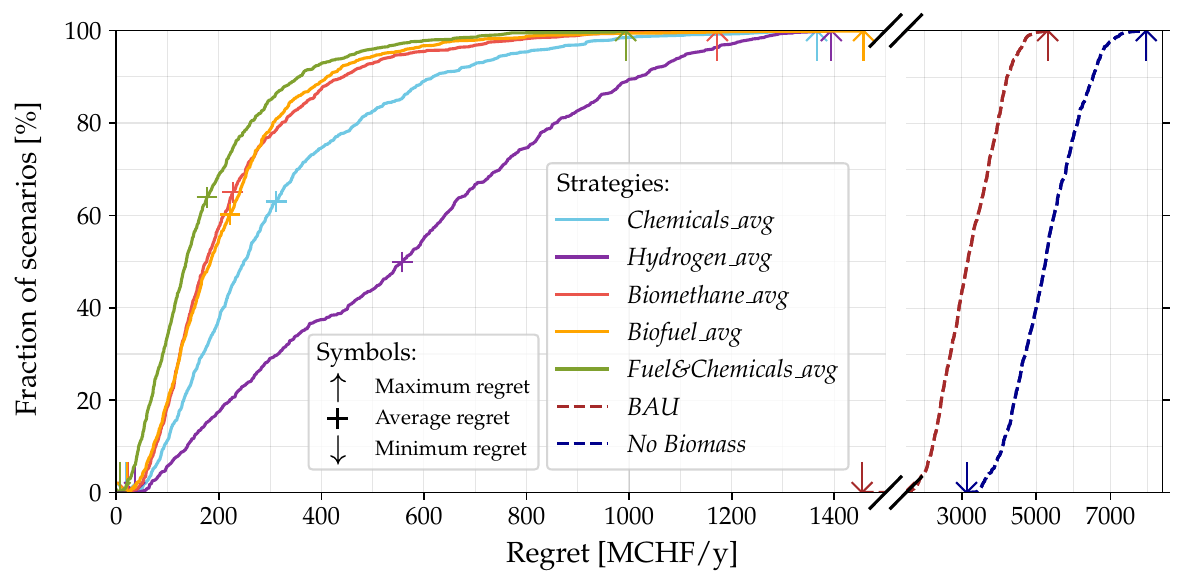}
    \caption{The cumulative regret curves display the regret of the biomass strategies across all 1000 scenarios. Downward arrows mark the minimal regret for strategies that are never optimal, while upward arrows show the maximum regret for each strategy. Crosses represent the average regret across all scenarios. The value-at-risk \emph{VaR}$_{\alpha}$ for a strategy is determined by selecting a confidence level $\alpha\in[0,100]$ on the y-axis and identifying the corresponding regret value on the x-axis. This regret value represents the level that will be exceeded in (100-$\alpha$)\% of scenarios. Solid lines represent the cumulative regret of strategies derived from the decision tree, while dashed lines show the regret curves for the additionally evaluated \emph{Business-as-Usual} (\emph{BAU}) and \emph{No Biomass} strategies. This graph for average-based strategies is related to Fig.~3 in the main paper, which reports the results for threshold-based strategies.} 
    \label{fig: regret_curve}
\end{figure}

As expected, the stricter strategy definition leads to several changes in the regret curves (Fig.~\ref{fig: regret_curve}) when compared to the threshold-based strategy definition: First, the strategies exhibit a non-zero minimum regret across all scenarios, as they cannot adapt to the optimal solution. Second, the average regret increases substantially, by a factor of three to five for each strategy (Tab.~\ref{tab: regret_table}). Third, the maximum regret doubles in comparison to the threshold-based definition.

However, overall, the findings confirm those presented in the main paper, Fig.~3, which used the threshold-based strategy definitions. Notably, the strategy \emph{Fuel\&Chemicals\_avg} continues to perform best and is even more dominant under the average-based strategy definition. It is followed by \emph{Biofuel\_avg} and \emph{Biomethane\_avg}, which perform similarly. In contrast, the \emph{Chemicals\_avg} strategy shows slightly weaker results compared to its threshold-based counterpart. Finally, the \emph{Hydrogen\_avg} strategy remains the least favorable of the biomass-based strategies, in line with the main paper’s conclusions.

\vspace{-20pt}

\begin{table}[h]
    \centering
    \caption{Summary of key results from the regret evaluation of biomass strategies. The best-performing strategy with respect to each criterion is highlighted by an asterisk(*). Related to Table 1. in the main paper but for average-based strategies.}
    \resizebox{\textwidth}{!}{%
    \begin{tabular}{l c c c c c}
        \toprule
        Strategy & Minimum & \emph{VaR}$_{50}$ & Average & \emph{VaR}$_{90}$ & Maximum \\
                 & regret  &                   & regret  &                   & regret \\
                 &  [MCHF/y] & [MCHF/y] & [MCHF/y] & [MCHF/y] & [MCHF/y] \\
        \midrule
        \emph{Fuel\&Chemicals\_avg} & 7*   & 136*  & 177*  & 354*  & 993* \\
        \emph{Chemicals\_avg}       & 19   & 248   & 312   & 617   & 1366 \\
        \emph{Biofuel\_avg}         & 21   & 183   & 222   & 411   & 1457 \\
        \emph{Biomethane\_avg}      & 23   & 176   & 227   & 441   & 1172 \\
        \emph{Hydrogen\_avg}        & 36   & 558   & 558   & 1016  & 1394 \\
        \emph{BAU}                  & 1454 & 3155  & 3206  & 4215  & 5310 \\
        \emph{No Biomass}           & 3127 & 5254  & 5268  & 6532  & 7963 \\
        \bottomrule
    \end{tabular}}
    \label{tab: regret_table}
\end{table}

\clearpage

\begin{figure}[h]
    \centering
    \includegraphics[width=\textwidth]{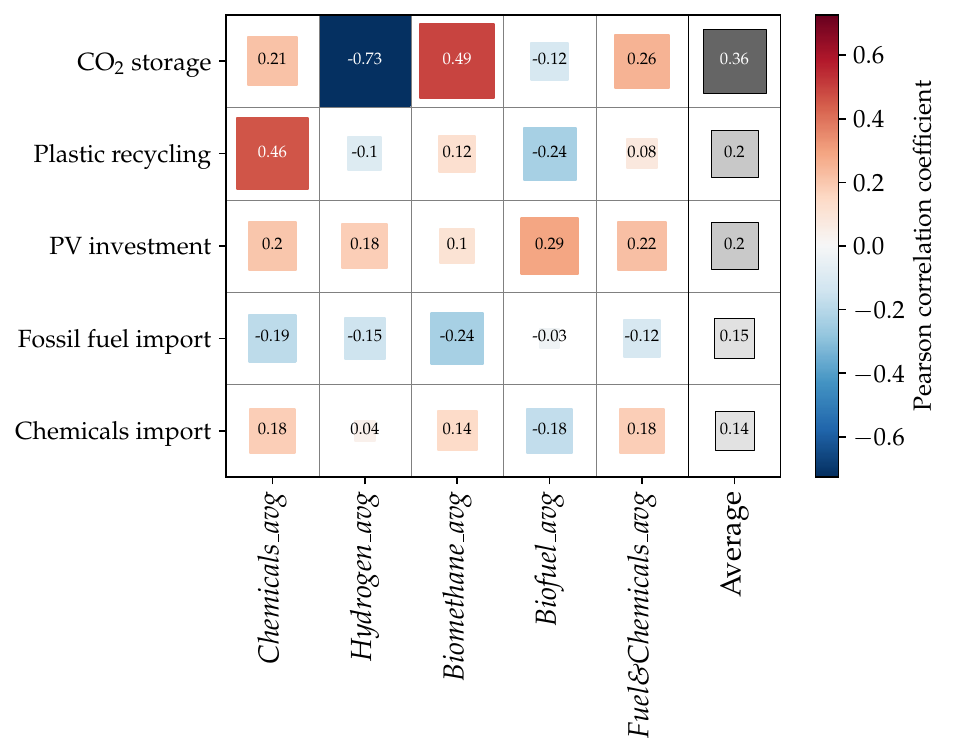}
    \caption{The Pearson correlation coefficients indicate the sensitivity of different strategies' regret to variations in input parameter values. Positive correlations (red) suggest that higher input parameter values increase regret, whereas negative correlations (blue) imply that higher values reduce regret. The five input parameters with the highest correlations (ordered by the average of absolute values across all strategies) are displayed for the five strategies derived from the decision tree. This graph for average-based strategies is related to Fig.~4 in the main paper, which reports the results for threshold-based strategies.}
    \label{fig: correlation}
\end{figure}

Overall, the sensitivity of strategy regret does not change significantly under the average-based strategy definition (Fig.~\ref{fig: correlation}). Four out of the five key drivers of regret remain the same, indicating that the sensitivity of the biomass decisions is robust to the exact strategy formulation.

 CO$_2$ storage availability continues to be the most influential factor. However, its impact changes slightly for strategies: the \emph{Biomethane\_avg strategy} becomes more sensitive to changes in CO$_2$ storage availability, while the \emph{Chemicals\_avg strategy} becomes less sensitive.

Among the observed changes, the plastic recycling rate increases in importance and now ranks as the second most influential parameter, particularly affecting the \emph{Chemicals\_avg} strategy. Additionally, PV investment cost newly appears as the third most important driver of regret. Higher PV costs lead to increasing regrets across all biomass strategies, with the \emph{Biofuel\_avg} strategy being especially sensitive to this parameter.

\begin{figure}[h]
\centering
    \includegraphics[width=\textwidth]{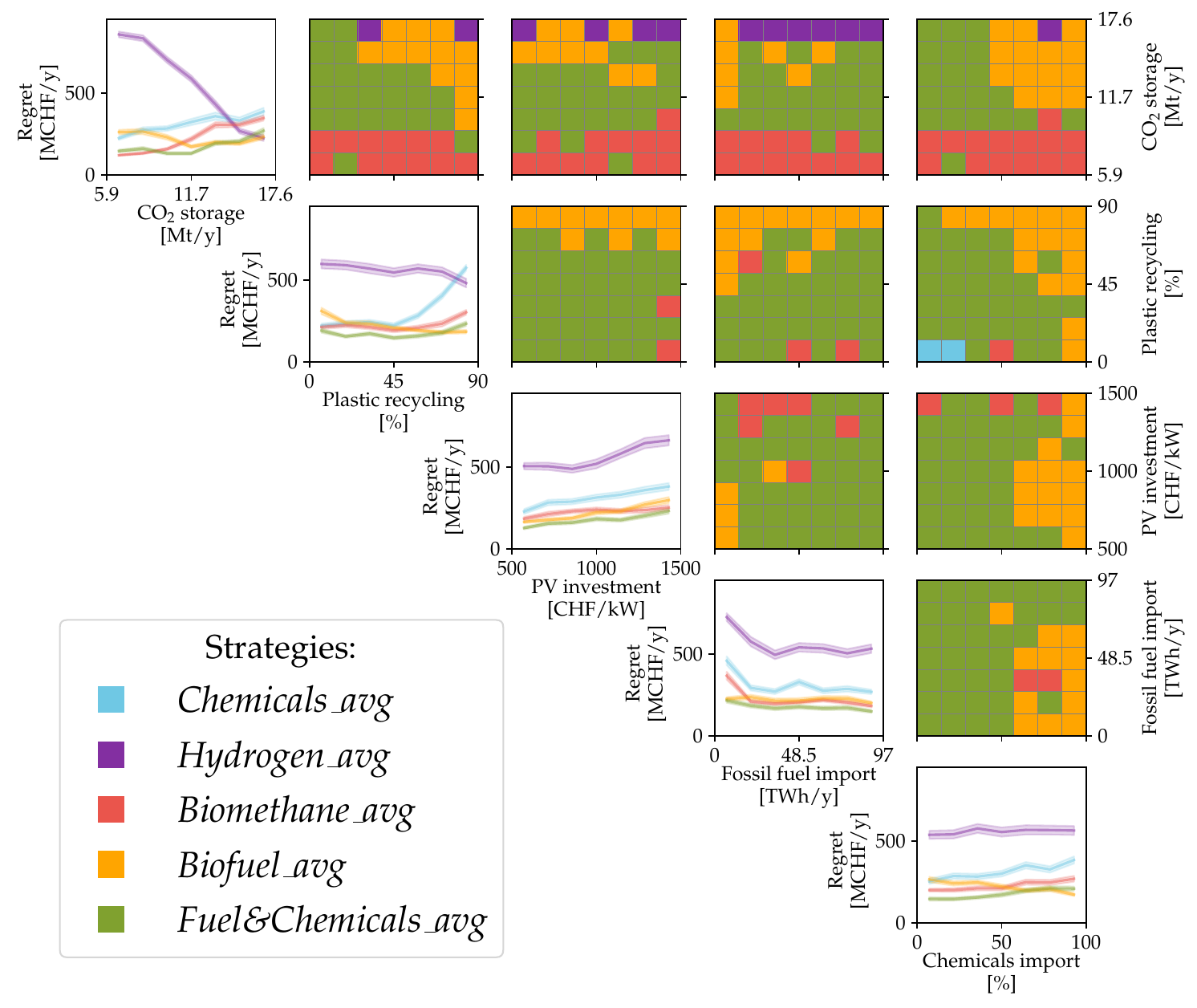}
    \caption{The figure illustrates the dependence of regret for the biomass strategies on the five most influential parameters. The diagonal panels display the dependence on individual parameters, while the upper triangular panels present decision maps for pairwise parameter combinations. These decision maps highlight the conditions under which one strategy outperforms the others, with the color of a pixel representing the strategy with the lowest mean regret under the given conditions. Each parameter range is divided into $n=7$ equally sized bins, resulting in an average of $N/n = 1000/7 \approx 143$ scenarios per bin and $N/(n\times n) = 1000/49 \approx 20.4$ scenarios per pixel. In the diagonal panels, the mean regret (solid line) and the standard error of the mean (shaded area) are shown for each bin. This graph for average-based strategies is related to Fig.~5 in the main paper, which reports the results for threshold-based strategies.}
    \label{fig: regret_map}
\end{figure}

Compared to the threshold-based strategy in the decision maps in Fig.~5 of the main paper, the average-based \emph{Fuel\&Chemicals\_avg} strategy becomes even more dominant in Fig.~\ref{fig: regret_map}. Particularly for parameter combinations that do not deviate too strongly from the central values of the uncertainty ranges, \emph{Fuel\&Chemicals\_avg} proves as the lowest-regret choice, suggesting that the strategy is robust across a broad spectrum of plausible future conditions.

The \emph{Hydrogen\_avg} strategy remains the dominant choice only for very high CO$_2$ storage availabilities, highlighting that its competitiveness highly depends on a fast and large-scale deployment of the carbon capture, transport, and storage infrastructure.

Interestingly, under average-based strategy definition, the  \emph{Biomethane\_avg} replaces \emph{Chemicals\_avg} as the preferred strategy in scenarios where CO$_2$ storage availability drops below 10 Mt/y. \emph{Chemicals\_avg} remains the lowest-regret option only when both plastic recycling rates and the allowed share of imported chemicals are very low.

In contrast, the \emph{Biofuel\_avg} strategy emerges as the dominant choice when high shares of imported chemicals are permitted or plastic recycling rates exceed 80\%, indicating that this strategy is favored in systems with either a well-established circular plastic economy or reliable trade relations for chemical and plastic imports.

\clearpage
\section{Supplementary Data}
\noindent\fbox{%
    \parbox{\textwidth}{%
        \textbf{Disclaimer:}\newline
        Our study has a primarily methodological focus. Nonetheless, to allow full reproducibility of our results, we report all data and models used in our case study, together with their sources. The model and all data are also openly accessible in the repository \url{https://gitlab.ethz.ch/epse/systems-design-public/low-regret-strategies}. For any questions or clarification, please reach out to the first author (gwiest@ethz.ch) or the corresponding author (morets@ethz.ch).
    }%
}

\vspace{5pt}

\noindent The following section documents the input data for the energy system model. For all uncertain parameters, we assume uniform distributions, defined either by a lower and upper bound (i.e., \emph{LB}–\emph{UB}), by a percentage range applied to a nominal value (i.e., (\emph{LB}\%–\emph{UB}\%)$\times$ \emph{NV}), or by a symmetric percentage deviation from a nominal value (i.e., \emph{NV}$\pm$\emph{X}\%).

\subsection{General parameters}

General system parameters can not be attributed to an individual demand, resource or technology but impact the overall scope of the system. 

\begin{table}[h]
    
    \centering
    \begin{threeparttable}
    \label{tab: general system parameters}
    \centering
    \caption{Overview of general parameters.}
    \begin{tabular}{l c c c }
    \toprule
    \textbf{Name} & \textbf{Symbol} & \textbf{Value} & \textbf{Unit}\\
    \midrule
    Discount rate & $i_{\mathrm{rate}}$ & 1.73 - 4.7\tnote{a} & \%\\
    Plastic recycling rate & $\%_{\mathrm{recycling}}$ & 0 - 90\tnote{b} & \%\\
    Share of public mobility & $\%_{\mathrm{public}}$ & 22.3 - 28.7\tnote{c} & \%\\
    Share of district heating & $\%_{\mathrm{DHN}}$ & 10 - 30\tnote{d} & \%\\
    Net transfer capacity & $NTC$ & (30\% - 100\%)$\times$12.9\tnote{e} & GW\\
    Unabatable emissions & $CO_{2,\mathrm{extra}}$ & 5.7\tnote{f} $\pm$ 20\% & Mt${_{\mathrm{CO}_2}}$/y\\
    
    \bottomrule
    \end{tabular}
    \begin{tablenotes}
    	\item[a] Taken from \cite{moret_characterization_2017}.
        \item[b] A plastic recycling rate of 0\% corresponds to all plastic waste getting incinerated. The upper limit of 90\% plastic recycling is assumed from discussions with experts.
        \item[c] Own calculation based on \cite{marcucci_adriana_assumptions_2022}.
        \item[d] Taken from \cite{limpens_energyscope_2019}.
        \item[e] Sum of net transfer capacities to Switzerland given in Table 6.9 for the TYNDP22 scenario "Distributed Energy" in the year 2050. The lower bound corresponds to low and the upper bound to high electricity market integration \cite{marcucci_cross_2023}.
        \item[f] Based on \cite{marcucci_cross_2023}, unabatable emissions, primarily from agriculture, are emissions that are hard-to-avoid and can not be captured.
    \end{tablenotes}
    \end{threeparttable}
\end{table}
\newpage
\subsection{Resources}

We differ between imported resources that enter the energy system from outside, local resources that are within the system's boundaries and exported resources that are removed from the system. Resources are parameterized by their cost, their yearly availability and their carbon intensity determined with respect to their lower heating value (LHV).

\subsubsection{Imported Resources}

Imported resources are external goods that enter the energy system from outside. We enforce some imports to be constant over the year due to infrastructure considerations, e.g., the natural gas grid, while other imports can vary over time.

\begin{table}[h]
\centering
\begin{threeparttable}
\caption{Import cost, availability, and carbon intensity for various energy sources.}
\begin{tabular}{l c c c}
\toprule
\textbf{Resource} & \textbf{Import Cost} & \textbf{Availability} & \textbf{Carbon Intensity} \\
                  & [CHF/kWh]    & [TWh/y]     & [kg${_{\mathrm{CO}_2}}$/kWh] \\
\midrule
Electricity      & 0.1 - 0.2\tnote{a}               & 31.278\tnote{b} $\pm$ 20\%    & 0         \\ 
Hydrogen         & 0.1342 - 0.1876\tnote{c}  & 0 - 11.$\overline{\mathrm{11}}$\tnote{d}  & 0         \\ 
Natural Gas      & 0.0131 - 0.0313\tnote{c}  & 0 - 35.88$\overline{\mathrm{4}}$\tnote{e}  & 0.2       \\ 
Fuel             & 0.0138 - 0.0497\tnote{c} & 0 - 96.882\tnote{f} & 0.265 \\ 
Sustainable Fuel & 0.1442 - 0.2549\tnote{c}   & 0 - 17.$\overline{\mathrm{77}}$\tnote{d}   & 0.265\tnote{g} \\ 
Ammonia          & 0.0515\tnote{h} $\pm$ 50\%       & 0 - 100\%\tnote{i} & 0  \\ 
Methanol         & 0.0722\tnote{j} $\pm$ 50\%       &                         & 0.2457 \\ 
+ HVC            & 0.1041\tnote{j} $\pm$ 50\%       & 0 - 100\%\tnote{k}             & 0.26      \\ 
+ Plastic        & 0.1278\tnote{j} $\pm$ 50\%       &                         & 0.2722 \\ 
\bottomrule
\end{tabular}
\begin{tablenotes}
\item[a] Taken from Table 3 in \cite{guidati_gianfranco_value_2022}.
\item[b] Average electricity import into Switzerland from 2014 to 2023 \cite{noauthor_switzerland_nodate}.
\item[c] Bounds given by ``Low'' and ``High'' scenario from \cite{marcucci_adriana_assumptions_2022}.
\item[d] Taken from Table~2.3 in \cite{marcucci_cross_2023}.
\item[e] Upper limit given by imports of natural gas to Switzerland in 2019 \cite{noauthor_switzerland_nodate-1}. 
\item[f] Upper limit given by imports of oil to Switzerland in 2019 \cite{noauthor_switzerland_nodate-2}.
\item[g] We assume that the carbon contained in the sustainable fuel previously has been removed as CO$_2$ from the atmosphere; therefore, if captured, it results in net negative emissions.
\item[h] Own calculation based on the average import price for ammonia in the years 2013-2022 obtained from \cite{noauthor_swiss-impex_nodate}.
\item[i] The share of the yearly ammonia demand that can be imported.
\item[j] Own calculations based on \cite{stadler_carbon_2019}.
\item[k] Since methanol can be used to produce high-value chemicals (HVC), which in turn can be used for plastic production, we limit the combined share of imports for these three products relative to their total demand.
\end{tablenotes}
\end{threeparttable}
\end{table}

\subsubsection{Local Resources}

For the local resources, we further differentiate between resources with time-independent availability over the year, e.g., geothermal, and time-dependent resources, such as wind, solar, and hydro energy.

\paragraph{Time-independent Local Resources}

These resources are modelled to have a constant availability throughout the whole year. Often, they are related to population, e.g., waste, sewage sludge, or animal stock, e.g., manure. Some resources might have a seasonal variability, such as wet/dry biomass; however, due to a lack of data for seasonal variability, we assume constant availability throughout the year as well.

\begin{table}[h]
\centering
\begin{threeparttable}
\caption{Cost, availability, and carbon intensity of local resources.}
\begin{tabular}{ l c c c }
\toprule
\textbf{Resource} & \textbf{Cost} & \textbf{Availability} & \textbf{Carbon Intensity} \\
                  & [CHF/kWh]    & [TWh/y]     & [kg${_{\mathrm{CO}_2}}$/kWh] \\
\midrule
Dry Biomass & 0.0299\tnote{a} $\pm$ 20\% & 16.738 - 17.549\tnote{a} & 0.39\tnote{b} \\ 
Wet Biomass & 0.0133\tnote{a} $\pm$ 20\% & 3.034 - 3.467\tnote{a} & 0.36\tnote{b} \\ 
Animal Manure & 0.0041\tnote{a} $\pm$ 20\% & 7.318\tnote{a} $\pm$ 20\% & 0.36\tnote{b} \\ 
Sewage Sludge & 0.0156\tnote{a} $\pm$ 20\% & 1.545 - 1.836\tnote{a} & 0.36\tnote{b} \\ 
Municipal Waste & 0 & 19.069 - 22.600\tnote{a,c} & 0.33\tnote{d} \\ 
Anergy Ground & 0 & 8.0\tnote{e} $\pm$ 20\% & 0 \\
Anergy Water & 0 & 3.0$\overline{\mathrm{55}}$ - 21.$\overline{\mathrm{11}}$\tnote{f} & 0 \\
Geothermal & 0 & 10.0\tnote{g} $\pm$ 20\% & 0 \\
\bottomrule
\label{tab:local resources}
\end{tabular}
\begin{tablenotes}
    \item[a] Own calculation based on \cite{marcucci_adriana_assumptions_2022}.
    \item[b] We assume that all the biogenic carbon was previously removed as $\mathrm{CO_2}$ from the atmosphere.
    \item[c] From this value of waste availability, we subtract the yearly demand for plastic (see Table~\ref{tab: end-use demands}) due to the implementation of recycling pathways.
    \item[d] We assume that 50\% of the carbon in municipal waste is biogenic.
    \item[e] p.68 Table 14 ``Erdwärmesonden (bodennah, ohne Regeneration)" \cite{prognos_ag_energieperspektiven_2021}.
    \item[f] p.68 Table 14, sum of ``Seen, Flüsse" and ``Grundwasser" for ``monovalent" use \cite{prognos_ag_energieperspektiven_2021}.
    \item[g] From Table 2 in \cite{guidati_gianfranco_value_2022}.
\end{tablenotes}
\end{threeparttable}
\end{table}

\clearpage
\paragraph{Time-dependent Local Resources}

Including time variability is essential to represent renewable energy technologies accurately in the model. Hydro, wind and solar power all are weather-dependent and strongly vary throughout the day and year. 

\begin{figure}[H]
    \centering
    \includegraphics[width=1\linewidth]{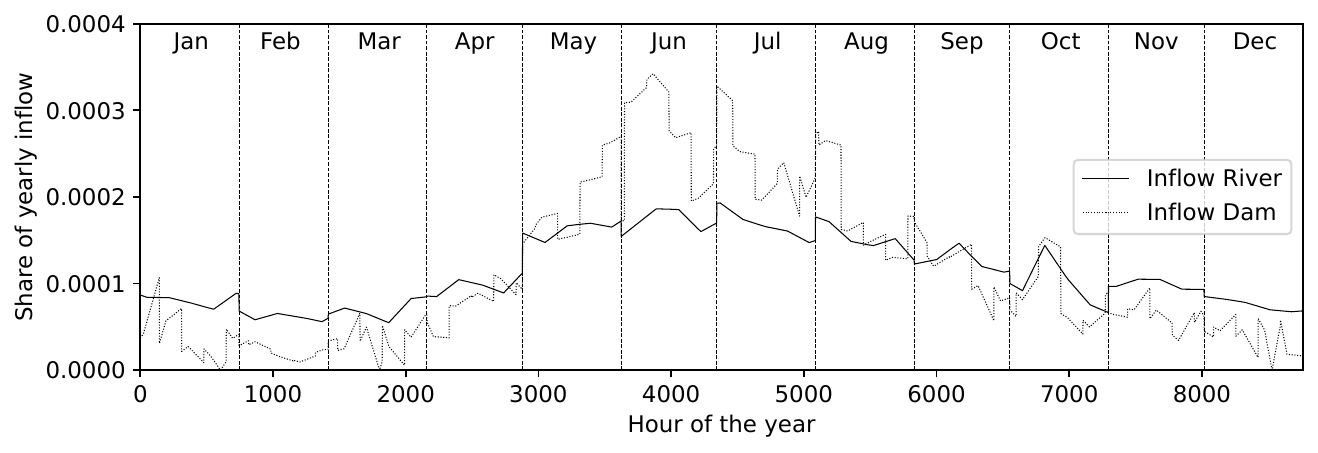}
    \caption{Share of yearly Hydro Inflow to Dams and Rivers.}
    \label{fig:hydro inflow}
\end{figure}

We estimate between 34.8 and 38.4 TWh \cite{marcucci_adriana_assumptions_2022} of hydro energy potential available in Switzerland per year, with a share of 55\% \cite{boes_swiss_2021} being utilized by hydro dam powerplants and the remaining 45\% in run-of-river power plants.
The inflow of water into lakes and rivers depends on seasonal precipitation and snowmelt patterns. The timeseries for both the inflow into rivers and into hydro reservoirs (Inflow dam) are shown in Figure \ref{fig:hydro inflow}. The timeseries are taken from \cite{limpens_energyscope_2019} and have been normalized to represent the share of the total yearly inflow. Note that all timeseries data are also openly available in the repository \url{https://gitlab.ethz.ch/epse/systems-design-public/low-regret-strategies} under Data/Raw\_Data/Timeseries.xlsx.

The availability of wind for power production fluctuates between days and seasons. In Switzerland, the capacity factors for wind production are,  on average, higher during winter months and lower during summer months. The timeseries that we use in our model for the capacity factor of wind turbines is taken from \cite{limpens_energyscope_2019}. Over the whole year, the capacity factors sum up to 2014.8 full load hours.

\begin{figure}[H]
    \centering
    \includegraphics[width=1\linewidth]{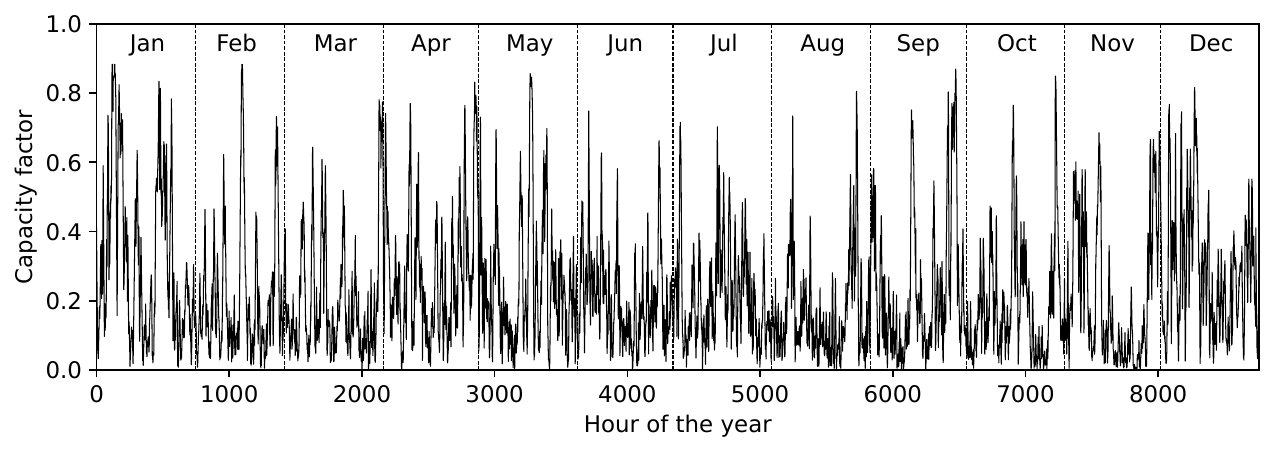}
    \caption{Capacity factors Wind.}
    \label{fig:capacity factor WIND_ONSHORE}
\end{figure}

\begin{figure}[H]
    \centering
    \includegraphics[width=1\linewidth]{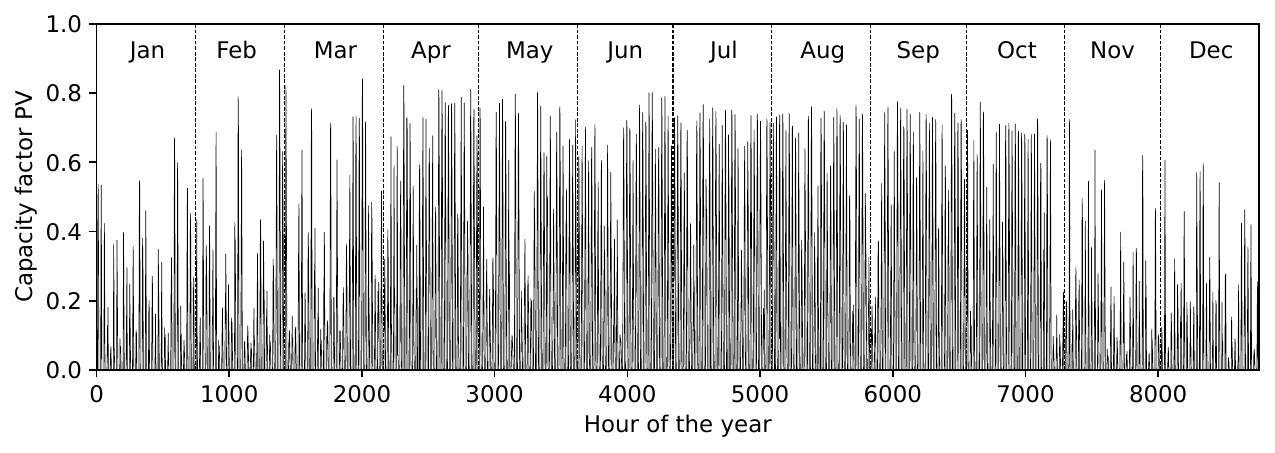}
    \caption{Capacity factor PV.}
    \label{fig:capacity factor PV}
\end{figure}

The production of electricity from photovoltaics varies not only over the day, with no production during night and maximum production at mid-day, but also seasonally, with lower production during winter months and higher production in summer. Furthermore, denser cloud coverage reduces electricity production. To avoid local clouds impacting the PV capacity factors of the timeseries for the whole country too strongly, we take an average over the PV timeseries for the five largest cities of Switzerland. Specifically, we use the tool Photovoltaic Geographical Information System
\texttt{PVGis 5.3} \cite{noauthor_jrc_nodate} from the European Commission to obtain the timeseries for the locations Zurich (47.378°N, 8.539°E), Geneva (46.199°N, 6.160°E), Basel (47.547°N, 7.589°E), Lausanne (46.531°N, 6.618°E), Bern (46.949°N, 7.448°E). We pick PVGIS-SARAH3 as the solar radiation database, a fixed mounting type using crystalline silicon as PV technology with 14\% system losses and optimize the slope and azimuth for PV power. Furthermore, we select the year 2018 to be consistent with the timeseries for Alpine PV. Averaging over the five locations, we obtain a timeseries for Switzerland (Fig. \ref{fig:capacity factor PV}) that has 1226.8 full load hours per year. 

The capacity factors of Alpine PV differ from standard PV due to several factors: the higher altitude reduces atmospheric absorption and leads lower cloud coverage, particularly in winter, both resulting in higher electricity production; furthermore, are the slope and azimuth of alpine PV installations typically optimized to maximize winter production instead of yearly production. We average the timeseries over the three given locations of the non-subsidized Alpine PV technology for the weather year 2018 from \cite{mellot_mitigating_2024}, as we have found its 1697.6 full load hours per year closely resembling those of currently planned Alpine PV projects, e.g., Sedrun Solar \cite{noauthor_maps_nodate} with 1617 full load hours per year.

\begin{figure}[H]
    \centering
    \includegraphics[width=1\linewidth]{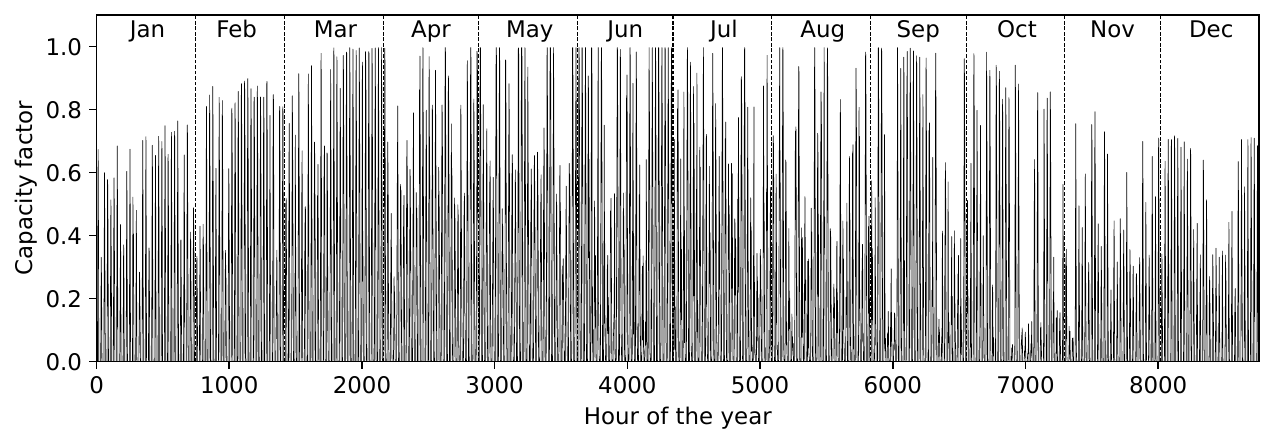}
    \caption{Capacity factor Alpine PV.}
    \label{fig:capacity factor ALPINE_PV}
\end{figure}
    
The capacity factor of solar thermal production differs from that of photovoltaics because solar thermal collectors utilize nearly the entire solar spectrum to generate heat, whereas photovoltaics can only convert the high-energy portion of the spectrum into electricity. Since atmospheric absorption and reflection vary with wavelength, changes in solar inclination throughout the day and year lead to differences in capacity factors between PV and solar thermal systems. 
The timeseries for the capacity factor of solar thermal used in our model has been obtained from Rapperswil OST during the SolTherm2050 project \cite{girardin_chancen_2021}.

\begin{figure}[H]
    \centering
    \includegraphics[width=1\linewidth]{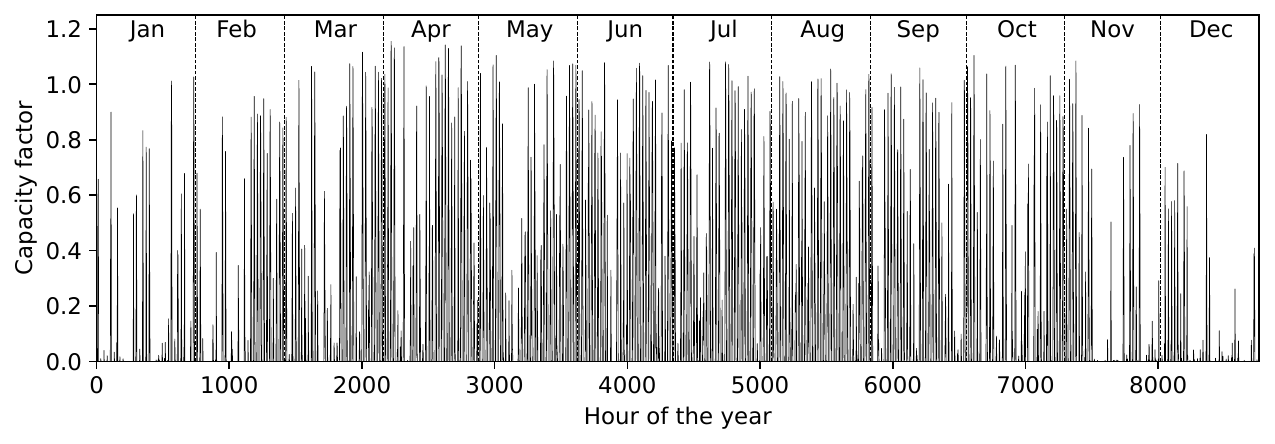}
    \caption{Capacity factor Solar Thermal.}
    \label{fig:capacity factor SOLAR THERMAL}
\end{figure}

\subsubsection{Export and Removal}

In addition to importing resources or using local resources, our energy model can export/remove resources from the system. This is particularly relevant for removing excess carbon to achieve a net-zero system. Furthermore, the model permits electricity exports, assuming zero cost/revenue to represent power curtailment. Thus, we do not allow the system to generate revenue with exports, as our current model can not realistically capture the price fluctuations in cross-border electricity trade. 

\begin{table}[h]
\centering
\begin{threeparttable}
\caption{Cost and availability export and removal options from the system.}
\begin{tabular}{l c c}
\toprule
 & \textbf{Cost}  & \textbf{Availability}  \\
 & [CHF/t$_{\mathrm{CO}_2}$] & [$\mathrm{Mt_{CO_2}}$/y] \\
\midrule
$\mathrm{CO}_2$ transport\,\&\,removal & 100 - 200\tnote{a} & 11.7\tnote{b} $\pm$ 50\% \\
(Bio-)Char removal & 0 & 1.785 - 2.045\tnote{c} \\ 
$\mathrm{CO}_2$ compensation &  1300\tnote{d} & $\infty$\tnote{e} \\ 
\midrule
 & [CHF/kWh] & [TWh/y] \\
Electricity export/curtailment & 0\tnote{f} & $\infty$\tnote{f} \\
\bottomrule
\end{tabular}
\begin{tablenotes}
    \item[a] Taken from Table 3 in \cite{guidati_gianfranco_value_2022}.
    \item[b]Negative emissions in the Zero Basis scenario in 2050 \cite{prognos_ag_energieperspektiven_2021}
    \item[c] Potential use in agriculture and urban green areas 1.35 - 1.61 Mt$_{\mathrm{CO}_2}$ plus 0.435 Mt$_{\mathrm{CO}_2}$ for asphalt, concrete, and plastic \cite{schonborn_biochar_2023}.
    \item[d] We set the price based on the cost of direct air capture and storage in 2024 from \url{https://climeworks.com}. This cost assumption is conservative, as the case study focuses on achieving a net-zero energy system nationally. 
    \item[e] We assume unlimited availability of CO$_2$ compensation outside Switzerland.
    \item[f] As we do not model the European electricity market accurately, we assume that all excess electricity is curtailed. 
\end{tablenotes}
\end{threeparttable}
\end{table}

\subsection{End-use Demands}

Our model considers various demands for the energy system, including electricity, heating and cooling, mobility, and non-energy demands such as cement, plastic and chemicals (Tab.~\ref{tab: end-use demands}).

For the model to be feasible, every demand has to be met at every hour of the year. We further distinguish between time-dependent and time-independent demands.
\begin{table}[h]
    
    \centering
    \resizebox{\textwidth}{!}{%
    \begin{threeparttable}
    
    \centering
    \caption{Overview of yearly end-use demands in the year 2050.}
    \begin{tabular}{l c c c c c c}
    \toprule
     & \textbf{Households} & \textbf{Services} & \textbf{Industry} & \textbf{Transport} & \textbf{Units} & \textbf{Constant} \\
    \midrule

    Electricity & 10.02 - 11.86\tnote{a} & 12.89 - 15.52\tnote{a} & 11.25 - 13.54\tnote{a} & 0 & TWh/y & X \\
    Space heating & 33.16 - 35.77\tnote{a} & 14.46 - 15.46\tnote{a} & 2.45 - 2.61\tnote{a} & 0 & TWh/y & X \\
    Hot water  & 6.35 - 7.51\tnote{a} & 2.01 - 2.38\tnote{a} & 0.34 - 0.41\tnote{a} & 0 & TWh/y & X \\ 
    Cooling  & 5.50 - 8.60\tnote{b} & 0 & 0 & 0 & TWh/y & X \\
    Process heat  & 0 & 0.67 - 0.81\tnote{a} & 16.32 - 19.65\tnote{a} & 0 & TWh/y & X \\
    Passenger mobility  & 0 & 0 & 0 & 124.75 - 146.75\tnote{a} & Bpkm/y & X \\
    Freight mobility  & 0 & 0 & 0 & 31.63 - 38.07\tnote{a} & Btkm/y & \checkmark \\
    Aviation mobility  & 0 & 0 & 0 & 20.59\tnote{c} $\pm$ 20\% & TWh/y & \checkmark \\
    Plastic  & 0 & 0 & 5.86\tnote{d} $\pm$ 20\% & 0 & TWh/y & \checkmark \\
    HVC  & 0 & 0 & 1.80\tnote{d} $\pm$ 20\% & 0 & TWh/y & \checkmark \\
    Cement  & 0 & 0 & 4.30\tnote{e} $\pm$ 20\% & 0 & Mt/y & \checkmark \\
    Ammonia & 0 & 0 & 0.12\tnote{f} $\pm$ 20\% & 0 & TWh/y & \checkmark \\
    Methanol & 0 & 0 & 0.20\tnote{d} $\pm$ 20\% & 0 & TWh/y & \checkmark \\
    
    \bottomrule
    \label{tab: end-use demands}
    \end{tabular}
    \begin{tablenotes}
    	\item[a] Own calculation based on \cite{marcucci_adriana_assumptions_2022}.
        \item[b] From \cite{mutschler_benchmarking_2021}, assuming high cooling device uptake for the decade 2050 - 2059.
        \item[c] From \cite{bundesamt_fur_zivilluftfahrt_schweizerische_2024}, average over the years 2015 - 2019 to avoid effects from the coronavirus pandemic.
        \item[d] Own calculations based on \cite{stadler_carbon_2019}.
        \item[e] Average production of cement in Switzerland based on \cite{cemsuisse_yearly_2023}
        \item[f] Average import of ammonia in the years 2013-2022 based on \cite{noauthor_swiss-impex_nodate}.
    \end{tablenotes}
    \end{threeparttable}}
\end{table}


The time variability in demands can follow seasonal patterns, e.g., higher space heating demand in winter and lower demand in summer; weekly patterns, e.g., higher process heat demand during weekdays than on weekends; and daily patterns, e.g., lower mobility demand during night and peak demand during commuting times. 

The electricity base demand varies significantly throughout the day and week, with higher electricity demand occurring especially during daytime hours and on weekdays. The electricity demand timeseries in our model is taken from \cite{limpens_energyscope_2019}.

\begin{figure}[H]
    \centering
    \includegraphics[width=1\linewidth]{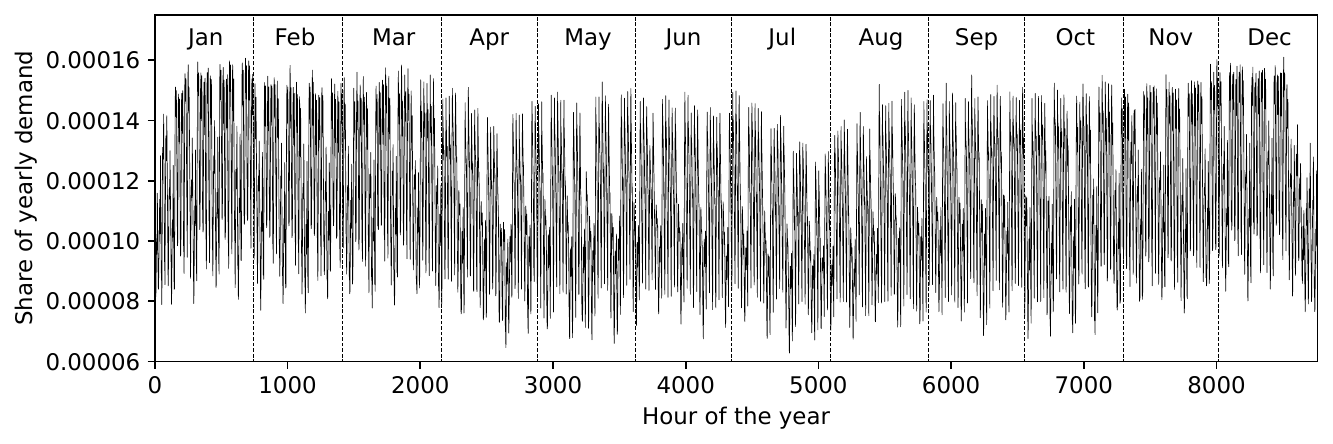}
    \caption{Electricity demand.}
    \label{fig: electricity demand}
\end{figure}

The space heating demand follows a strong seasonal pattern, with high demand in the winter months due to colder outside temperatures and little demand in the summer months. We use the space heating timeseries from \cite{limpens_energyscope_2019}. 

\begin{figure}[H]
    \centering
    \includegraphics[width=1\linewidth]{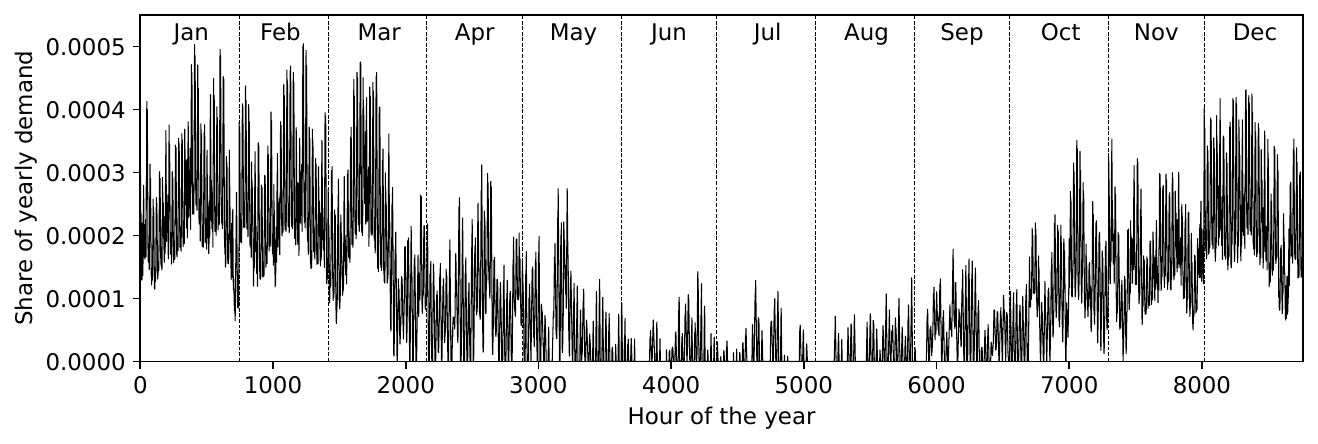}
    \caption{Space heating demand.}
    \label{fig: space heat demand}
\end{figure}
    
\begin{figure}[H]
    \centering
    \includegraphics[width=1\linewidth]{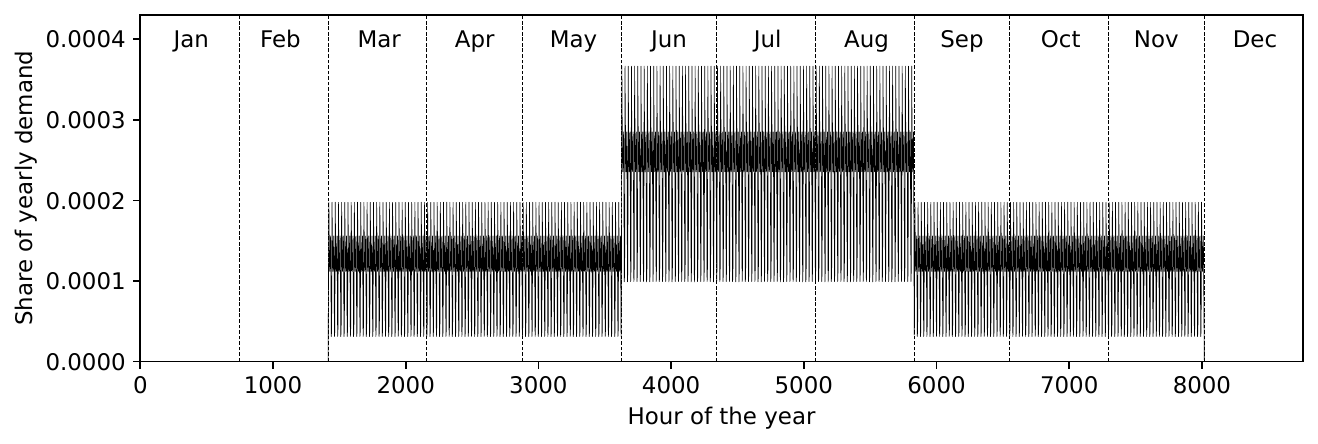}
    \caption{Cooling demand.}
    \label{fig:cooling}
\end{figure}

The cooling demand varies throughout the year, with no demand in winter and peak demand during the summer months. To construct the time series for our model, we use the air conditioning demand profile of the ``demand hourly profile in STEM" (package version 2019-02-27) from the JASM - Data platform \cite{noauthor_jasm_nodate}, which provides daily profiles for winter, intermediate seasons, and summer, differentiated by weekdays and weekends. Based on the daily profiles for residential buildings, we generate a full-year timeseries (Fig.~\ref{fig:cooling}), defining winter as December to February, intermediate seasons as March to May and September to November, and summer as June to August. Furthermore, we ensure that the weekdays and weekends of the cooling demand timeseries are aligned with the electricity demand timeseries.

\begin{figure}[H]
    \centering
    \includegraphics[width=1\linewidth]{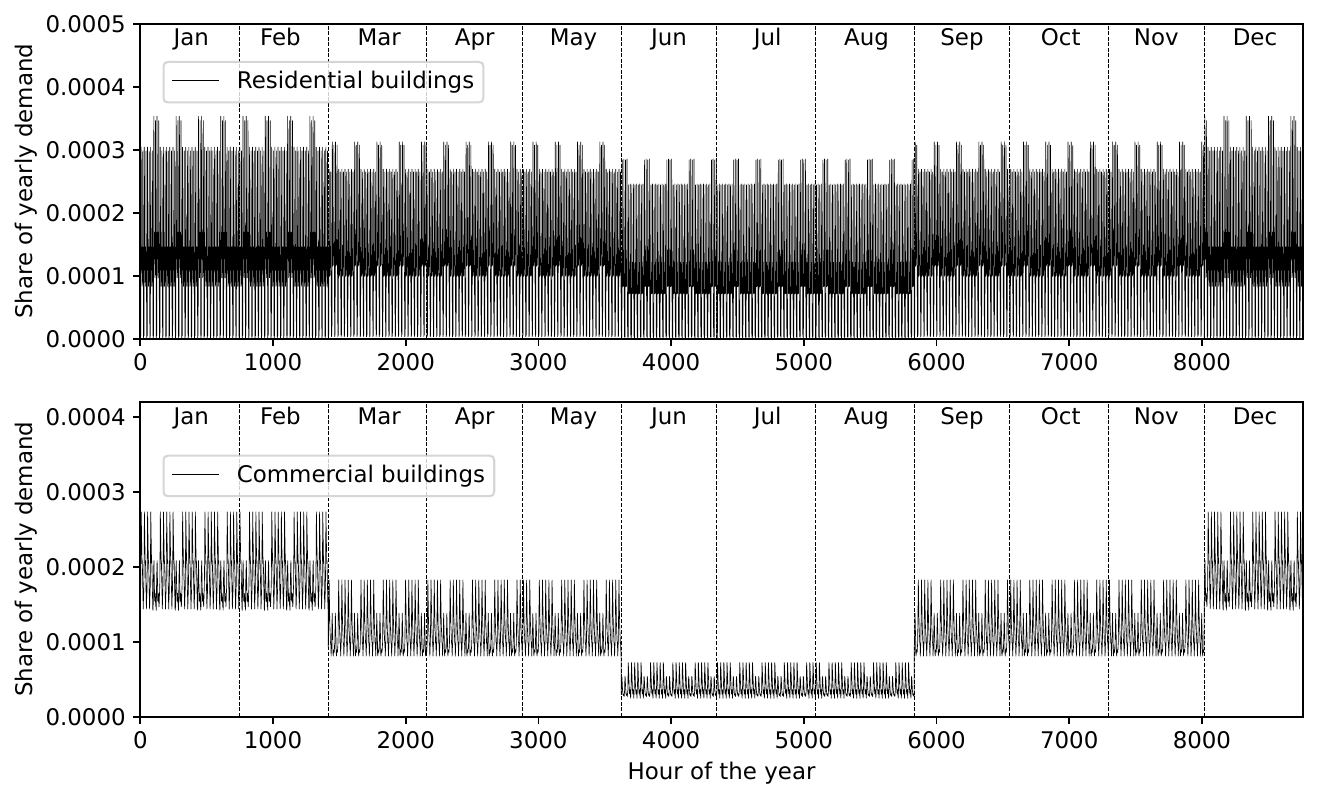}
    \caption{Hot water demand.}
    \label{fig:warm water}
\end{figure}

\begin{figure}[H]
    \centering
    \includegraphics[width=1\linewidth]{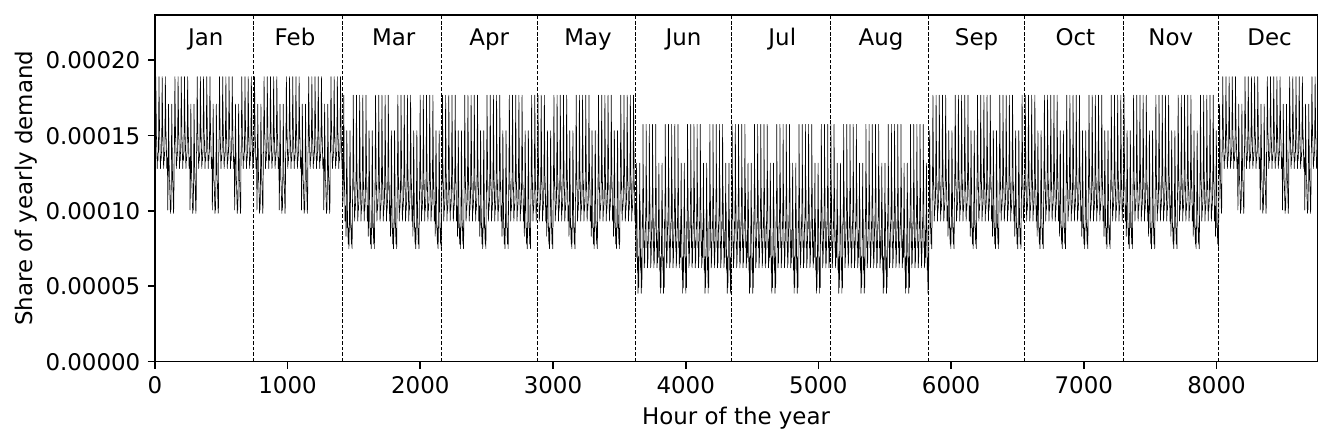}
    \caption{Process heat demand.}
    \label{fig:process heat}
\end{figure}

Following the same approach used for the cooling demand time series, we construct the hot water demand time series (Fig. \ref{fig:warm water}) and the process heat demand time series (Fig. \ref{fig:process heat}) using the daily profiles available on the JASM - Data platform \cite{noauthor_jasm_nodate}.

\begin{figure}[H]
    \centering
    \includegraphics[width=1\linewidth]{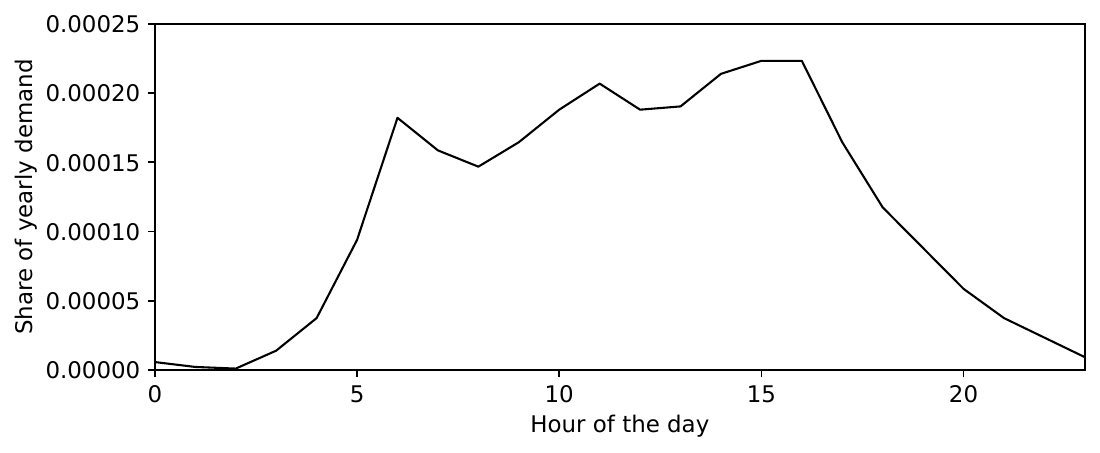}
    \caption{Hourly profile of the private mobility demand throughout one day.}
    \label{fig:mobility}
\end{figure}

The demand for mobility of passengers varies strongly throughout the day.
From the day profile (Fig.~\ref{fig:mobility}) given in \cite{limpens_energyscope_2019}, we construct the timeseries for passenger mobility by concatenating this profile for every day of the year. 

\subsection{Conversion Technologies}

Conversion technologies are the core of an energy system model.

\begin{figure}[H]
\centering
\begin{minipage}{.48\textwidth}
    \centering
    \includegraphics[width=0.9\linewidth]{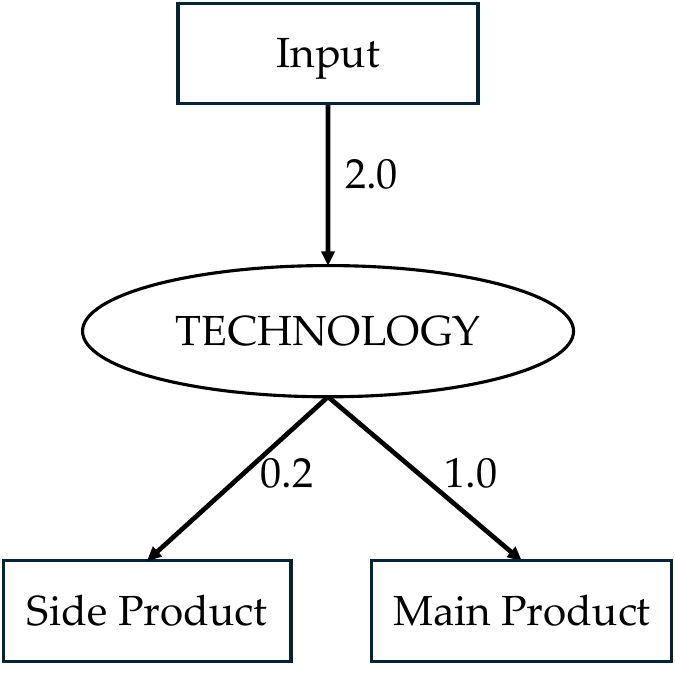}
\end{minipage}%
\begin{minipage}{.5\textwidth}
    \centering
    \resizebox{\textwidth}{!}{%
    \begin{threeparttable}
        \begin{tabular}{l c c }
            \toprule
            \textbf{Parameter} & \textbf{Value} & \textbf{Unit}\\
            \midrule
            $c_{\mathrm{inv}}$ & 1000 $\pm$ 30\% & CHF/capacity\tnote{a}\\
            $c_{\mathrm{maint}}$ & (2\% - 5\%)$\times  c_{\mathrm{inv}}$ & CHF/capacity/y\tnote{b}\\
            $\tau$ & 20 & year(s)\\
            $c_p$ & 100 & \% \\
            $f_{\mathrm{perc}}$ & 0 - 100 & \% \\
            $f_{\mathrm{min}}$ & 0 & capacity \\
            $f_{\mathrm{max}}$ & $\infty$ & capacity \\
            \bottomrule
        \end{tabular}
        \begin{tablenotes}
        	\item[a] Investment costs are given in CHF per unit of installed capacity. The capacity is always given with reference to the unitary flow (=1) in the input/output flow diagram, i.e., here, the ``Main Product''. Note: For technologies that co-produce outputs (e.g., CHP), the installed capacity is typically referenced to the input flow.
            \item[b] The yearly maintenance costs are referenced to the same capacity as the investment cost.
        \end{tablenotes}
    \end{threeparttable}
}

\end{minipage}
\caption{Exemplary input and output flows and economic parameters.}
\label{fig:EXAMPLE}
\end{figure}

Figure S15 illustrates an example of input and output flows, as well as the associated economic parameters, for a generic conversion technology. For each technology considered in the model, a similar figure is provided. The left-hand side of the diagram shows the input and output flows of the technology. These flows are expressed in units of energy (kWh), using the lower heating value for fuels.

There are a few exceptions to this convention:
\begin{itemize}
    \item Carbon dioxide flows are expressed in kilograms of CO$_2$ (kg$_{\mathrm{CO}_2}$)
    \item Cement flows are given in kilograms of cement (kg$_{\mathrm{Cem}}$)
    \item Passenger mobility is measured in passenger-kilometers (pkm)
    \item Freight transport is measured in tonne-kilometers (tkm).
\end{itemize}

From the input and output flows, the conversion efficiency relative to the main product can be determined. In the illustrated example, an output of 1.0 from an input of 2.0 yields an efficiency of 50\%.
    
The right-hand side gives the economic parameters of the technology. They are defined by their investment costs \( c_{\mathrm{inv}} \), maintenance costs \( c_{\mathrm{maint}} \), lifetime \( \tau \), available operation time per year \( c_{\mathrm{p}} \), share of production for a specific end-use \( f_{\mathrm{perc}} \), and constraints on capacity with \( f_{\mathrm{min}} \) representing existing capacity and \( f_{\mathrm{max}} \) representing  maximum installable capacity.

If not specified otherwise, the available operation time per year is set by default to \( c_{\mathrm{p}} = 100\% \). The share of production for a specific end-use is unconstrained, i.e., \( f_{\mathrm{perc}} = 0\%-100\% \), and the installable capacity is also unconstrained, i.e., \( f_{\mathrm{min}} = 0 \) GW and \( f_{\mathrm{max}} = \infty \) GW. If no data for maintenance cost could be obtained, we assume \(2\%-5\%\) of the mean investment cost. If no uncertainty range was found in the literature, we assume \(\pm30\%\) for investment and maintenance costs.
    
In the following, the conversion technologies are categorized by the main product that they provide.

\subsubsection{Electricity}

In this section, technologies, having electricity as their main output, are listed. 

Photovoltaic (PV) panels capture energy from sunlight and convert it into electricity. Consequently, electricity production from PV systems is weather-dependent. This dependence is modeled by multiplying the installed PV capacity by the capacity factor timeseries, \( c_{p,t} \) (see Fig.~\ref{fig:capacity factor PV} ), to determine electricity production for each hour of the year. 

\begin{figure}[H]
\centering
\begin{minipage}{.48\textwidth}
    \centering
    \includegraphics[width=0.55\linewidth]{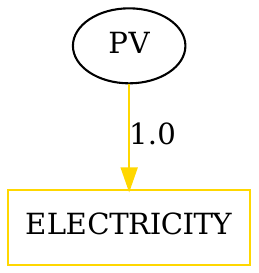}
\end{minipage}%
\begin{minipage}{.5\textwidth}
    \centering
    \centering
\resizebox{\textwidth}{!}{%
\begin{threeparttable}
\begin{tabular}{l c c }
\toprule
\textbf{Parameter} & \textbf{Value} & \textbf{Unit}\\
\midrule
$c_{\mathrm{inv}}$ & 500 - 1500\tnote{a} & CHF/kW\\
$c_{\mathrm{maint}}$ & (2\% - 5\%)$\times  c_{\mathrm{inv}}$ & CHF/kW/y\\
$\tau$ & 25\tnote{a} & year(s)\\
\bottomrule
\end{tabular}
\begin{tablenotes}
	\item[a] Taken from the photovoltaics in \cite{guidati_gianfranco_value_2022}.
\end{tablenotes}
\end{threeparttable}
}

\end{minipage}
\caption{Input and output flows and economic parameters of PV.}
\label{fig:PV}
\end{figure}

Alpine Photovoltaics (PV\_ALPINE) are photovoltaic panels specifically installed in alpine regions. Due to lower cloud coverage, lower air density, and enhanced reflectivity from snow, PV at higher altitudes yields a higher electricity production than standard PV installations, especially during winter months. Furthermore, the inclination of the panels is typically chosen to maximize electricity production in winter. Consequently, the capacity factor timeseries alpine PV differs from standard PV (see Fig. \ref{fig:capacity factor ALPINE_PV}). The more remote locations and technically more challenging installation of alpine PV result in higher investment and maintenance costs.  

\begin{figure}[H]
\begin{minipage}[H]{.48\textwidth}
    \centering
    \includegraphics[width=0.62\linewidth]{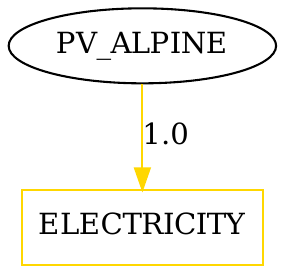}
\end{minipage}
\begin{minipage}{.5\textwidth}
\centering
\resizebox{\textwidth}{!}{%
\begin{threeparttable}
\begin{tabular}{l c c }
\toprule
\textbf{Parameter} & \textbf{Value} & \textbf{Unit}\\
\midrule
$c_{\mathrm{inv}}$ & 2000\tnote{a} $\pm$ 30\% & CHF/kW\\
$c_{\mathrm{maint}}$ & (2\% - 5\%)$\times c_{\mathrm{inv}}$ & CHF/kW/y\\
$\tau$ & 25\tnote{a} & year(s)\\
$f_{\mathrm{max}}$ & 0.0 - 17.67\tnote{b} & GW\\
\bottomrule
\end{tabular}
\begin{tablenotes}
	\item[a] Taken from Alpine photovoltaics in \cite{guidati_gianfranco_value_2022}.
    \item[b] We divide the maximum production of 30 TWh/y \cite{marcucci_cross_2023} by the full load hours to get the capacity limit.
\end{tablenotes}
\end{threeparttable}
}
\end{minipage}
\caption{Input and output flows and economic parameters of PV\_ALPINE.}
\label{fig:ALPINE_PV}
\end{figure}

Wind turbines (WIND\_ONSHORE) convert the kinetic energy of moving air into electricity. Analogous to PV, the electricity production of wind turbines is calculated by multiplying the installed capacity by a capacity factor timeseries (see Fig.~\ref{fig:capacity factor WIND_ONSHORE}). The potential for wind turbine installations in Switzerland is limited due to technical and social constraints. The 2050 production potential is estimated to be between 1.7 and 15.0~TWh/y~\cite{marcucci_cross_2023}.

\begin{figure}[H]
\begin{minipage}[H]{.48\textwidth}
\centering
\includegraphics[width=0.83\linewidth]{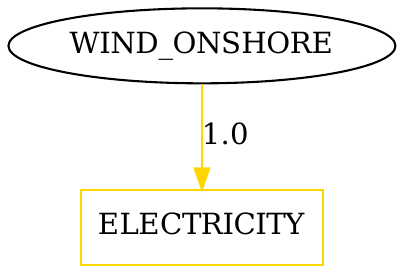}
\end{minipage}
\begin{minipage}{.5\textwidth}
\centering
\resizebox{\textwidth}{!}{%
\begin{threeparttable}
\begin{tabular}{l c c }
\toprule
\textbf{Parameter} & \textbf{Value} & \textbf{Unit}\\
\midrule
$c_{\mathrm{inv}}$ & 2000\tnote{a} $\pm$ 30\% & CHF/kW\\
$c_{\mathrm{maint}}$ & (2\% - 5\%)$\times c_{\mathrm{inv}}$ & CHF/kW/y\\
$\tau$ & 20\tnote{a} & year(s)\\
$f_{\mathrm{max}}$ & 0.84 - 7.44\tnote{b} & GW\\
\bottomrule
\end{tabular}
\begin{tablenotes}
	\item[a] Based on the wind power technology from \cite{guidati_gianfranco_value_2022}.
    \item[b] We divide the production potential \cite{marcucci_cross_2023} by the full load hours to get the capacity limit.
\end{tablenotes}
\end{threeparttable}
}
\end{minipage}
\caption{Input and output flows and economic parameters of WIND\_ONSHORE.}
\label{fig:WIND_ONSHORE}
\end{figure}

Run-of-river hydropower plants (HYDRO\_RIVER) utilize the energy of flowing river water to generate electricity using turbines. The river flow rates vary throughout the year due to seasonal changes and weather conditions. To model the variability, the installed capacity of run-of-river hydropower plants is multiplied by a capacity factor timeseries (see Fig.~\ref{fig:hydro inflow}) to calculate electricity production at an hourly resolution. Run-of-river hydropower is a well-established technology that has already been largely utilized in Switzerland, and it is not expected to increase significantly due to environmental and technical constraints. Therefore, we fix the capacity in our model to the level of 2012, i.e., $f_{\mathrm{min}}=f_{\mathrm{max}}$ = 3.8 GW \cite{limpens_energyscope_2019}.

\begin{figure}[H]
\begin{minipage}[H]{.48\textwidth}
\centering
\includegraphics[width=0.65\linewidth]{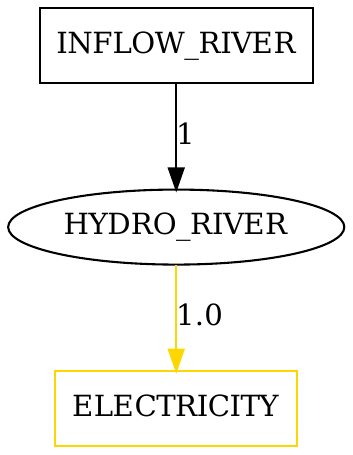}
\end{minipage}
\begin{minipage}{.5\textwidth}
\centering
\resizebox{\textwidth}{!}{%
\begin{threeparttable}
\begin{tabular}{l c c }
\toprule
\textbf{Parameter} & \textbf{Value} & \textbf{Unit}\\
\midrule
$c_{\mathrm{inv}}$ & 6000\tnote{a} $\pm$ 30\% & CHF/kW/y\\
$c_{\mathrm{maint}}$ & (2\% - 5\%)$\times c_{\mathrm{inv}}$ & CHF/kW/y\\
$\tau$ & 40\tnote{a} & year(s)\\
$f_{\mathrm{min}}$ & 3.8\tnote{b} & GW\\
$f_{\mathrm{max}}$ & 3.8\tnote{b} & GW\\
\bottomrule
\end{tabular}
\begin{tablenotes}
	\item[a] Based on the run-of-river hydro power from \cite{guidati_gianfranco_value_2022}.
    \item[b] Installed capacity in 2012 \cite{limpens_energyscope_2019}.
\end{tablenotes}
\end{threeparttable}
}

\end{minipage}
\caption{Input and output flows and economic parameters of HYDRO\_RIVER.}
\label{fig:HYDRO_RIVER}
\end{figure}

Dam hydropower plants (HYDRO\_DAM) generate electricity by storing water in reservoirs and releasing it through turbines to produce power. The reservoir levels increase due to seasonal-dependent inflows from rainfall and snowmelt (see timeseries in Fig. \ref{fig:hydro inflow}) and decrease when the hydropower plants generate electricity. The potential for dam hydropower in Switzerland is largely developed, with limited opportunities for expansion. In our model, we assume that the capacity is fixed to the level of 2012 at $f_{\mathrm{min}} = f_{\mathrm{max}} = 8.08$ GW~\cite{limpens_energyscope_2019}.

\begin{figure}[H]
\begin{minipage}[H]{.48\textwidth}
\centering
\includegraphics[width=0.65\linewidth]{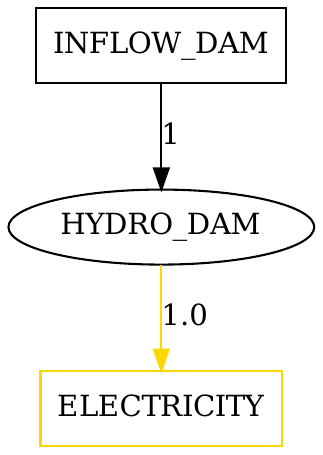}
\end{minipage}
\begin{minipage}{.5\textwidth}
\centering
\resizebox{\textwidth}{!}{%
\begin{threeparttable}
\begin{tabular}{l c c }
\toprule
\textbf{Parameter} & \textbf{Value} & \textbf{Unit}\\
\midrule
$c_{\mathrm{inv}}$ & 6000\tnote{a} $\pm$ 30\% & CHF/kW\\
$c_{\mathrm{maint}}$ & (2\% - 5\%)$\times c_{\mathrm{inv}}$ & CHF/kW/y\\
$\tau$ & 40\tnote{a} & year(s)\\
$f_{\mathrm{min}}$ & 8.08\tnote{b} & GW\\
$f_{\mathrm{max}}$ & 8.08\tnote{b} & GW\\
\bottomrule
\end{tabular}
\begin{tablenotes}
	\item[a] Based on the regulated hydro power from \cite{guidati_gianfranco_value_2022}.
    \item[b] Installed capacity in 2012 \cite{limpens_energyscope_2019}.
\end{tablenotes}
\end{threeparttable}
}
\end{minipage}
\caption{Input and output flows and economic parameters of HYDRO\_DAM.}
\label{fig:HYDRO_DAM}
\end{figure}

Geothermal power plants (GEOTHERMAL) generate electricity by using the heat from below the earth's surface. This heat is extracted by drilling wells into geothermal reservoirs and is then utilized to drive turbines connected to electricity generators. 

\begin{figure}[H]
\begin{minipage}[H]{.48\textwidth}
\centering
\includegraphics[width=0.65\linewidth]{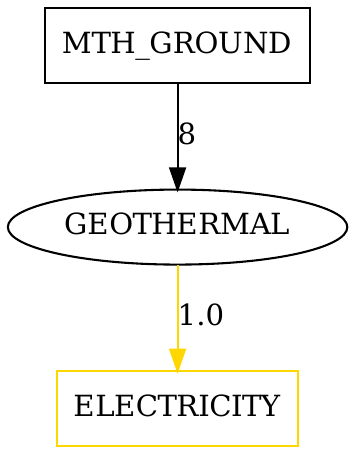}
\end{minipage}
\begin{minipage}{.5\textwidth}
\centering
\resizebox{\textwidth}{!}{%
\begin{threeparttable}
\begin{tabular}{l c c }
\toprule
\textbf{Parameter} & \textbf{Value} & \textbf{Unit}\\
\midrule
$c_{\mathrm{inv}}$ & 10000\tnote{a} $\pm$ 30\% & CHF/kW\\
$c_{\mathrm{maint}}$ & (2\% - 5\%)$\times c_{\mathrm{inv}}$ & CHF/kW/y\\
$\tau$ & 30\tnote{a} & year(s)\\
\bottomrule
\end{tabular}
\begin{tablenotes}
    \item[a] Based on the geothermal power plant from \cite{guidati_gianfranco_value_2022}.
	\item Note: The production is limited by the geothermal potential (Tab.~\ref{tab:local resources}).
\end{tablenotes}
\end{threeparttable}
}
\end{minipage}
\caption{Input and output flows and economic parameters of GEOTHERMAL.}
\label{fig:GEOTHERMAL}
\end{figure}

\subsubsection{Low-temperature Heat}
Low-temperature heat is required to heat residential and commercial buildings as well as provide them with hot water. We differentiate between technologies that produce heat centrally, which is then distributed via a district heating network to the buildings and technologies that directly provide the heat at the buildings (decentralized heating).

\paragraph{Decentralized Heating}

A residential oil boiler (DEC\_BOILER\_OIL) burns heating oil to heat residential and commercial buildings.


\begin{figure}[H]
\begin{minipage}[H]{.48\textwidth}
\centering
\includegraphics[width=0.8\linewidth]{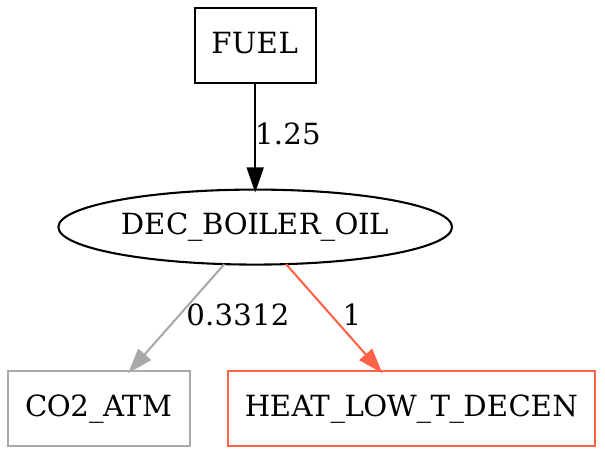}
\end{minipage}
\begin{minipage}{.5\textwidth}
\centering
\resizebox{\textwidth}{!}{%
\begin{threeparttable}
\begin{tabular}{l c c }
\toprule
\textbf{Parameter} & \textbf{Value} & \textbf{Unit}\\
\midrule
$c_{\mathrm{inv}}$ & 900\tnote{a} $\pm$ 30\% & CHF/kW\\
$c_{\mathrm{maint}}$ & (2\% - 5\%)$\times c_{\mathrm{inv}}$ & CHF/kW/y\\
$\tau$ & 25\tnote{a} & year(s)\\
\bottomrule
\end{tabular}
\begin{tablenotes}
	\item[a] Based on the residential oil boiler from \cite{guidati_gianfranco_value_2022}.
\end{tablenotes}
\end{threeparttable}
}
\end{minipage}
\caption{Input and output flows and economic parameters of DEC\_BOILER\_OIL.}
\label{fig:DEC_BOILER_OIL}
\end{figure}

A gas boiler (DEC\_BOILER\_GAS) burns natural gas to heat residential and commercial buildings.
 
\begin{figure}[H]
\begin{minipage}[H]{.48\textwidth}
\centering
\includegraphics[width=0.8\linewidth]{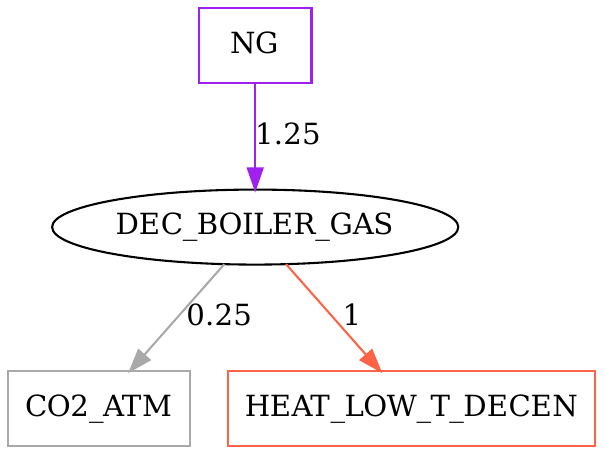}
\end{minipage}
\begin{minipage}{.5\textwidth}
\centering
\resizebox{\textwidth}{!}{%
\begin{threeparttable}
\begin{tabular}{l c c }
\toprule
\textbf{Parameter} & \textbf{Value} & \textbf{Unit}\\
\midrule
$c_{\mathrm{inv}}$ & 1000\tnote{a} $\pm$ 30\% & CHF/kW\\
$c_{\mathrm{maint}}$ & (2\% - 5\%)$\times c_{\mathrm{inv}}$ & CHF/kW/y\\
$\tau$ & 25\tnote{a} & year(s)\\
\bottomrule
\end{tabular}
\begin{tablenotes}
	\item[a] Based on the residential gas boiler from \cite{guidati_gianfranco_value_2022}.
\end{tablenotes}
\end{threeparttable}
}
\end{minipage}
\caption{Input and output flows and economic parameters of DEC\_BOILER\_GAS.}
\label{fig:DEC_BOILER_GAS}
\end{figure}

Wood boilers (DEC\_BOILER\_WOOD) burn wood to provide heating for residential and commercial buildings.
 
\begin{figure}[H]
\begin{minipage}[H]{.48\textwidth}
\centering
\includegraphics[width=0.8\linewidth]{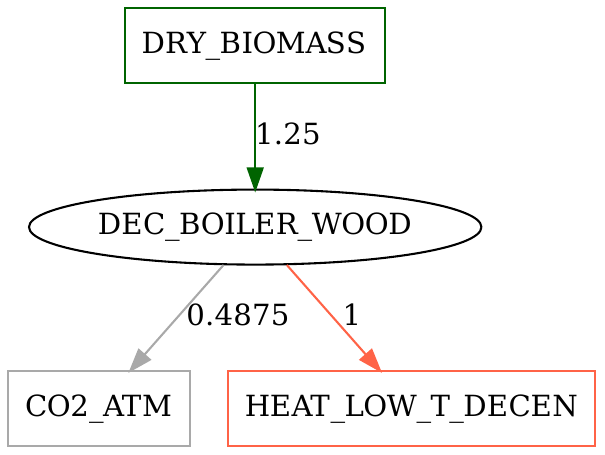}
\end{minipage}
\begin{minipage}{.5\textwidth}
\centering
\resizebox{\textwidth}{!}{%
\begin{threeparttable}
\begin{tabular}{l c c }
\toprule
\textbf{Parameter} & \textbf{Value} & \textbf{Unit}\\
\midrule
$c_{\mathrm{inv}}$ & 950\tnote{a} $\pm$ $30\%$ & CHF/kW\\
$c_{\mathrm{maint}}$ & (2\% - 5\%)$\times c_{\mathrm{inv}}$ & CHF/kW/y\\
$\tau$ & 25\tnote{a} & year(s)\\
\bottomrule
\end{tabular}
\begin{tablenotes}
	\item \item[a] Based on the residential wood boiler from \cite{guidati_gianfranco_value_2022}.
\end{tablenotes}
\end{threeparttable}
}
\end{minipage}
\caption{Input and output flows and economic parameters of DEC\_BOILER\_WOOD.}
\label{fig:DEC_BOILER_WOOD}
\end{figure}

Residential electric heaters (DEC\_ELEC) use resistive elements to directly convert electricity into heat.

\begin{figure}[H]
\begin{minipage}[H]{.48\textwidth}
\centering
\includegraphics[width=0.65\linewidth]{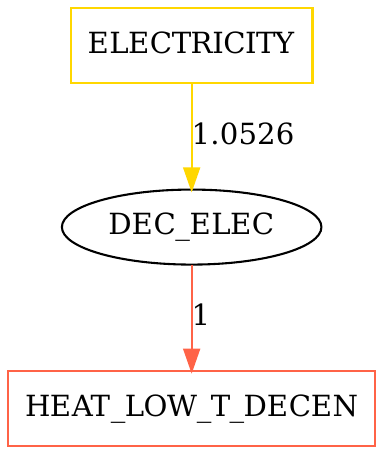}
\end{minipage}
\begin{minipage}{.5\textwidth}
\centering
\resizebox{\textwidth}{!}{%
\begin{threeparttable}
\begin{tabular}{l c c }
\toprule
\textbf{Parameter} & \textbf{Value} & \textbf{Unit}\\
\midrule
$c_{\mathrm{inv}}$ & 650\tnote{a} $\pm$ 30\% & CHF/kW\\
$c_{\mathrm{maint}}$ & (2\% - 5\%)$\times c_{\mathrm{inv}}$ & CHF/kW/y\\
$\tau$ & 25\tnote{a} & year(s)\\
\bottomrule
\end{tabular}
\begin{tablenotes}
	\item[a] Based on the residential electrical heater from \cite{guidati_gianfranco_value_2022}.
\end{tablenotes}
\end{threeparttable}
}
\end{minipage}
\caption{Input and output flows and economic parameters of DEC\_ELEC.}
\label{fig:DEC_ELEC}
\end{figure}

Residential air heat pumps (DEC\_AHP) do not generate heat directly from electricity but instead use it to transfer heat from the ambient air. This process makes them significantly more energy-efficient than direct electric heating. 

\begin{figure}[H]
\begin{minipage}[H]{.48\textwidth}
\centering
\includegraphics[width=0.65\linewidth]{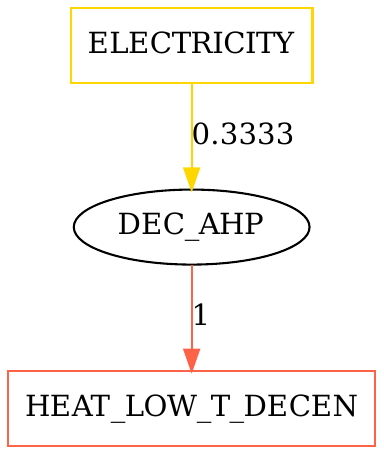}
\end{minipage}
\begin{minipage}{.5\textwidth}
\centering
\resizebox{\textwidth}{!}{%
\begin{threeparttable}
\begin{tabular}{l c c }
\toprule
\textbf{Parameter} & \textbf{Value} & \textbf{Unit}\\
\midrule
$c_{\mathrm{inv}}$ & 2000 - 3000\tnote{a} & CHF/kW\\
$c_{\mathrm{maint}}$ & (2\% - 5\%)$\times c_{\mathrm{inv}}$ & CHF/kW/y\\
$\tau$ & 25\tnote{a} & year(s)\\
\bottomrule
\end{tabular}
\begin{tablenotes}
	\item[a] Based on the residential air source heat pump from \cite{guidati_gianfranco_value_2022}.
\end{tablenotes}
\end{threeparttable}
}
\end{minipage}
\caption{Input and output flows and economic parameters of DEC\_AHP.}
\label{fig:DEC_AHP}
\end{figure}

Ground source heat pumps (DEC\_GHP) instead use electricity to transfer heat from the ground to heat the building. As the temperatures in the ground are higher compared to the ambient air during winter months, which have the highest demand for residential heat, ground source heat pumps typically have a higher yearly averaged energetic efficiency compared to air source heat pumps.

\begin{figure}[H]
\begin{minipage}[H]{.48\textwidth}
\centering
\includegraphics[width=0.9\linewidth]{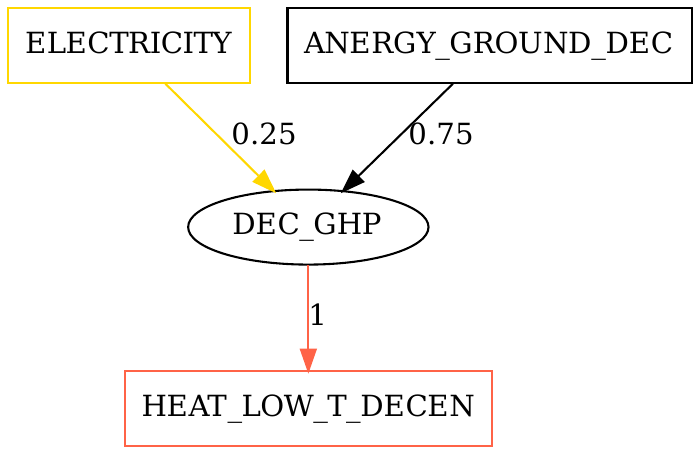}
\end{minipage}
\begin{minipage}{.5\textwidth}
\centering
\resizebox{\textwidth}{!}{%
\begin{threeparttable}
\begin{tabular}{l c c }
\toprule
\textbf{Parameter} & \textbf{Value} & \textbf{Unit}\\
\midrule
$c_{\mathrm{inv}}$ & 4000 - 6000\tnote{a} & CHF/kW\\
$c_{\mathrm{maint}}$ & (2\% - 5\%)$\times c_{\mathrm{inv}}$ & CHF/kW/y\\
$\tau$ & 25\tnote{a} & year(s)\\
\bottomrule
\end{tabular}
\begin{tablenotes}
	\item[a] Based on the residential ground source heat pump from \cite{guidati_gianfranco_value_2022}.
    \item Note: the deployment is limited by the potential of Anergy Ground (see Tab.~\ref{tab:local resources}).
\end{tablenotes}
\end{threeparttable}
}
\end{minipage}
\caption{Input and output flows and economic parameters of DEC\_GHP.}
\label{fig:DEC_GHP}
\end{figure}

Decentralized solar thermal (DEC\_SOLAR) refers to solar thermal systems installed at individual residential or commercial buildings, to capture and convert sunlight into heat.
Consequently, heat production from solar thermal is weather-dependent. This dependence is modeled by multiplying the installed solar thermal capacity by the capacity factor timeseries, \( c_{p,t} \) (see Fig.~\ref{fig:capacity factor SOLAR THERMAL} ).

\begin{figure}[H]
\begin{minipage}[H]{.48\textwidth}
\centering
\includegraphics[width=0.7\linewidth]{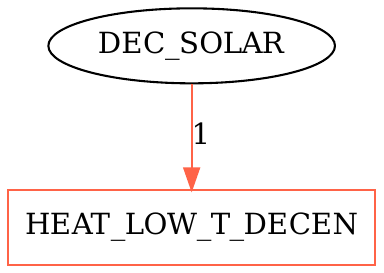}
\end{minipage}
\begin{minipage}{.5\textwidth}
\centering
\resizebox{\textwidth}{!}{%
\begin{threeparttable}
\begin{tabular}{l c c }
\toprule
\textbf{Parameter} & \textbf{Value} & \textbf{Unit}\\
\midrule
$c_{\mathrm{inv}}$ & 1200 - 1700\tnote{a} & CHF/kW\\
$c_{\mathrm{maint}}$ & (2\% - 5\%)$\times c_{\mathrm{inv}}$ & CHF/kW/y\\
$\tau$ & 25\tnote{a} & year(s)\\
\bottomrule
\end{tabular}
\begin{tablenotes}
	\item[a] Based on the residential solar thermal from \cite{guidati_gianfranco_value_2022}.
\end{tablenotes}
\end{threeparttable}
}
\end{minipage}
\caption{Input and output flows and economic parameters of DEC\_SOLAR.}
\label{fig:DEC_SOLAR}
\end{figure}

\paragraph{District Heating}

A district heating gas boiler (DHN\_BURNER\_CH4) generates heat for district heating networks by burning natural gas. Thus, it provides a dispatchable heat supply.

\begin{figure}[H]
\begin{minipage}[H]{.48\textwidth}
\centering
\includegraphics[width=0.9\linewidth]{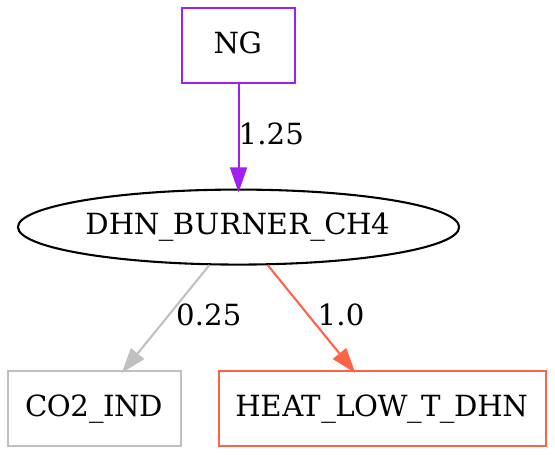}
\end{minipage}
\begin{minipage}{.5\textwidth}
\centering
\resizebox{\textwidth}{!}{%
\begin{threeparttable}
\begin{tabular}{l c c }
\toprule
\textbf{Parameter} & \textbf{Value} & \textbf{Unit}\\
\midrule
$c_{\mathrm{inv}}$ & 150\tnote{a} $\pm$ 30\% & CHF/kW\\
$c_{\mathrm{maint}}$ & (2\% - 5\%)$\times c_{\mathrm{inv}}$ & CHF/kW/y\\
$\tau$ & 25\tnote{a} & year(s)\\
\bottomrule
\end{tabular}
\begin{tablenotes}
	\item[a] Based on the district heating gas burner from \cite{guidati_gianfranco_value_2022}.
\end{tablenotes}
\end{threeparttable}
}
\end{minipage}
\caption{Input and output flows and economic parameters of DHN\_BURNER\_CH4.}
\label{fig:DHN_BURNER_CH4}
\end{figure}

DHN\_BIOGAS\_MOTOR represents a district heating biogas plant where animal manure is anaerobically digested by bacteria to produce biogas. Instead of upgrading it to biomethane for injection into the natural gas grid, the biogas is directly combusted in a gas engine to generate both electricity and heat, which is supplied to a district heating network.

\begin{figure}[H]
\begin{minipage}[H]{.48\textwidth}
\centering
\includegraphics[width=0.9\linewidth]{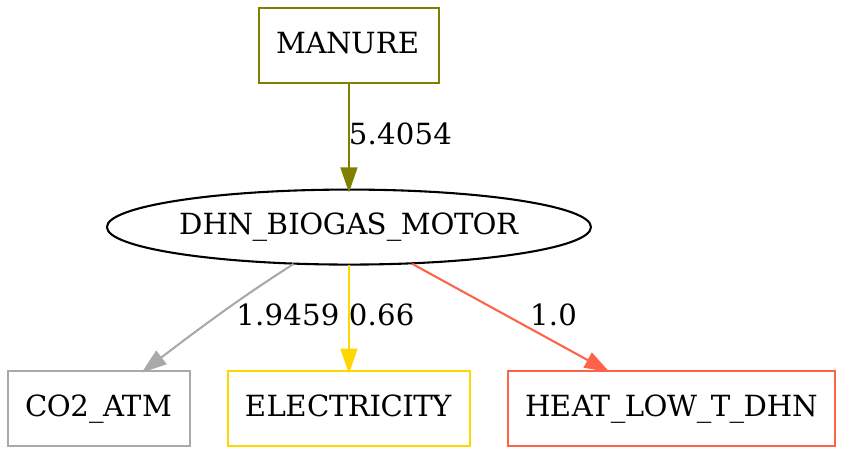}
\end{minipage}
\begin{minipage}{.5\textwidth}
\centering
\resizebox{\textwidth}{!}{%
\begin{threeparttable}
\begin{tabular}{l c c }
\toprule
\textbf{Parameter} & \textbf{Value} & \textbf{Unit}\\
\midrule
$c_{\mathrm{inv}}$ & 6000\tnote{a} $\pm$ 30\% & CHF/kW\\
$c_{\mathrm{maint}}$ & (2\% - 5\%)$\times c_{\mathrm{inv}}$ & CHF/kW/y\\
$\tau$ & 20\tnote{a} & year(s)\\
\bottomrule
\end{tabular}
\begin{tablenotes}
	\item[a] Based on the biogas combined heat \& power plant from \cite{guidati_gianfranco_value_2022}.
    \item Note: The efficiencies reported in \cite{guidati_gianfranco_value_2022} do not account for losses during the conversion of manure to biogas. We assume a 37\% efficiency for this step.
\end{tablenotes}
\end{threeparttable}
}
\end{minipage}
\caption{Input and output flows and economic parameters of DHN\_BIOGAS\_MOTOR.}
\label{fig:DHN_BIOGAS_MOTOR}
\end{figure}

A direct electrical district heater (DHN\_DIRECT\_ELEC) converts electricity directly into heat for district heating networks using resistive heating elements. 

\begin{figure}[H]
\begin{minipage}[H]{.48\textwidth}
\centering
\includegraphics[width=0.7\linewidth]{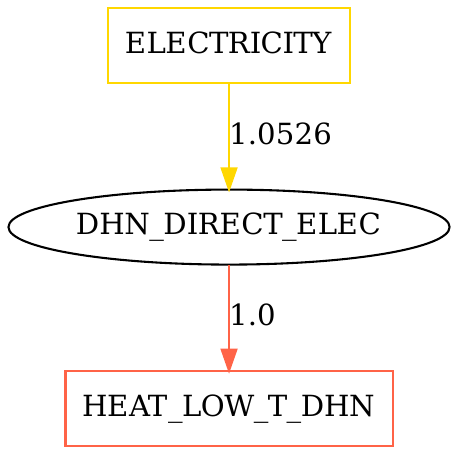}
\end{minipage}
\begin{minipage}{.5\textwidth}
\centering
\resizebox{\textwidth}{!}{%
\begin{threeparttable}
\begin{tabular}{l c c }
\toprule
\textbf{Parameter} & \textbf{Value} & \textbf{Unit}\\
\midrule
$c_{\mathrm{inv}}$ & 325\tnote{a} $\pm$ 30\% & CHF/kW\\
$c_{\mathrm{maint}}$ & (2\% - 5\%)$\times c_{\mathrm{inv}}$ & CHF/kW/y\\
$\tau$ & 25\tnote{a} & year(s)\\
\bottomrule
\end{tabular}
\begin{tablenotes}
	\item[a] Based on the district heating electrical heater from \cite{guidati_gianfranco_value_2022}.
\end{tablenotes}
\end{threeparttable}
}
\end{minipage}
\caption{Input and output flows and economic parameters of DHN\_DIRECT\_ELEC.}
\label{fig:DHN_DIRECT_ELEC}
\end{figure}

Water source heat pumps (DHN\_WHP) use a water source, such as a river or lake, as a heat reservoir for typically large-scale heat pumps that then supply district heating networks. Compared to ambient air, water reservoirs generally have more stable and, in winter, higher temperatures, improving the efficiency and performance of the heat pump. Therefore, we model the water source heat pump with a coefficient of performance (COP) of 4.

\begin{figure}[H]
\begin{minipage}[H]{.48\textwidth}
\centering
\includegraphics[width=0.9\linewidth]{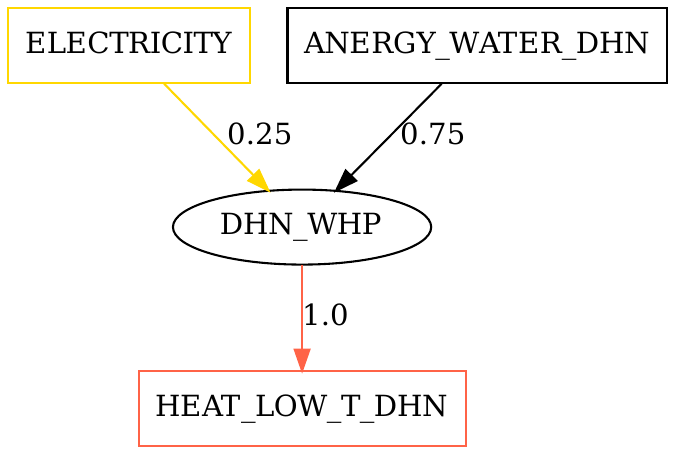}
\end{minipage}
\begin{minipage}{.5\textwidth}
\centering
\resizebox{\textwidth}{!}{%
\begin{threeparttable}
\begin{tabular}{l c c }
\toprule
\textbf{Parameter} & \textbf{Value} & \textbf{Unit}\\
\midrule
$c_{\mathrm{inv}}$ & 1500 - 2500\tnote{a} & CHF/kW\\
$c_{\mathrm{maint}}$ & (2\% - 5\%)$\times c_{\mathrm{inv}}$ & CHF/kW/y\\
$\tau$ & 25\tnote{a} & year(s)\\
\bottomrule
\end{tabular}
\begin{tablenotes}
	\item[a] Based on the district heating heat pump from \cite{guidati_gianfranco_value_2022}.
    \item Note: the deployment is limited by the potential of Anergy Water (see Tab.~\ref{tab:local resources}).
\end{tablenotes}
\end{threeparttable}
}
\end{minipage}
\caption{Input and output flows and economic parameters of DHN\_WHP.}
\label{fig:DHN_WHP}
\end{figure}

A geothermal heat plant (DHN\_DEEP\_GEO) harnesses thermal energy from underground reservoirs to provide heat for district heating networks. By extracting heat from deep geothermal sources, the geothermal plant offers a stable and renewable heat supply over the whole year.

\begin{figure}[H]
\begin{minipage}[H]{.48\textwidth}
\centering
\includegraphics[width=0.55\linewidth]{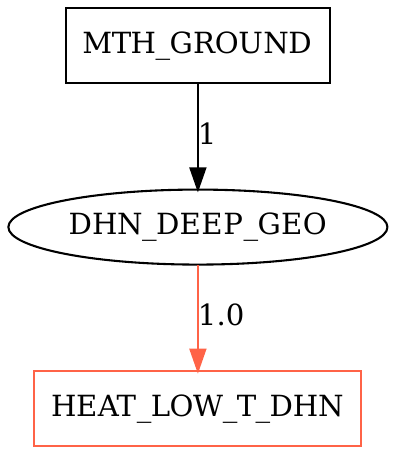}
\end{minipage}
\begin{minipage}{.5\textwidth}
\centering
\resizebox{\textwidth}{!}{%
\begin{threeparttable}
\begin{tabular}{l c c }
\toprule
\textbf{Parameter} & \textbf{Value} & \textbf{Unit}\\
\midrule
$c_{\mathrm{inv}}$ & 2000 - 4000\tnote{a} & CHF/kW\\
$c_{\mathrm{maint}}$ & (2\% - 5\%)$\times c_{\mathrm{inv}}$ & CHF/kW/y\\
$\tau$ & 30\tnote{a} & year(s)\\
\bottomrule
\end{tabular}
\begin{tablenotes}
	\item[a] Based on the geothermal heat generation technology from \cite{guidati_gianfranco_value_2022}.
    \item Note: The production is limited by the geothermal potential (Tab.~\ref{tab:local resources}).
\end{tablenotes}
\end{threeparttable}
}
\end{minipage}
\caption{Input and output flows and economic parameters of DHN\_DEEP\_GEO.}
\label{fig:DHN_DEEP_GEO}
\end{figure}

Solar thermal district heating (DHN\_SOLAR\_LOW\_T) uses solar collectors to capture sunlight and convert it into heat for district heating networks. As decentral solar thermal, it is weather-dependent. The heat production is obtained by multiplying the installed solar thermal capacity by the capacity factor timeseries, \( c_{p,t} \) (see Fig.~\ref{fig:capacity factor SOLAR THERMAL} ), to determine heat production for each hour of the year. Typically, solar thermal heating is combined with seasonal thermal storage to balance the higher supply during summer with the higher demand during winter. 

\begin{figure}[H]
\begin{minipage}[H]{.48\textwidth}
\centering
\includegraphics[width=0.7\linewidth]{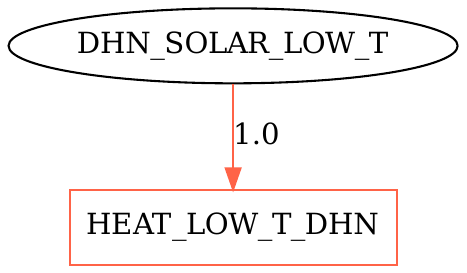}
\end{minipage}
\begin{minipage}{.5\textwidth}
\centering
\resizebox{\textwidth}{!}{%
\begin{threeparttable}
\begin{tabular}{l c c }
\toprule
\textbf{Parameter} & \textbf{Value} & \textbf{Unit}\\
\midrule
$c_{\mathrm{inv}}$ & 500 - 750\tnote{a} & CHF/kW\\
$c_{\mathrm{maint}}$ & (2\% - 5\%)$\times c_{\mathrm{inv}}$ & CHF/kW/y\\
$\tau$ & 20\tnote{a} & year(s)\\
\bottomrule
\end{tabular}
\begin{tablenotes}
	\item[a] Based on the district heating/industrial solar thermal from \cite{guidati_gianfranco_value_2022}.
\end{tablenotes}
\end{threeparttable}
}
\end{minipage}
\caption{Input and output flows and economic parameters of DHN\_SOLAR\_LOW\_T.}
\label{fig:DHN_SOLAR_LOW_T}
\end{figure}

\subsubsection{Medium-temperature Heat}
Medium-temperature heat, typically ranging from 100°C to 400°C, is required for various industrial processes, including food processing, paper production, and chemical manufacturing. 

IND\_DEEP\_GEO represents deep geothermal energy used to supply medium-temperature heat for processes. This technology harnesses heat from underground, accessed through deep drilling.

\begin{figure}[H]
\begin{minipage}[H]{.48\textwidth}
\centering
\includegraphics[width=0.7\linewidth]{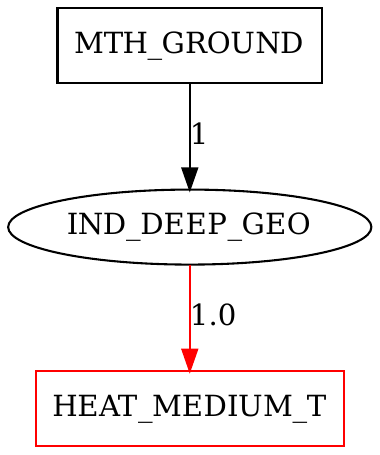}
\end{minipage}
\begin{minipage}{.5\textwidth}
\centering
\resizebox{\textwidth}{!}{%
\begin{threeparttable}
\begin{tabular}{l c c }
\toprule
\textbf{Parameter} & \textbf{Value} & \textbf{Unit}\\
\midrule
$c_{\mathrm{inv}}$ & 2000 - 4000\tnote{a} & CHF/kW\\
$c_{\mathrm{maint}}$ & (2\% - 5\%)$\times c_{\mathrm{inv}}$ & CHF/kW/y\\
$\tau$ & 30\tnote{a} & year(s)\\
\bottomrule
\end{tabular}
\begin{tablenotes}
	\item[a] Based on the geothermal heat generation from \cite{guidati_gianfranco_value_2022}.
    \item Note: The production is limited by the geothermal potential (Tab.~\ref{tab:local resources}).
\end{tablenotes}
\end{threeparttable}
}
\end{minipage}
\caption{Input and output flows and economic parameters of IND\_DEEP\_GEO.}
\label{fig:IND_DEEP_GEO}
\end{figure}

Medium-temperature solar thermal (DHN\_SOLAR\_MEDIUM\_T) uses solar collectors to generate heat at moderate temperatures. Its output depends on weather conditions, with heat production determined by the installed capacity of the technology and the capacity factor timeseries $c_{p,t}$ (see Fig.~\ref{fig:capacity factor SOLAR THERMAL}).

\begin{figure}[H]
\begin{minipage}[H]{.48\textwidth}
\centering
\includegraphics[width=0.9\linewidth]{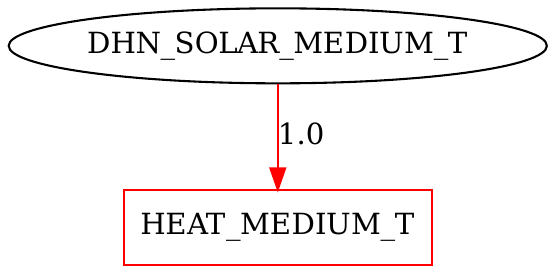}
\end{minipage}
\begin{minipage}{.5\textwidth}
\centering
\resizebox{\textwidth}{!}{%
\begin{threeparttable}
\begin{tabular}{l c c }
\toprule
\textbf{Parameter} & \textbf{Value} & \textbf{Unit}\\
\midrule
$c_{\mathrm{inv}}$ & 500 - 750\tnote{a} & CHF/kW\\
$c_{\mathrm{maint}}$ & (2\% - 5\%)$\times c_{\mathrm{inv}}$ & CHF/kW/y\\
$\tau$ & 20\tnote{a} & year(s)\\
\bottomrule
\end{tabular}
\begin{tablenotes}
	\item Based on the district heating/industrial solar thermal from \cite{guidati_gianfranco_value_2022}.
\end{tablenotes}
\end{threeparttable}
}
\end{minipage}
\caption{Input and output flows and economic parameters of IND\_DEEP\_GEO.}
\label{fig:DHN_SOLAR_MEDIUM_T}
\end{figure}

DIRECT\_ELEC\_MEDIUM\_T represents medium-temperature heat generated directly from electricity by using resistance heaters.

\begin{figure}[H]
\begin{minipage}[H]{.48\textwidth}
\centering
\includegraphics[width=\linewidth]{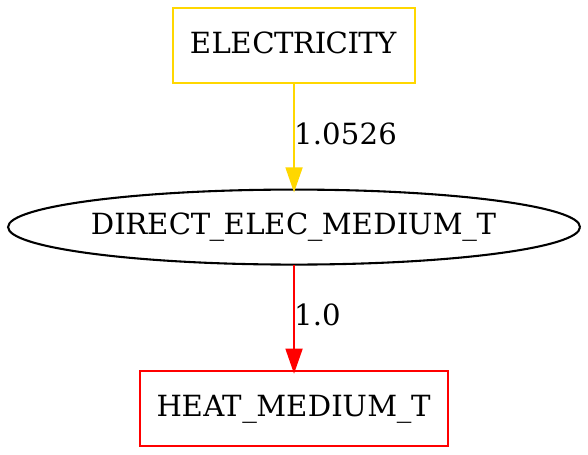}
\end{minipage}
\begin{minipage}{.5\textwidth}
\centering
\resizebox{\textwidth}{!}{%
\begin{threeparttable}
\begin{tabular}{l c c }
\toprule
\textbf{Parameter} & \textbf{Value} & \textbf{Unit}\\
\midrule
$c_{\mathrm{inv}}$ & 275\tnote{a} $\pm$ 30\% & CHF/kW\\
$c_{\mathrm{maint}}$ & (2\% - 5\%)$\times c_{\mathrm{inv}}$ & CHF/kW/y\\
$\tau$ & 25\tnote{a} & year(s)\\
\bottomrule
\end{tabular}
\begin{tablenotes}
	\item[a] Based on the industrial electric heater from \cite{guidati_gianfranco_value_2022}.
\end{tablenotes}
\end{threeparttable}
}
\end{minipage}
\caption{Input and output flows and economic parameters of DIRECT\_ELEC\_MEDIUM\_T.}
\label{fig:DIRECT_ELEC_MEDIUM_T}
\end{figure}

\subsubsection{Combined Heat and Power}

Combined heat and power (CHP) plants co-generate electricity together with heat.

Plastic waste incineration plant (PLASTIC\_ENERGY\_RECOVERY) utilizes the energy content in the plastic waste thermally to produce medium-temperature heat and electricity.

\begin{figure}[H]
\begin{minipage}[H]{.48\textwidth}
\centering
\includegraphics[width=0.95\linewidth]{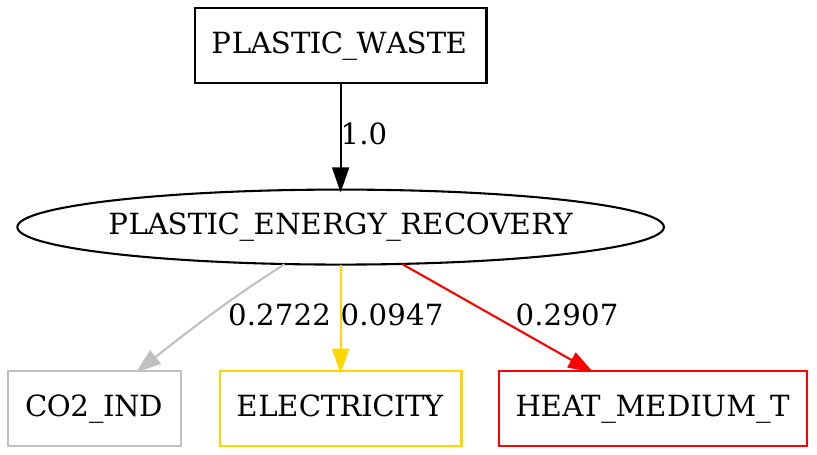}
\end{minipage}
\begin{minipage}{.5\textwidth}
\centering
\resizebox{\textwidth}{!}{%
\begin{threeparttable}
\begin{tabular}{l c c }
\toprule
\textbf{Parameter} & \textbf{Value} & \textbf{Unit}\\
\midrule
$c_{\mathrm{inv}}$ & 106.8\tnote{a} $\pm$ 30\% & CHF/kW\\
$c_{\mathrm{maint}}$ & 12.9\tnote{a} $\pm$ 30\% & CHF/kW/y\\
$\tau$ & 35\tnote{a} & year(s)\\
\bottomrule
\end{tabular}
\begin{tablenotes}
	\item[a] Own calculations based on \cite{meys_achieving_2021}, \cite{bachmann_towards_2023} and own estimates.
\end{tablenotes}
\end{threeparttable}
}
\end{minipage}
\caption{Input and output flows and economic parameters of PLASTIC\_ENERGY\_RECOVERY.}
\label{fig:PLASTIC_ENERGY_RECOVERY}
\end{figure}

The leftover digestate from anaerobic digestion of wet biomass, animal manure, and sewage sludge can be utilized in the digestate-powered CHP plant (IND\_CHP\_DIGESTATE) to generate medium-temperature and electricity.

\begin{figure}[H]
\begin{minipage}[H]{.48\textwidth}
\centering
\includegraphics[width=0.95\linewidth]{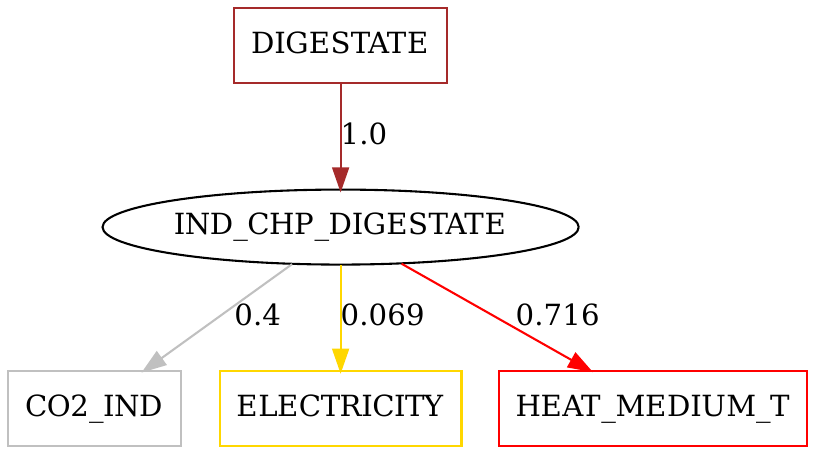}
\end{minipage}
\begin{minipage}{.5\textwidth}
\centering
\resizebox{\textwidth}{!}{%
\begin{threeparttable}
\begin{tabular}{l c c }
\toprule
\textbf{Parameter} & \textbf{Value} & \textbf{Unit}\\
\midrule
$c_{\mathrm{inv}}$ & 2000 - 3000\tnote{a} & CHF/kW\\
$c_{\mathrm{maint}}$ & (2\% - 5\%)$\times c_{\mathrm{inv}}$ & CHF/kW/y\\
$\tau$ & 25\tnote{a} & year(s)\\
\bottomrule
\end{tabular}
\begin{tablenotes}
	\item[a] Based on the waste combined heat \& power plant from \cite{guidati_gianfranco_value_2022}.
    \item Note: The efficiencies are based on the CHP\_WASTE plant operated to produce medium-temperature heat.
\end{tablenotes}
\end{threeparttable}
}
\end{minipage}
\caption{Input and output flows and economic parameters of IND\_CHP\_DIGESTATE.}
\label{fig:IND_CHP_DIGESTATE}
\end{figure}

Hydrogen-based combined heat and power (IND\_CHP\_H2) uses hydrogen to cogenerate electricity and medium-temperature heat without producing CO$_2$.

\begin{figure}[H]
\begin{minipage}[H]{.48\textwidth}
\centering
\includegraphics[width=0.8\linewidth]{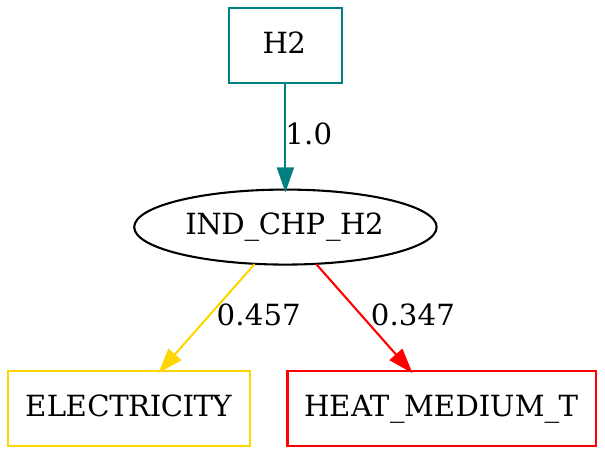}
\end{minipage}
\begin{minipage}{.5\textwidth}
\centering
\resizebox{\textwidth}{!}{%
\begin{threeparttable}
\begin{tabular}{l c c }
\toprule
\textbf{Parameter} & \textbf{Value} & \textbf{Unit}\\
\midrule
$c_{\mathrm{inv}}$ & 400 - 600\tnote{a} & CHF/kW\\
$c_{\mathrm{maint}}$ & (2\% - 5\%)$\times c_{\mathrm{inv}}$ & CHF/kW/y\\
$\tau$ & 25\tnote{a} & year(s)\\
\bottomrule
\end{tabular}
\begin{tablenotes}
	\item[a] Based on the gas combined heat \& power plant from \cite{guidati_gianfranco_value_2022}.
\end{tablenotes}
\end{threeparttable}
}
\end{minipage}
\caption{Input and output flows and economic parameters of IND\_CHP\_H2.}
\label{fig:IND_CHP_H2}
\end{figure}

Additionally, we model several combined heat and power plants with operational flexibility, either in the resources they consume or the outputs they generate. This flexibility allows a single plant to operate in different modes, optimizing performance based on demand and resource availability. The key advantage is that capacity only needs to be installed once, yet it can be shared across multiple operational modes. For instance, the plant can prioritize heat and electricity production during winter when heating demand is high and shift to increased electricity generation in summer when heat demand is lower.

\newpage
PP\_CH4 represents a gas turbine combined cycle that primarily uses methane as its fuel source but can also operate with hydrogen or liquid fuels for combustion. It is primarily designed for electricity generation, but it also has the capability to provide both electricity and medium-temperature heat.

\begin{table}[h]

\centering
\begin{threeparttable}
\caption{Economic parameters of PP\_CH4.}
\begin{tabular}{l c c }
\toprule
\textbf{Parameter} & \textbf{Value} & \textbf{Unit}\\
\midrule
$c_{\mathrm{inv}}$ & 400 - 600\tnote{a} & CHF/kW\\
$c_{\mathrm{maint}}$ & (2\% - 5\%)$\times c_{\mathrm{inv}}$ & CHF/kW/y\\
$\tau$ & 25\tnote{a} & year(s)\\
\bottomrule
\end{tabular}
\begin{tablenotes}
	\item[a] Based on the gas turbine combined cycle from \cite{guidati_gianfranco_value_2022}.
\end{tablenotes}
\end{threeparttable}

\end{table}

Table \ref{tab: operational modes PP_CH4} shows the input and output flows for the different operational modes. Inputs are indicated by negative values, while positive values express output flows. Each row corresponds to a different operational mode.

\begin{table}[h]
    \centering
    \resizebox{\textwidth}{!}{%
    \begin{threeparttable}
        \caption{Operational modes for PP\_CH4.}
        \begin{tabular}{c c c c c c}
            \toprule
            NG    & H2    & FUEL  & ELECTRICITY & HEAT\_MEDIUM\_T & CO2\_IND  \\
            $[\mathrm{kWh}]$ & [kWh] & [kWh] & [kWh] & [kWh] & [kg$_{\mathrm{CO}_2}$] \\
            \midrule
            -1\tnote{a} &  0 &  0 & 0.6\tnote{a}  & 0    & 0.2 \\
            -1\tnote{b} &  0 &  0 & 0.54\tnote{b} & 0.26\tnote{b} & 0.2 \\
             0 & -1\tnote{c} &  0 & 0.6\tnote{c}  & 0    & 0 \\
             0 &  0 & -1\tnote{c} & 0.6\tnote{c}  & 0    & 0.265 \\
             0 &  0 & -1\tnote{b,c} & 0.54\tnote{b,c} & 0.26\tnote{b,c} & 0.265 \\
            \bottomrule   
            \label{tab: operational modes PP_CH4}
        \end{tabular}
        \begin{tablenotes}
            \item[a] Based on the gas turbine combined cycle from \cite{guidati_gianfranco_value_2022}.
            \item[b] Assuming an 80\% total efficiency for the combined heat \& power operational mode. 
            \item[c] Assuming the same efficiency when using hydrogen/liquid fuels to generate power.
        \end{tablenotes}
    \end{threeparttable}}
\end{table}

\clearpage

CHP\_CH4 represents a combined heat and power (CHP) plant that primarily uses methane as its fuel source. This plant is designed to simultaneously generate both electricity and low-temperature heat, however, it can operate to purely produce electricity or alternatively provide electricity combined with medium-temperature heat. Furthermore, the technology can operate with alternative fuels like hydrogen or liquid fuels.

\begin{table}[h]
    \centering
\begin{threeparttable}
\caption{Economic parameters for CHP\_CH4.}
\begin{tabular}{l c c }
\toprule
\textbf{Parameter} & \textbf{Value} & \textbf{Unit}\\
\midrule
$c_{\mathrm{inv}}$ & 400 - 600\tnote{a} & CHF/kW\\
$c_{\mathrm{maint}}$ & (2\% - 5\%)$\times c_{\mathrm{inv}}$ & CHF/kW/y\\
$\tau$ & 25\tnote{a} & year(s)\\
\bottomrule
\end{tabular}
\begin{tablenotes}
	\item[a] Based on the gas combined heat \& power plant from \cite{guidati_gianfranco_value_2022}.
\end{tablenotes}
\end{threeparttable}

\end{table}

\begin{table}[h]
    \centering
    \resizebox{\textwidth}{!}{%
    \begin{threeparttable}
        \caption{Operational modes for CHP\_CH4.}
        
        \begin{tabular}{c c c c c c c}
            \toprule
            NG & H2 & FUEL & ELECTRICITY & HEAT\_LOW\_T\_DHN & HEAT\_MEDIUM\_T & CO2\_IND   \\
            $\mathrm{[kWh]}$ & [kWh] & [kWh] & [kWh] & [kWh] & [kWh] & [kg$_{\mathrm{CO}_2}$] \\
            \midrule
            -1\tnote{a} &  0 &  0 & 0.457\tnote{a}  & 0     & 0.347\tnote{a} & 0.2 \\
            -1\tnote{b} &  0 &  0 & 0.478\tnote{b}  & 0.358\tnote{b} & 0     & 0.2 \\
            -1\tnote{b} &  0 &  0 & 0.539\tnote{b}  & 0     & 0     & 0.2 \\
             0 & -1\tnote{c} &  0 & 0.457\tnote{c}  & 0     & 0.347\tnote{c} & 0 \\
             0 & -1\tnote{b,c} &  0 & 0.478\tnote{b,c}  & 0.358\tnote{b,c} & 0     & 0 \\
             0 & -1\tnote{b,c} &  0 & 0.539\tnote{b,c}  & 0     & 0     & 0 \\
             0 &  0 & -1\tnote{c} & 0.457\tnote{c}  & 0     & 0.347\tnote{c} & 0.265 \\
             0 &  0 & -1\tnote{b,c} & 0.478\tnote{b,c}  & 0.358\tnote{b,c} & 0     & 0.265 \\
             0 &  0 & -1\tnote{b,c} & 0.539\tnote{b,c}  & 0     & 0     & 0.265 \\

            \bottomrule
        \end{tabular}
        \begin{tablenotes}
            \item[a] Based on the gas combined heat \& power plant from \cite{guidati_gianfranco_value_2022}.
            \item[b] Assumptions taken from a newer version of \cite{marcucci_swiss_2021}.
            \item[c] Assuming the same efficiency when using hydrogen/liquid fuels to operate the plant.
        \end{tablenotes}
    \end{threeparttable}}
\end{table}

\clearpage

CHP\_WOOD represents a combined heat and power (CHP) plant that uses wood as its fuel source. This plant is designed to generate both electricity and low-temperature heat simultaneously but can also operate in different modes, such as solely producing electricity or providing electricity combined with medium-temperature heat.

\begin{table}[h]
    \centering
\begin{threeparttable}
\caption{Economic Parameters for CHP\_WOOD.}
\begin{tabular}{l c c }
\toprule
\textbf{Parameter} & \textbf{Value} & \textbf{Unit}\\
\midrule
$c_{\mathrm{inv}}$ & 2000 - 3000\tnote{a} & CHF/kW\\
$c_{\mathrm{maint}}$ & (2\% - 5\%)$\times c_{\mathrm{inv}}$ & CHF/kW/y\\
$\tau$ & 25\tnote{a} & year(s)\\
\bottomrule
\end{tabular}
\begin{tablenotes}
	\item[a] Based on the wood combined heat \& power plant from \cite{guidati_gianfranco_value_2022}.
\end{tablenotes}
\end{threeparttable}

\end{table}

\begin{table}[h]
    \centering
    \resizebox{\textwidth}{!}{%
    \begin{threeparttable}
        \caption{Operational modes for CHP\_WOOD.}
        
        \begin{tabular}{c c c c c}
            \toprule
            DRY\_BIOMASS & ELECTRICITY & HEAT\_LOW\_T\_DHN & HEAT\_MEDIUM\_T & CO2\_IND   \\
            $\mathrm{[kWh]}$ & [kWh]       & [kWh]             & [kWh]           & [kg$_{\mathrm{CO}_2}$] \\
            \midrule
            -1\tnote{a} & 0.164\tnote{a}  & 0.676\tnote{a} & 0     & 0.39 \\
            -1\tnote{b} & 0.234\tnote{b}  & 0     & 0     & 0.39 \\
            -1 & 0.116\tnote{b}  & 0     & 0.727\tnote{b} & 0.39 \\
            \bottomrule
        \end{tabular}
        \begin{tablenotes}
            \item[a] Based on the wood combined heat \& power plant from \cite{guidati_gianfranco_value_2022}.
            \item[b] Assumptions taken from a newer version of \cite{marcucci_swiss_2021}.
        \end{tablenotes}
    \end{threeparttable}}
\end{table}

\clearpage

CHP\_WASTE represents a combined heat and power plant that uses municipal waste as its fuel source. As for the CHP\_WOOD technology, it can operate by producing electricity alone or combined with either low- or medium-temperature heat.

\begin{table}[h]
    \centering
\begin{threeparttable}
\caption{Economic parameters for CHP\_WASTE.}
\begin{tabular}{l c c }
\toprule
\textbf{Parameter} & \textbf{Value} & \textbf{Unit}\\
\midrule
$c_{\mathrm{inv}}$ & 2000 - 3000\tnote{a} & CHF/kW\\
$c_{\mathrm{maint}}$ & (2\% - 5\%)$\times c_{\mathrm{inv}}$ & CHF/kW/y\\
$\tau$ & 25\tnote{a} & year(s)\\
\bottomrule
\end{tabular}
\begin{tablenotes}
	\item[a] Based on the waste combined heat \& power plant from \cite{guidati_gianfranco_value_2022}.
\end{tablenotes}
\end{threeparttable}

\end{table}

\begin{table}[h]
    \centering
    \resizebox{\textwidth}{!}{%
    \begin{threeparttable}
        \caption{Operational modes for CHP\_WASTE.}
        
        \begin{tabular}{c c c c c}
            \toprule
            WASTE & ELECTRICITY & HEAT\_LOW\_T\_DHN & HEAT\_MEDIUM\_T & CO2\_IND   \\
            $\mathrm{[kWh]}$ & [kWh]       & [kWh]             & [kWh]           & [kg$_{\mathrm{CO}_2}$] \\
            \midrule
            -1\tnote{a} & 0.119\tnote{a} & 0.663\tnote{a} & 0     & 0.33 \\
            -1\tnote{b} & 0.19\tnote{b}  & 0     & 0     & 0.33 \\
            -1\tnote{b} & 0.069\tnote{b} & 0     & 0.716\tnote{b} & 0.33 \\
            \bottomrule
        \end{tabular}
        \begin{tablenotes}
            \item[a] Based on the waste combined heat \& power plant from \cite{guidati_gianfranco_value_2022}.
            \item[b] Assumptions taken from a newer version of \cite{marcucci_swiss_2021}.
        \end{tablenotes}
    \end{threeparttable}}
    \label{tab:CHP_WASTE operational modes}
\end{table}

\clearpage

\subsubsection{High-temperature Heat}
High-temperature heat is essential for many industrial processes, including clinker production for cement, steelmaking, and chemical manufacturing. These processes require high temperatures that can be supplied using industrial burners powered by various fuels, including fossil sources such as coal, oil, and natural gas, as well as renewable alternatives like biomass (e.g., wood and digestate), waste, and hydrogen.

IND\_BURNER\_COAL represents an industrial burner that uses coal as combustion fuel. 

\begin{figure}[H]
\begin{minipage}[H]{.48\textwidth}
\centering
\includegraphics[width=0.7\linewidth]{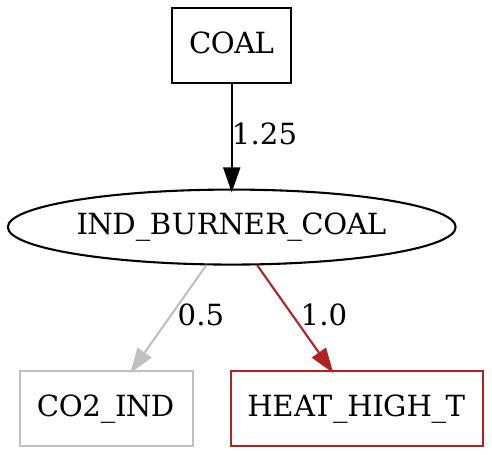}
\end{minipage}
\begin{minipage}{.5\textwidth}
\centering
\resizebox{\textwidth}{!}{%
\begin{threeparttable}
\begin{tabular}{l c c }
\toprule
\textbf{Parameter} & \textbf{Value} & \textbf{Unit}\\
\midrule
$c_{\mathrm{inv}}$ & 500 - 800\tnote{a} & CHF/kW\\
$c_{\mathrm{maint}}$ & (2\% - 5\%)$\times c_{\mathrm{inv}}$ & CHF/kW/y\\
$\tau$ & 25\tnote{a} & year(s)\\
\bottomrule
\end{tabular}
\begin{tablenotes}
	\item[a] Based on the industrial coal burner from \cite{guidati_gianfranco_value_2022}.
\end{tablenotes}
\end{threeparttable}
}
\end{minipage}
\caption{Input and output flows and economic parameters of IND\_BURNER\_COAL.}
\label{fig:IND_BURNER_COAL}
\end{figure}

IND\_BURNER\_DIESEL burns liquid fuel such as oil/diesel to provide high-temperature heat.

\begin{figure}[H]
\begin{minipage}[H]{.48\textwidth}
\centering
\includegraphics[width=0.7\linewidth]{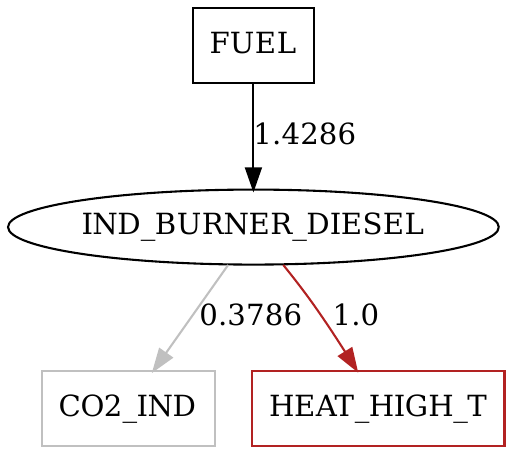}
\end{minipage}
\begin{minipage}{.5\textwidth}
\centering
\resizebox{\textwidth}{!}{%
\begin{threeparttable}
\begin{tabular}{l c c }
\toprule
\textbf{Parameter} & \textbf{Value} & \textbf{Unit}\\
\midrule
$c_{\mathrm{inv}}$ & 80\tnote{a} $\pm$ 30\% & CHF/kW\\
$c_{\mathrm{maint}}$ & (2\% - 5\%)$\times c_{\mathrm{inv}}$ & CHF/kW/y\\
$\tau$ & 25\tnote{a} & year(s)\\
\bottomrule
\end{tabular}
\begin{tablenotes}
	\item[a] Based on the industrial fuel burner from \cite{guidati_gianfranco_value_2022}.
\end{tablenotes}
\end{threeparttable}
}
\end{minipage}
\caption{Input and output flows and economic parameters of IND\_BURNER\_DIESEL.}
\label{fig:IND_BURNER_DIESEL}
\end{figure}

IND\_BURNER\_CH4 represents an industrial burner that combusts natural gas to produce heat.

\begin{figure}[H]
\begin{minipage}[H]{.48\textwidth}
\centering
\includegraphics[width=0.7\linewidth]{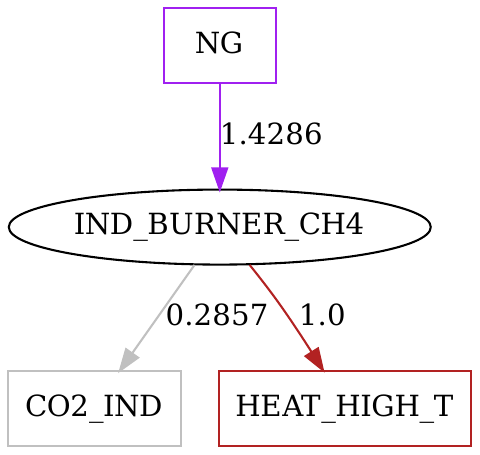}
\end{minipage}
\begin{minipage}{.5\textwidth}
\centering
\resizebox{\textwidth}{!}{%
\begin{threeparttable}
\begin{tabular}{l c c }
\toprule
\textbf{Parameter} & \textbf{Value} & \textbf{Unit}\\
\midrule
$c_{\mathrm{inv}}$ & 90\tnote{a} $\pm$ 30\% & CHF/kW\\
$c_{\mathrm{maint}}$ & (2\% - 5\%)$\times c_{\mathrm{inv}}$ & CHF/kW/y\\
$\tau$ & 25\tnote{a} & year(s)\\
\bottomrule
\end{tabular}
\begin{tablenotes}
	\item[a] Based on the industrial gas burner from \cite{guidati_gianfranco_value_2022}.
\end{tablenotes}
\end{threeparttable}
}
\end{minipage}
\caption{Input and output flows and economic parameters of IND\_BURNER\_CH4.}
\label{fig:IND_BURNER_CH4}
\end{figure}

IND\_BURNER\_WASTE burns waste to provide high-temperature heat for the industry.

\begin{figure}[H]
\begin{minipage}[H]{.48\textwidth}
\centering
\includegraphics[width=0.75\linewidth]{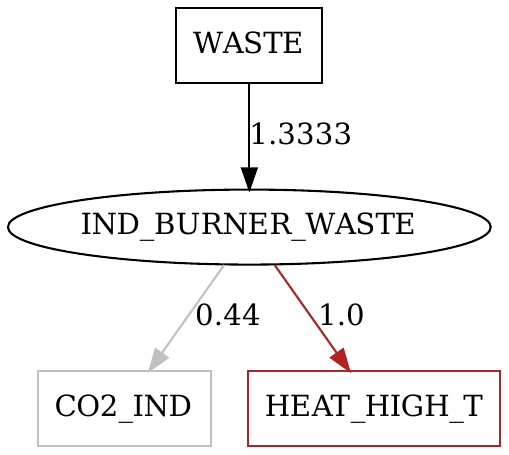}
\end{minipage}
\begin{minipage}{.5\textwidth}
\centering
\resizebox{\textwidth}{!}{%
\begin{threeparttable}
\begin{tabular}{l c c }
\toprule
\textbf{Parameter} & \textbf{Value} & \textbf{Unit}\\
\midrule
$c_{\mathrm{inv}}$ & 500 - 800\tnote{a} & CHF/kW\\
$c_{\mathrm{maint}}$ & (2\% - 5\%)$\times c_{\mathrm{inv}}$ & CHF/kW/y\\
$\tau$ & 25\tnote{a} & year(s)\\
\bottomrule
\end{tabular}
\begin{tablenotes}
	\item[a] Based on the industrial waste burner from \cite{guidati_gianfranco_value_2022}.
\end{tablenotes}
\end{threeparttable}
}
\end{minipage}
\caption{Input and output flows and economic parameters of IND\_BURNER\_WASTE.}
\label{fig:IND_BURNER_WASTE}
\end{figure}

IND\_BURNER\_DIGESTATE represents an industrial burner that burns dry digestate to produce heat.

\begin{figure}[H]
\begin{minipage}[H]{.48\textwidth}
\centering
\includegraphics[width=0.9\linewidth]{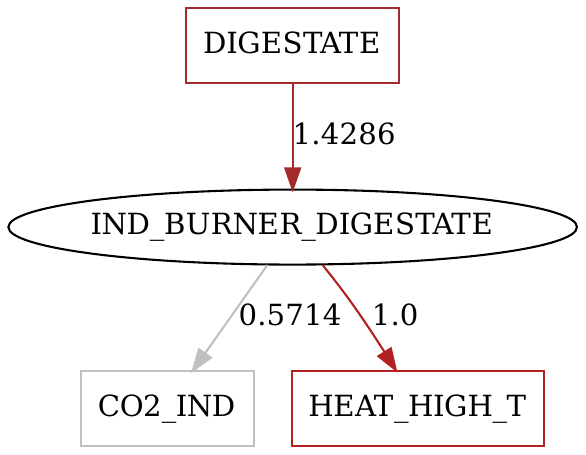}
\end{minipage}
\begin{minipage}{.5\textwidth}
\centering
\resizebox{\textwidth}{!}{%
\begin{threeparttable}
\begin{tabular}{l c c }
\toprule
\textbf{Parameter} & \textbf{Value} & \textbf{Unit}\\
\midrule
$c_{\mathrm{inv}}$ & 800\tnote{a} $\pm$ 30\% & CHF/kW\\
$c_{\mathrm{maint}}$ & (2\% - 5\%)$\times c_{\mathrm{inv}}$ & CHF/kW/y\\
$\tau$ & 25\tnote{b} & year(s)\\
\bottomrule
\end{tabular}
\begin{tablenotes}
	\item[a] Assumed to be at the upper bound of the industrial waste burner.
    \item[b] Assumed to be equivalent to the industrial waste burner.
    \item Note: We assume an energetic efficiency of 70\%. 
\end{tablenotes}
\end{threeparttable}
}
\end{minipage}
\caption{Input and output flows and economic parameters of IND\_BURNER\_DIGESTATE.}
\label{fig:IND_BURNER_DIGESTATE}
\end{figure}

IND\_BURNER\_WOOD burns wood to provide high-temperature heat.

\begin{figure}[H]
\begin{minipage}[H]{.48\textwidth}
\centering
\includegraphics[width=0.75\linewidth]{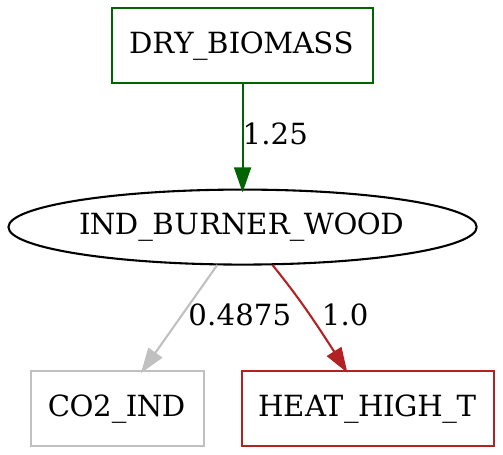}
\end{minipage}
\begin{minipage}{.5\textwidth}
\centering
\resizebox{\textwidth}{!}{%
\begin{threeparttable}
\begin{tabular}{l c c }
\toprule
\textbf{Parameter} & \textbf{Value} & \textbf{Unit}\\
\midrule
$c_{\mathrm{inv}}$ & 500 - 800\tnote{a} & CHF/kW\\
$c_{\mathrm{maint}}$ & (2\% - 5\%)$\times c_{\mathrm{inv}}$ & CHF/kW/y\\
$\tau$ & 25\tnote{a} & year(s)\\
\bottomrule
\end{tabular}
\begin{tablenotes}
	\item[a] Based on the industrial wood burner from \cite{guidati_gianfranco_value_2022}.
\end{tablenotes}
\end{threeparttable}
}
\end{minipage}
\caption{Input and output flows and economic parameters of IND\_BURNER\_WOOD.}
\label{fig:IND_BURNER_WOOD}
\end{figure}

IND\_BURNER\_H2 represents an industrial burner that combusts hydrogen to produce heat.

\begin{figure}[H]
\begin{minipage}[H]{.48\textwidth}
\centering
\includegraphics[width=0.7\linewidth]{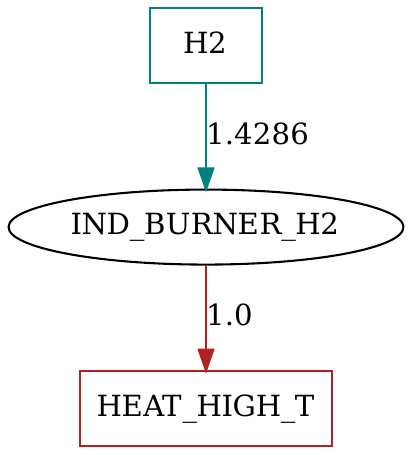}
\end{minipage}
\begin{minipage}{.5\textwidth}
\centering
\resizebox{\textwidth}{!}{%
\begin{threeparttable}
\begin{tabular}{l c c }
\toprule
\textbf{Parameter} & \textbf{Value} & \textbf{Unit}\\
\midrule
$c_{\mathrm{inv}}$ & 90\tnote{a} $\pm$ 30\% & CHF/kW\\
$c_{\mathrm{maint}}$ & (2\% - 5\%)$\times c_{\mathrm{inv}}$ & CHF/kW/y\\
$\tau$ & 25\tnote{a} & year(s)\\
\bottomrule
\end{tabular}
\begin{tablenotes}
	\item[a] Based on the industrial gas burner from \cite{guidati_gianfranco_value_2022}.
\end{tablenotes}
\end{threeparttable}
}
\end{minipage}
\caption{Input and output flows and economic parameters of IND\_BURNER\_H2.}
\label{fig:IND_BURNER_H2}
\end{figure}

DIRECT\_ELEC\_HIGH\_T represents high-temperature heat generated directly from electricity, typically using electric heating technologies such as resistance heaters or electric arc furnaces.

\begin{figure}[H]
\begin{minipage}[H]{.48\textwidth}
\centering
\includegraphics[width=0.8\linewidth]{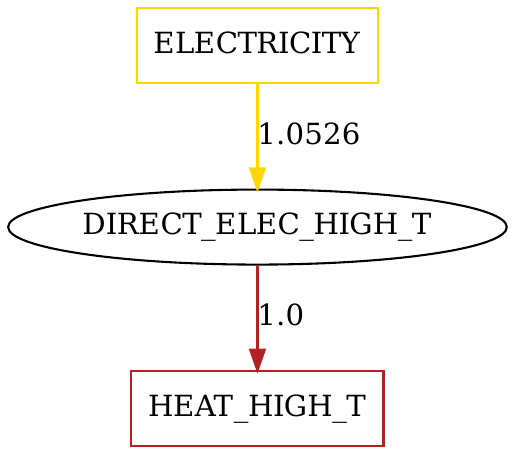}
\end{minipage}
\begin{minipage}{.5\textwidth}
\centering
\resizebox{\textwidth}{!}{%
\begin{threeparttable}
\begin{tabular}{l c c }
\toprule
\textbf{Parameter} & \textbf{Value} & \textbf{Unit}\\
\midrule
$c_{\mathrm{inv}}$ & 275\tnote{a} $\pm$ 30\% & CHF/kW\\
$c_{\mathrm{maint}}$ & (2\% - 5\%)$\times c_{\mathrm{inv}}$ & CHF/kW/y\\
$\tau$ & 25\tnote{a} & year(s)\\
\bottomrule
\end{tabular}
\begin{tablenotes}
	\item[a] Based on the industrial electric heater from \cite{guidati_gianfranco_value_2022}.
\end{tablenotes}
\end{threeparttable}
}
\end{minipage}
\caption{Input and output flows and economic parameters of DIRECT\_ELEC\_HIGH\_T.}
\label{fig:DIRECT_ELEC_HIGH_T}
\end{figure}

\subsubsection{Cooling}
With rising ambient air temperatures, ACTIVE\_COOLING, e.g., air conditioning, becomes increasingly important to maintain comfortable temperatures during summer months. We model the ACTIVE\_COOLING as an air source heat pump with a coefficient of performance of 3.

\begin{figure}[H]
\begin{minipage}[H]{.48\textwidth}
\centering
\includegraphics[width=0.7\linewidth]{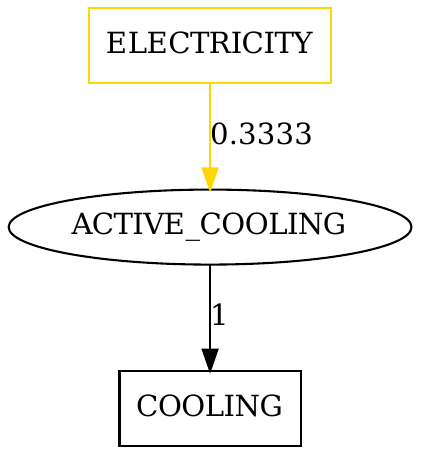}
\end{minipage}
\begin{minipage}{.5\textwidth}
\centering
\resizebox{\textwidth}{!}{%
\begin{threeparttable}
\begin{tabular}{l c c }
\toprule
\textbf{Parameter} & \textbf{Value} & \textbf{Unit}\\
\midrule
$c_{\mathrm{inv}}$ & 2000 - 3000\tnote{a} & CHF/kW\\
$c_{\mathrm{maint}}$ & (2\% - 5\%)$\times c_{\mathrm{inv}}$ & CHF/kW/y\\
$\tau$ & 25\tnote{a} & year(s)\\
\bottomrule
\end{tabular}
\begin{tablenotes}
	\item[a] Based on the residential air source heat pump from \cite{guidati_gianfranco_value_2022}.
\end{tablenotes}
\end{threeparttable}
}
\end{minipage}
\caption{Input and output flows and economic parameters of ACTIVE\_COOLING.}
\label{fig:ACTIVE_COOLING}
\end{figure}

\subsubsection{Endogenous Products}

In this section, technologies are listed that do not fulfil end-use demands but produce endogenous products that other technologies can further convert.

\paragraph{Methane}

Methane is a gaseous energy carrier that can be stored seasonally and can be converted to fulfil several end-use demands in the energy system, e.g., heat, electricity, mobility, and high-value chemicals.

In our model, wet biomass, e.g., green wastes, must undergo an anaerobic digestion (FERMENTATION\_CENTRAL) step. In this step, bacteria break down the wet biomass in the absence of oxygen and produce gaseous methane and CO$_2$ as well as solid digestate.

\begin{figure}[H]
\begin{minipage}[H]{.48\textwidth}
\centering
\includegraphics[width=0.95\linewidth]{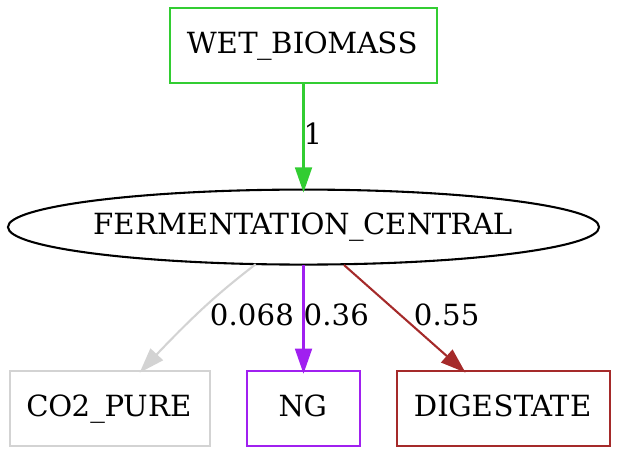}
\end{minipage}
\begin{minipage}{.5\textwidth}
\centering
\resizebox{\textwidth}{!}{%
\begin{threeparttable}
\begin{tabular}{l c c }
\toprule
\textbf{Parameter} & \textbf{Value} & \textbf{Unit}\\
\midrule
$c_{\mathrm{inv}}$ & 1200\tnote{a} $\pm$ 30\% & CHF/kW\\
$c_{\mathrm{maint}}$ & (2\% - 5\%)$\times c_{\mathrm{inv}}$ & CHF/kW/y\\
$\tau$ & 25\tnote{a} & year(s)\\
\bottomrule
\end{tabular}
\begin{tablenotes}
	\item[a] Based on the anaerobic digestion plant from \cite{guidati_gianfranco_value_2022}.
\end{tablenotes}
\end{threeparttable}
}
\end{minipage}
\caption{Input and output flows and economic parameters of FERMENTATION\_CENTRAL.}
\label{fig:FERMENTATION_CENTRAL}
\end{figure}

Analogous to wet biomass, animal manure must also be processed in anaerobic digestion (FERMENTATION\_RURAL). 

\begin{figure}[H]
\begin{minipage}[H]{.48\textwidth}
\centering
\includegraphics[width=0.95\linewidth]{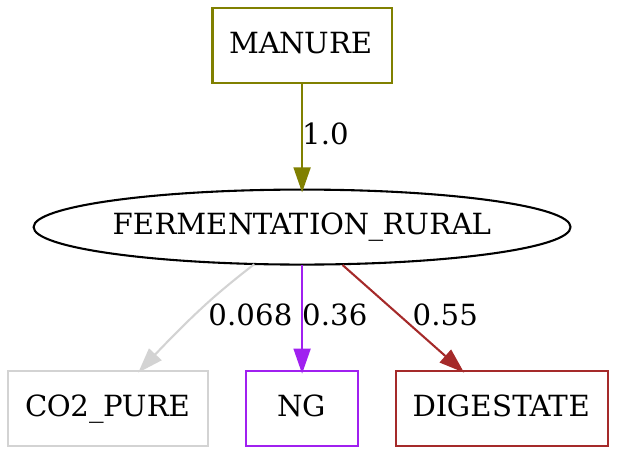}
\end{minipage}
\begin{minipage}{.5\textwidth}
\centering
\resizebox{\textwidth}{!}{%
\begin{threeparttable}
\begin{tabular}{l c c }
\toprule
\textbf{Parameter} & \textbf{Value} & \textbf{Unit}\\
\midrule
$c_{\mathrm{inv}}$ & 1200\tnote{a} $\pm$ 30\% & CHF/kW\\
$c_{\mathrm{maint}}$ & (2\% - 5\%)$\times c_{\mathrm{inv}}$ & CHF/kW/y\\
$\tau$ & 25\tnote{a} & year(s)\\
\bottomrule
\end{tabular}
\begin{tablenotes}
	\item[a] Based on the anaerobic digestion plant from \cite{guidati_gianfranco_value_2022}.
\end{tablenotes}
\end{threeparttable}
}
\end{minipage}
\caption{Input and output flows and economic parameters of FERMENTATION\_RURAL.}
\label{fig:FERMENTATION_RURAL}
\end{figure}

In wastewater treatment plants (FERMENTATION\_WWPLANT) sewage sludge is digested by bacteria in the absence of oxygen to produce methane, CO$_2$, and digestate. 

\begin{figure}[H]
\begin{minipage}[H]{.48\textwidth}
\centering
\includegraphics[width=0.95\linewidth]{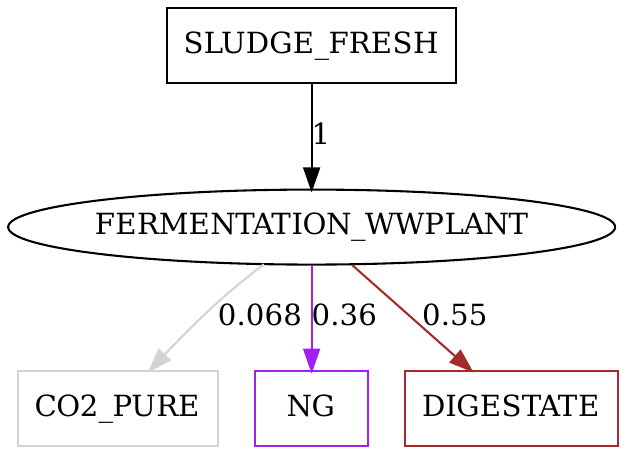}
\end{minipage}
\begin{minipage}{.5\textwidth}
\centering
\resizebox{\textwidth}{!}{%
\begin{threeparttable}
\begin{tabular}{l c c }
\toprule
\textbf{Parameter} & \textbf{Value} & \textbf{Unit}\\
\midrule
$c_{\mathrm{inv}}$ & 1200\tnote{a} $\pm$ 30\% & CHF/kW\\
$c_{\mathrm{maint}}$ & (2\% - 5\%)$\times c_{\mathrm{inv}}$ & CHF/kW/y\\
$\tau$ & 25\tnote{a} & year(s)\\
\bottomrule
\end{tabular}
\begin{tablenotes}
	\item[a] Based on the anaerobic digestion plant from \cite{guidati_gianfranco_value_2022}.
\end{tablenotes}
\end{threeparttable}
}
\end{minipage}
\caption{Input and output flows and economic parameters of FERMENTATION\_WWPLANT.}
\label{fig:FERMENTATION_WWPLANT}
\end{figure}

In the GASIFICATION-CH4 process, wood undergoes gasification to produce syngas (CO, H$_2$, CO$_2$). The syngas is then methanated, converting CO and H$_2$ into methane and water.

\begin{figure}[H]
\begin{minipage}[H]{.48\textwidth}
\centering
\includegraphics[width=0.98\linewidth]{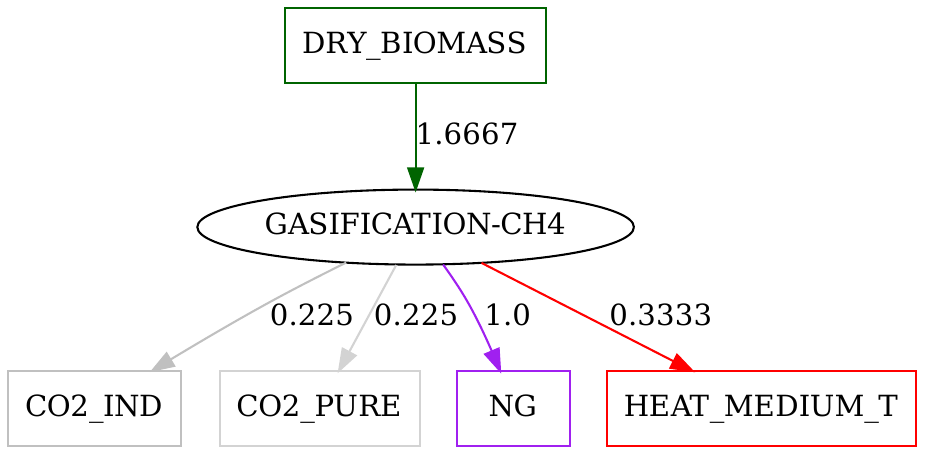}
\end{minipage}
\begin{minipage}{.5\textwidth}
\centering
\resizebox{\textwidth}{!}{%
\begin{threeparttable}
\begin{tabular}{l c c }
\toprule
\textbf{Parameter} & \textbf{Value} & \textbf{Unit}\\
\midrule
$c_{\mathrm{inv}}$ & 2000 - 3000\tnote{a} & CHF/kW\\
$c_{\mathrm{maint}}$ & (2\% - 5\%)$\times c_{\mathrm{inv}}$ & CHF/kW/y\\
$\tau$ & 25\tnote{a} & year(s)\\
\bottomrule
\end{tabular}
\begin{tablenotes}
	\item[a] Based on the gasification to methane technology from \cite{guidati_gianfranco_value_2022}.
    \item Note: The reported efficiency in \cite{guidati_gianfranco_value_2022} does not account for the utilization of waste heat; we assume a total energy efficiency of 80\%.
\end{tablenotes}
\end{threeparttable}
}
\end{minipage}
\caption{Input and output flows and economic parameters of GASIFICATION-CH4.}
\label{fig:GASIFICATION-CH4}
\end{figure}

Hydrothermal gasification (HTG) is a relatively new thermochemical technology that can convert digestate into gas at high pressure and temperature, producing primarily methane, hydrogen and CO$_2$.

\begin{figure}[H]
\begin{minipage}[H]{.48\textwidth}
\centering
\includegraphics[width=0.6\linewidth]{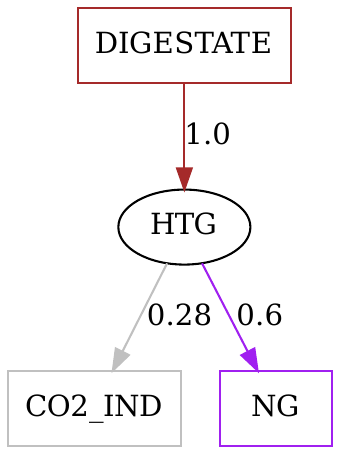}
\end{minipage}
\begin{minipage}{.5\textwidth}
\centering
\resizebox{\textwidth}{!}{%
\begin{threeparttable}
\begin{tabular}{l c c }
\toprule
\textbf{Parameter} & \textbf{Value} & \textbf{Unit}\\
\midrule
$c_{\mathrm{inv}}$ & 1200 $\pm$ 30\% & CHF/kW\\
$c_{\mathrm{maint}}$ & (2\% - 5\%)$\times c_{\mathrm{inv}}$ & CHF/kW/y\\
$\tau$ & 25 & year(s)\\
\bottomrule
\end{tabular}
\begin{tablenotes}
	\item Note: The efficiency, cost and lifetime are estimated from discussions with experts.
\end{tablenotes}
\end{threeparttable}
}
\end{minipage}
\caption{Input and output flows and economic parameters of HTG.}
\label{fig:HTG}
\end{figure}

In the SABATIER process, hydrogen reacts with pure carbon dioxide over a catalyst to produce methane and water.

\begin{figure}[H]
\begin{minipage}[H]{.48\textwidth}
\centering
\includegraphics[width=0.7\linewidth]{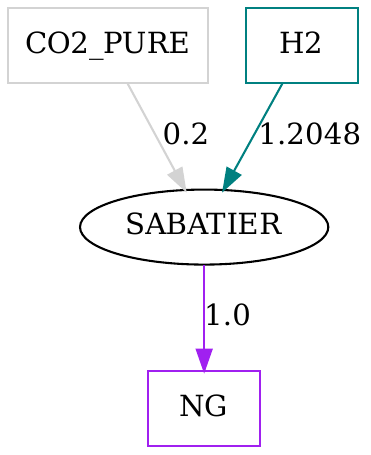}
\end{minipage}
\begin{minipage}{.5\textwidth}
\centering
\resizebox{\textwidth}{!}{%
\begin{threeparttable}
\begin{tabular}{l c c }
\toprule
\textbf{Parameter} & \textbf{Value} & \textbf{Unit}\\
\midrule
$c_{\mathrm{inv}}$ & 1000 - 2000\tnote{a} & CHF/kW\\
$c_{\mathrm{maint}}$ & (2\% - 5\%)$\times c_{\mathrm{inv}}$ & CHF/kW/y\\
$\tau$ & 25\tnote{a} & year(s)\\
\bottomrule
\end{tabular}
\begin{tablenotes}
	\item[a] Based on the methanation process from \cite{guidati_gianfranco_value_2022}.
\end{tablenotes}
\end{threeparttable}
}
\end{minipage}
\caption{Input and output flows and economic parameters of SABATIER.}
\label{fig:SABATIER}
\end{figure}

\paragraph{Liquid fuel}

Liquid fuel is an intermediary product, which subsequently can be used to provide various end-use demands, e.g., mobility, electricity, heat, and high-value chemicals. There are several options to produce liquid fuels:

Power-to-liquid (P2L) uses hydrogen together with pure CO$_2$ to produce liquid fuel exothermically.

\begin{figure}[H]
\begin{minipage}[H]{.48\textwidth}
\centering
\includegraphics[width=0.85\linewidth]{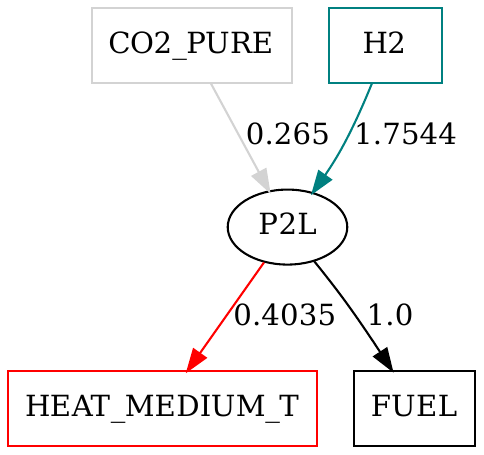}
\end{minipage}
\begin{minipage}{.5\textwidth}
\centering
\resizebox{\textwidth}{!}{%
\begin{threeparttable}
\begin{tabular}{l c c }
\toprule
\textbf{Parameter} & \textbf{Value} & \textbf{Unit}\\
\midrule
$c_{\mathrm{inv}}$ & 2500\tnote{a} $\pm$ 30\% & CHF/kW\\
$c_{\mathrm{maint}}$ & (2\% - 5\%)$\times c_{\mathrm{inv}}$ & CHF/kW/y\\
$\tau$ & 25\tnote{a} & year(s)\\
\bottomrule
\end{tabular}
\begin{tablenotes}
	\item[a] Based on the power-to-liquid technology from \cite{guidati_gianfranco_value_2022}.
    \item Note: The reported efficiency in \cite{guidati_gianfranco_value_2022} does not account for the utilization of waste heat; we assume a total energy efficiency of 80\%.
\end{tablenotes}
\end{threeparttable}
}
\end{minipage}
\caption{Input and output flows and economic parameters of P2L.}
\label{fig:P2L}
\end{figure}

Biomass-to-Liquid (B2L) uses the hydrocarbons from wood to produce liquid fuel. In the exothermic process, CO$_2$ is created as a byproduct.

\begin{figure}[H]
\begin{minipage}[H]{.48\textwidth}
\centering
\includegraphics[width=0.98\linewidth]{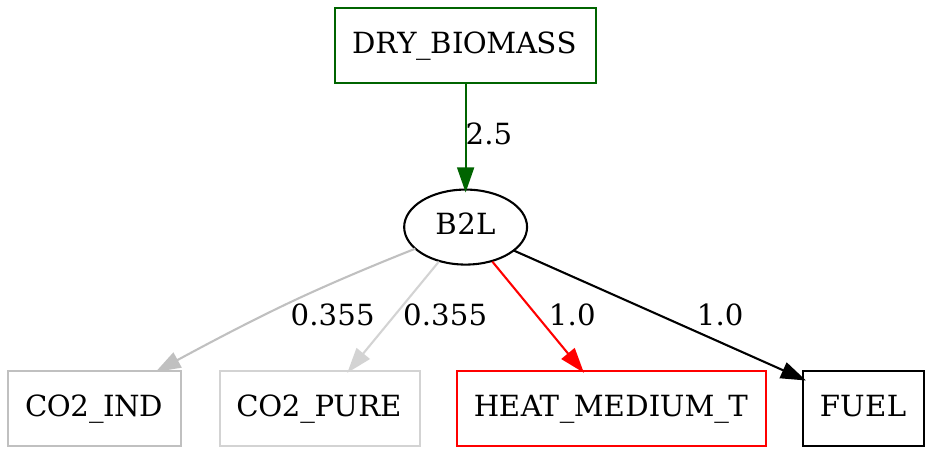}
\end{minipage}
\begin{minipage}{.5\textwidth}
\centering
\resizebox{\textwidth}{!}{%
\begin{threeparttable}
\begin{tabular}{l c c }
\toprule
\textbf{Parameter} & \textbf{Value} & \textbf{Unit}\\
\midrule
$c_{\mathrm{inv}}$ & 3500 $\pm$ 30\% & CHF/kW\\
$c_{\mathrm{maint}}$ & (2\% - 5\%)$\times c_{\mathrm{inv}}$ & CHF/kW/y\\
$\tau$ & 25 & year(s)\\
\bottomrule
\end{tabular}
\begin{tablenotes}
	\item[a] Based on the biomass-to-liquid technology from \cite{guidati_gianfranco_value_2022}.
    \item Note: The reported efficiency in \cite{guidati_gianfranco_value_2022} does not account for the utilization of waste heat; we assume a total energy efficiency of 80\%.
\end{tablenotes}
\end{threeparttable}
}
\end{minipage}
\caption{Input and output flows and economic parameters of B2L.}
\label{fig:B2L}
\end{figure}

Hydrothermal Liquefaction (HTL) transforms digestate into liquid fuels with CO$_2$ as a byproduct in the flue gas stream.  

\begin{figure}[H]
\begin{minipage}[H]{.48\textwidth}
\centering
\includegraphics[width=0.6\linewidth]{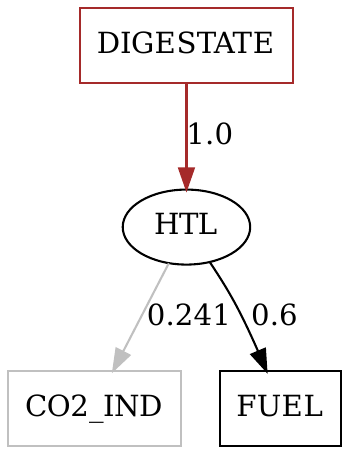}
\end{minipage}
\begin{minipage}{.5\textwidth}
\centering
\resizebox{\textwidth}{!}{%
\begin{threeparttable}
\begin{tabular}{l c c }
\toprule
\textbf{Parameter} & \textbf{Value} & \textbf{Unit}\\
\midrule
$c_{\mathrm{inv}}$ & 1200 $\pm$ 30\% & CHF/kW\\
$c_{\mathrm{maint}}$ & (2\% - 5\%)$\times c_{\mathrm{inv}}$ & CHF/kW/y\\
$\tau$ & 25 & year(s)\\
\bottomrule
\end{tabular}
\begin{tablenotes}
	\item Note: Assumed to have similar efficiency, cost and lifetime as HTG.
\end{tablenotes}
\end{threeparttable}
}
\end{minipage}
\caption{Input and output flows and economic parameters of HTL.}
\label{fig:HTL}
\end{figure}

\paragraph{Coal}
Coal is an energy-dense carrier that can be stored and transported efficiently. It can be converted to meet diverse energy demands, including industrial heat and electricity.

In PYROLYSIS\_SLOW\_WOOD, wood is pyrolyzed in the absence of oxygen, primarily producing biochar.

\begin{figure}[H]
\begin{minipage}[H]{.48\textwidth}
\centering
\includegraphics[width=\linewidth]{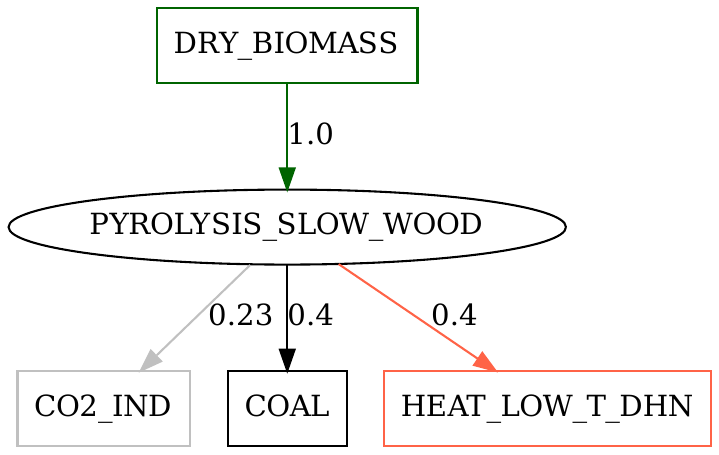}
\end{minipage}
\begin{minipage}{.5\textwidth}
\centering
\resizebox{\textwidth}{!}{%
\begin{threeparttable}
\begin{tabular}{l c c }
\toprule
\textbf{Parameter} & \textbf{Value} & \textbf{Unit}\\
\midrule
$c_{\mathrm{inv}}$ & 1200\tnote{a} $\pm$ 30\% & CHF/kW\\
$c_{\mathrm{maint}}$ & (2\% - 5\%)$\times c_{\mathrm{inv}}$ & CHF/kW/y\\
$\tau$ & 25\tnote{a} & year(s)\\
\bottomrule
\end{tabular}
\begin{tablenotes}
	\item[a] Based on the pyrolysis of wood technology from \cite{guidati_gianfranco_value_2022}.
    \item Note: The reported efficiency in \cite{guidati_gianfranco_value_2022} does not account for the utilization of waste heat; we assume a total energy efficiency of 80\%.
\end{tablenotes}
\end{threeparttable}
}
\end{minipage}
\caption{Input and output flows and economic parameters of PYROLYSIS\_SLOW\_WOOD.}
\label{fig:PYROLYSIS_SLOW_WOOD}
\end{figure}

Hydrothermal carbonization (HTC) can convert digestate exothermically at moderate temperature and pressure into a coal-like product and water.

\begin{figure}[H]
\begin{minipage}[H]{.48\textwidth}
\centering
\includegraphics[width=\linewidth]{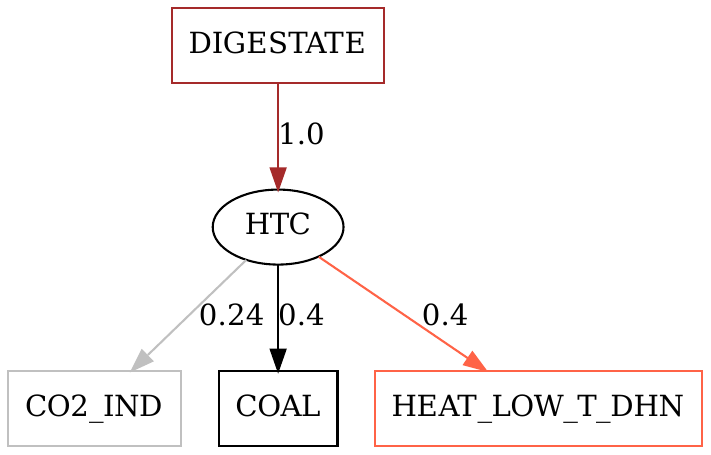}
\end{minipage}
\begin{minipage}{.5\textwidth}
\centering
\resizebox{\textwidth}{!}{%
\begin{threeparttable}
\begin{tabular}{l c c }
\toprule
\textbf{Parameter} & \textbf{Value} & \textbf{Unit}\\
\midrule
$c_{\mathrm{inv}}$ & 1200\tnote{a} $\pm$ 30\% & CHF/kW\\
$c_{\mathrm{maint}}$ & (2\% - 5\%)$\times c_{\mathrm{inv}}$ & CHF/kW/y\\
$\tau$ & 25\tnote{a} & year(s)\\
\bottomrule
\end{tabular}
\begin{tablenotes}
	\item[a] Based on the hydrothermal carbonization technology from \cite{guidati_gianfranco_value_2022}.
    \item Note: The reported efficiency in \cite{guidati_gianfranco_value_2022} does not account for the utilization of waste heat; we assume a total energy efficiency of 80\%.
\end{tablenotes}
\end{threeparttable}
}
\end{minipage}
\caption{Input and output flows and economic parameters of HTC.}
\label{fig:HTC}
\end{figure}

\paragraph{Hydrogen}
Hydrogen is a carbon-free energy carrier that can be stored and transported efficiently. It can be converted to meet diverse energy demands, including heat, electricity, and mobility.

In ELECTROLYSIS, water is split into hydrogen and oxygen using an electric current. If powered by renewable electricity, the process has the potential to produce clean hydrogen.

\begin{figure}[H]
\begin{minipage}[H]{.48\textwidth}
\centering
\includegraphics[width=0.7\linewidth]{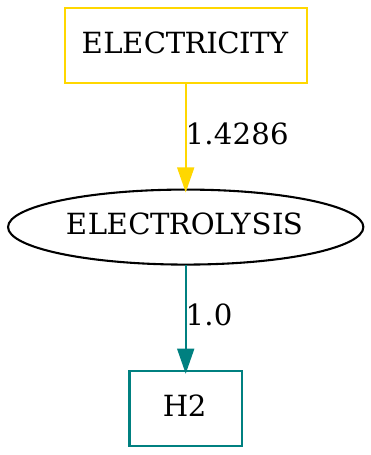}
\end{minipage}
\begin{minipage}{.5\textwidth}
\centering
\resizebox{\textwidth}{!}{%
\begin{threeparttable}
\begin{tabular}{l c c }
\toprule
\textbf{Parameter} & \textbf{Value} & \textbf{Unit}\\
\midrule
$c_{\mathrm{inv}}$ & 600 - 1500\tnote{a} & CHF/kW\\
$c_{\mathrm{maint}}$ & (2\% - 5\%)$\times c_{\mathrm{inv}}$ & CHF/kW/y\\
$\tau$ & 25\tnote{a} & year(s)\\
\bottomrule
\end{tabular}
\begin{tablenotes}
	\item[a] Based on the electrolysis technology from \cite{guidati_gianfranco_value_2022}.
\end{tablenotes}
\end{threeparttable}
}
\end{minipage}
\caption{Input and output flows and economic parameters of ELECTROLYSIS.}
\label{fig:ELECTROLYSIS}
\end{figure}

Wood can thermochemically be converted into syngas through gasification (H2-GASIFICATION) at high temperatures. The syngas is then processed, typically via water-gas shift reactions, to separate hydrogen. 

\begin{figure}[H]
\begin{minipage}[H]{.48\textwidth}
\centering
\includegraphics[width=\linewidth]{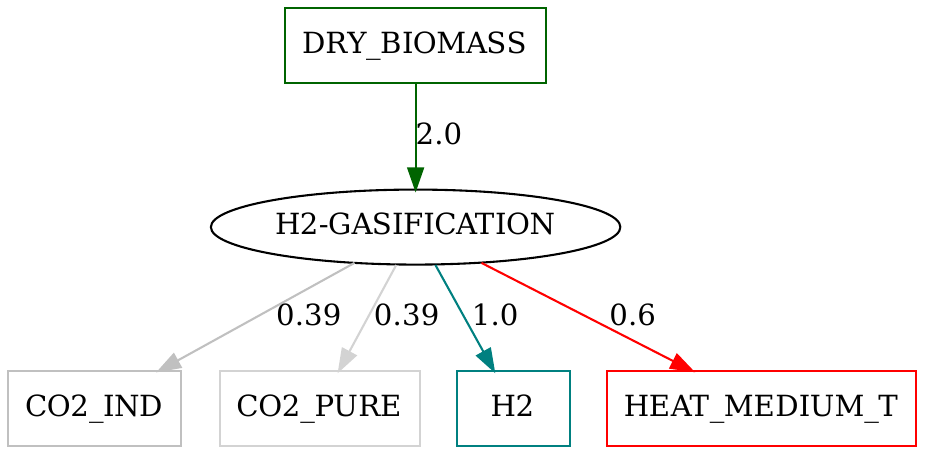}
\end{minipage}
\begin{minipage}{.5\textwidth}
\centering
\resizebox{\textwidth}{!}{%
\begin{threeparttable}
\begin{tabular}{l c c }
\toprule
\textbf{Parameter} & \textbf{Value} & \textbf{Unit}\\
\midrule
$c_{\mathrm{inv}}$ & 3000 - 4000\tnote{a} & CHF/kW\\
$c_{\mathrm{maint}}$ & (2\% - 5\%)$\times c_{\mathrm{inv}}$ & CHF/kW/y\\
$\tau$ & 25\tnote{a} & year(s)\\
\bottomrule
\end{tabular}
\begin{tablenotes}
	\item[a] Based on the gasification to hydrogen technology from \cite{guidati_gianfranco_value_2022}.
    \item Note: The reported efficiency in \cite{guidati_gianfranco_value_2022} does not account for the utilization of waste heat; we assume a total energy efficiency of 80\%.
\end{tablenotes}
\end{threeparttable}
}
\end{minipage}
\caption{Input and output flows and economic parameters of H2-GASIFICATION.}
\label{fig:H2-GASIFICATION}
\end{figure}

In STEAM\_METHANE\_REFORMING, methane reacts with steam at high temperatures to produce hydrogen and carbon monoxide. A subsequent water-gas shift reaction converts CO and steam into additional hydrogen and CO$_2$. This process is today the primary method for hydrogen production.

\begin{figure}[H]
\begin{minipage}[H]{.48\textwidth}
\centering
\includegraphics[width=\linewidth]{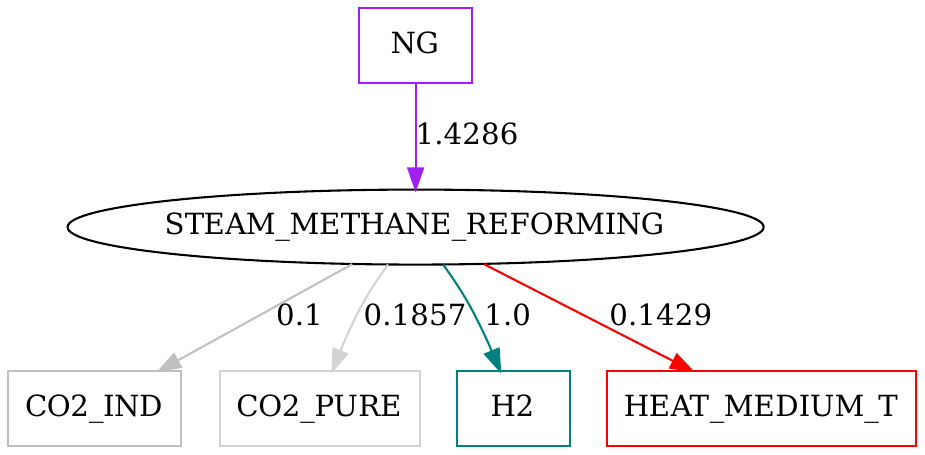}
\end{minipage}
\begin{minipage}{.5\textwidth}
\centering
\resizebox{\textwidth}{!}{%
\begin{threeparttable}
\begin{tabular}{l c c }
\toprule
\textbf{Parameter} & \textbf{Value} & \textbf{Unit}\\
\midrule
$c_{\mathrm{inv}}$ & 1000 - 2000\tnote{a} & CHF/kW\\
$c_{\mathrm{maint}}$ & (2\% - 5\%)$\times c_{\mathrm{inv}}$ & CHF/kW/y\\
$\tau$ & 25\tnote{a} & year(s)\\
\bottomrule
\end{tabular}
\begin{tablenotes}
	\item[a] Based on the steam methane reforming technology from \cite{guidati_gianfranco_value_2022}.
    \item Note: The reported efficiency in \cite{guidati_gianfranco_value_2022} does not account for the utilization of waste heat; we assume a total energy efficiency of 80\%.
\end{tablenotes}
\end{threeparttable}
}
\end{minipage}
\caption{Input and output flows and economic parameters of STEAM\_METHANE\_REFORMING.}
\label{fig:STEAM_METHANE_REFORMING}
\end{figure}

Methane can also be thermally decomposed in the absence of oxygen \linebreak (METHANE\_PYROLYSIS) to produce hydrogen and solid carbon. This process enables low-emission hydrogen production without generating CO$_2$ as a byproduct.

\begin{figure}[H]
\begin{minipage}[H]{.48\textwidth}
\centering
\includegraphics[width=0.8\linewidth]{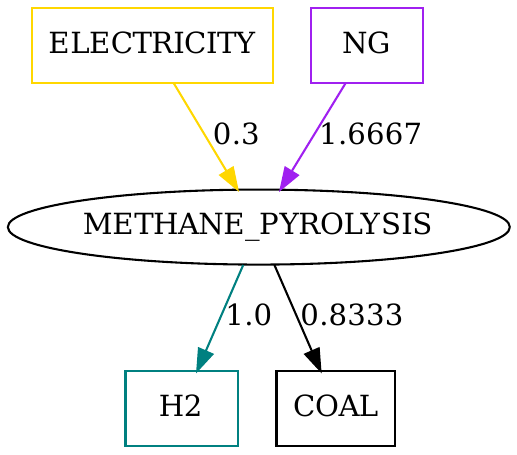}
\end{minipage}
\begin{minipage}{.5\textwidth}
\centering
\resizebox{\textwidth}{!}{%
\begin{threeparttable}
\begin{tabular}{l c c }
\toprule
\textbf{Parameter} & \textbf{Value} & \textbf{Unit}\\
\midrule
$c_{\mathrm{inv}}$ & 1200\tnote{a} $\pm$ 30\% & CHF/kW\\
$c_{\mathrm{maint}}$ & (2\% - 5\%)$\times c_{\mathrm{inv}}$ & CHF/kW/y\\
$\tau$ & 25\tnote{a} & year(s)\\
\bottomrule
\end{tabular}
\begin{tablenotes}
	\item[a] Assumed to be similar as for PYROLYSIS\_SLOW\_WOOD.
    \item Note: We assume a complete conversion from methane to coal and hydrogen. The required electricity for the conversion is an estimation.
\end{tablenotes}
\end{threeparttable}
}
\end{minipage}
\caption{Input and output flows and economic parameters of METHANE\_PYROLYSIS.}
\label{fig:METHANE_PYROLYSIS}
\end{figure}

In AMMONIA\_TO\_HYDROGEN, ammonia is cracked at high temperatures to produce hydrogen and nitrogen. This process enables hydrogen storage and transport in the form of ammonia, which can be converted back into hydrogen.

\begin{figure}[H]
\begin{minipage}[H]{.48\textwidth}
\centering
\includegraphics[width=0.8\linewidth]{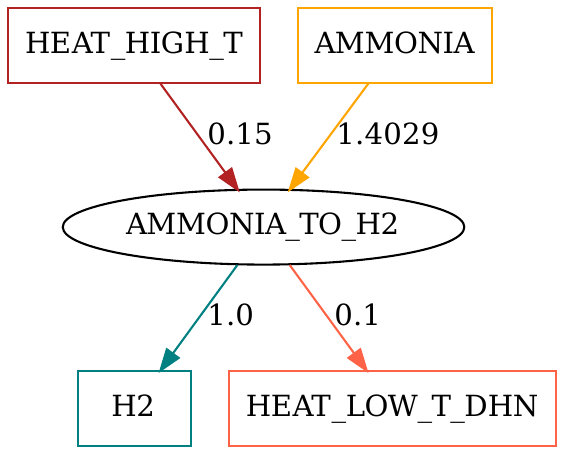}
\end{minipage}
\begin{minipage}{.5\textwidth}
\centering
\resizebox{\textwidth}{!}{%
\begin{threeparttable}
\begin{tabular}{l c c }
\toprule
\textbf{Parameter} & \textbf{Value} & \textbf{Unit}\\
\midrule
$c_{\mathrm{inv}}$ & 1365\tnote{a} $\pm$ 30\% & CHF/kW\\
$c_{\mathrm{maint}}$ & 38.0\tnote{a} $\pm$ 30\% & CHF/kW/y\\
$\tau$ & 25\tnote{a} & year(s)\\
$c_{\mathrm{p}}$ & 85.0\tnote{a} & \%\\
\bottomrule
\end{tabular}
\begin{tablenotes}
    \item Note: Conversion efficiency based on the ammonia to hydrogen process from \cite{limpens_generating_2021} in the year 2050.
	\item[a] Based on the Ammonia cracking technology from \cite{noauthor_input_nodate}.
    
\end{tablenotes}
\end{threeparttable}
}
\end{minipage}
\caption{Input and output flows and economic parameters of AMMONIA\_TO\_H2.}
\label{fig:AMMONIA_TO_H2}
\end{figure}

\subsubsection{Non-energy Demands}

Our model considers Cement, Ammonia, Methanol, High-value chemicals, and Plastic as non-energy demands in the energy system. 

\paragraph{Cement}

In a cement plant (CEM\_PLANT), limestone (CaCO$_3$) is calcined at high temperatures to produce clinker, the main component of cement. This energy-intensive process releases geogenic CO$_2$ emissions from limestone decomposition, making it a major source of industrial emissions.

\begin{figure}[H]
\begin{minipage}[H]{.45\textwidth}
\centering
\includegraphics[width=0.8\linewidth]{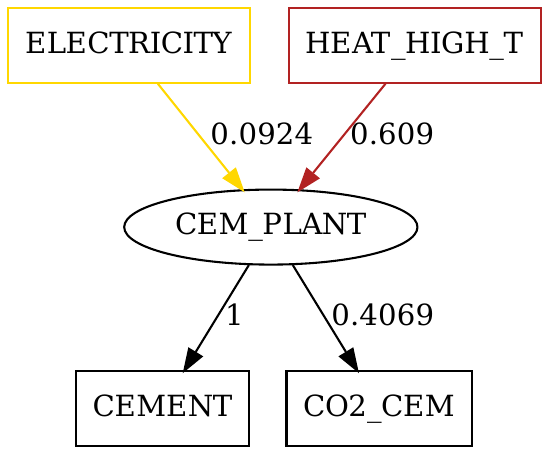}
\end{minipage}
\begin{minipage}{.5\textwidth}
\centering
\resizebox{\textwidth}{!}{%
\begin{threeparttable}
\begin{tabular}{l c c }
\toprule
\textbf{Parameter} & \textbf{Value} & \textbf{Unit}\\
\midrule
$c_{\mathrm{inv}}$ & 0\tnote{a} & CHF/kg/h\\
$c_{\mathrm{maint}}$ & 0\tnote{a} & CHF/kg/h/y\\
$\tau$ & 1 & year(s)\\
\bottomrule
\end{tabular}
\begin{tablenotes}
    \item Note: Conversion efficiencies are based on calculations using data from \cite{cemsuisse_yearly_2023}.
    \item[a] Due to lack of data, we do not assign costs to the cement plant. Since no competing technology exists, this assumption impacts only the objective value, not the solution.

\end{tablenotes}
\end{threeparttable}
}
\end{minipage}
\caption{Input and output flows and economic parameters of CEM\_PLANT.}
\label{fig:CEM_PLANT}
\end{figure}

\paragraph{Ammonia}

In the HABER\_BOSCH process, nitrogen and hydrogen react in the presence of a catalyst to produce ammonia. This energy-intensive process is today the primary method for industrial ammonia production.

\begin{figure}[H]
\begin{minipage}[H]{.48\textwidth}
\centering
\includegraphics[width=0.8\linewidth]{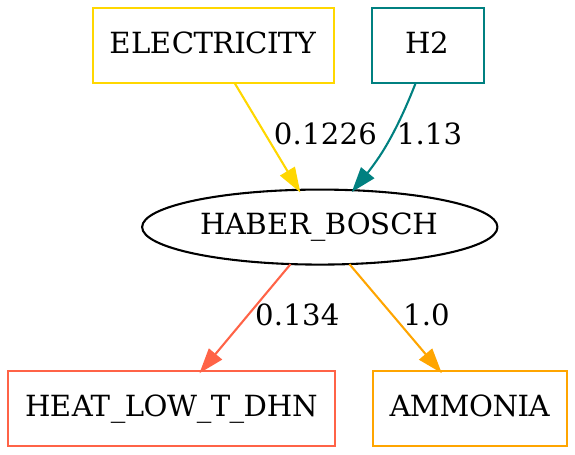}
\end{minipage}
\begin{minipage}{.5\textwidth}
\centering
\resizebox{\textwidth}{!}{%
\begin{threeparttable}
\begin{tabular}{l c c }
\toprule
\textbf{Parameter} & \textbf{Value} & \textbf{Unit}\\
\midrule
$c_{\mathrm{inv}}$ & 847\tnote{a} $\pm$ 30\% & CHF/kW\\
$c_{\mathrm{maint}}$ & 16.6\tnote{a} $\pm$ 30\% & CHF/kW/y\\
$\tau$ & 20\tnote{a} & year(s)\\
$c_{\mathrm{p}}$ & 85.0\tnote{a} & \%\\
\bottomrule
\end{tabular}
\begin{tablenotes}
	\item Note: The conversion efficiencies are based on \cite{limpens_generating_2021} in the year 2050.
    \item[a] Based on the Haber Bosch process from \cite{noauthor_input_nodate}.
\end{tablenotes}
\end{threeparttable}
}
\end{minipage}
\caption{Input and output flows and economic parameters of HABER\_BOSCH.}
\label{fig:HABER_BOSCH}
\end{figure}

\paragraph{Methanol}

In BIOMASS\_TO\_METHANOL, woody/dry biomass is first converted into syngas through gasification, which is subsequently synthesized into methanol. 

\begin{figure}[H]
\begin{minipage}[H]{.48\textwidth}
\centering
\includegraphics[width=\linewidth]{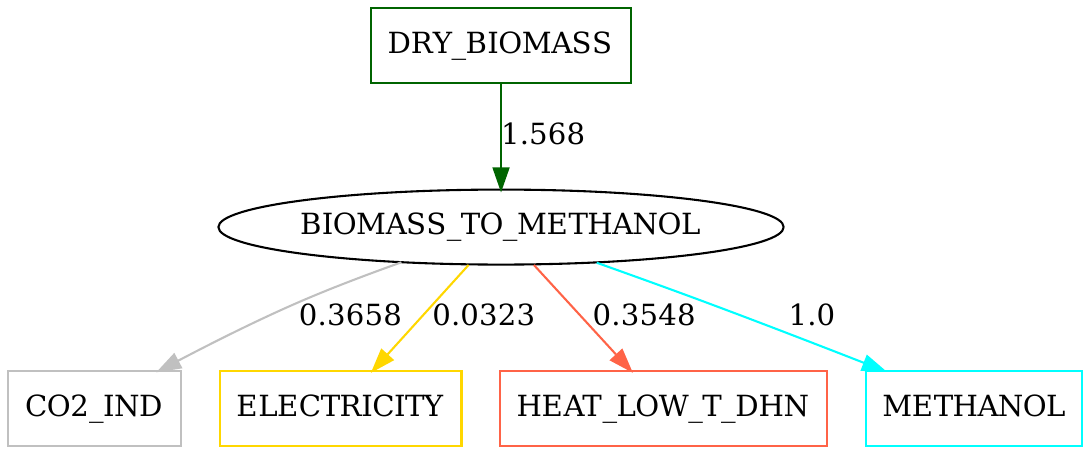}
\end{minipage}
\begin{minipage}{.5\textwidth}
\centering
\resizebox{\textwidth}{!}{%
\begin{threeparttable}
\begin{tabular}{l c c }
\toprule
\textbf{Parameter} & \textbf{Value} & \textbf{Unit}\\
\midrule
$c_{\mathrm{inv}}$ & 2520\tnote{a} $\pm$ 30\% & CHF/kW\\
$c_{\mathrm{maint}}$ & 38.5\tnote{a} $\pm$ 30\% & CHF/kW/y\\
$\tau$ & 20\tnote{a} & year(s)\\
$c_{\mathrm{p}}$ & 85.0\tnote{a} & \%\\
\bottomrule
\end{tabular}
\begin{tablenotes}
	\item Note: The conversion efficiencies are based on \cite{limpens_generating_2021} for the year 2050.
    \item[a] Based on the biomass to methanol process from \cite{noauthor_input_nodate}.
\end{tablenotes}
\end{threeparttable}
}
\end{minipage}
\caption{Input and output flows and economic parameters of BIOMASS\_TO\_METHANOL.}
\label{fig:BIOMASS_TO_METHANOL}
\end{figure}

METHANE\_TO\_METHANOL is a two-step process to generate methanol. First, methane is used to produce synthesis gas, which is subsequently synthesized into methanol.

\begin{figure}[H]
\begin{minipage}[H]{.45\textwidth}
\centering
\includegraphics[width=0.8\linewidth]{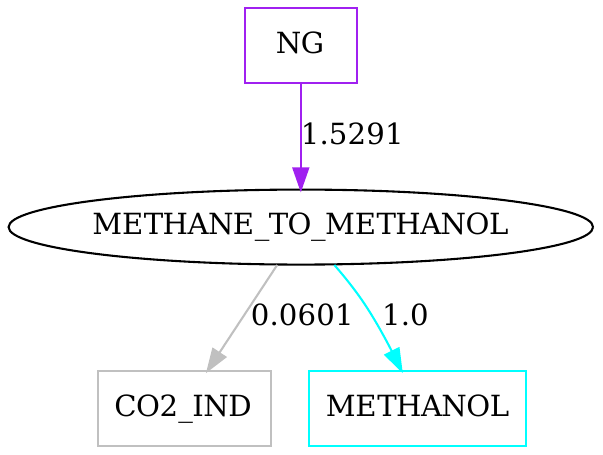}
\end{minipage}
\begin{minipage}{.5\textwidth}
\centering
\resizebox{\textwidth}{!}{%
\begin{threeparttable}
\begin{tabular}{l c c }
\toprule
\textbf{Parameter} & \textbf{Value} & \textbf{Unit}\\
\midrule
$c_{\mathrm{inv}}$ & 958.6\tnote{a} $\pm$ 30\% & CHF/kW\\
$c_{\mathrm{maint}}$ & 47.9\tnote{a} $\pm$ 30\% & CHF/kW/y\\
$\tau$ & 20\tnote{a} & year(s)\\
\bottomrule
\end{tabular}
\begin{tablenotes}
	\item[a] Based on the biomass to methanol process from \cite{noauthor_input_nodate}.
\end{tablenotes}
\end{threeparttable}
}
\end{minipage}
\caption{Input and output flows and economic parameters of METHANE\_TO\_METHANOL.}
\label{fig:METHANE_TO_METHANOL}
\end{figure}

Methanol synthesis (SYN\_METHANOLATION) converts pure carbon dioxide with hydrogen into methanol.

\begin{figure}[H]
\begin{minipage}[H]{.45\textwidth}
\centering
\includegraphics[width=0.9\linewidth]{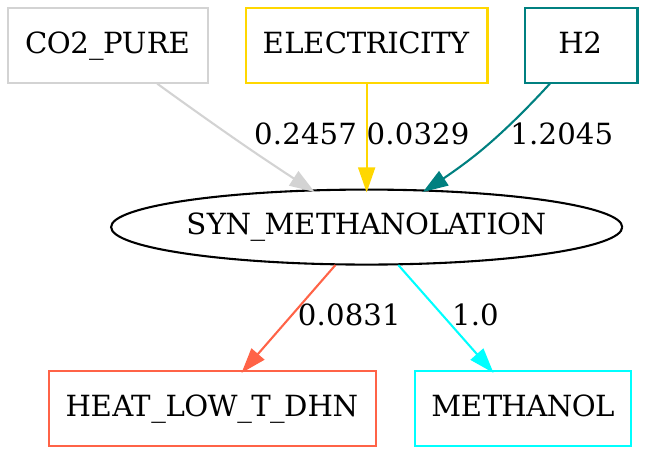}
\end{minipage}
\begin{minipage}{.5\textwidth}
\centering
\resizebox{\textwidth}{!}{%
\begin{threeparttable}
\begin{tabular}{l c c }
\toprule
\textbf{Parameter} & \textbf{Value} & \textbf{Unit}\\
\midrule
$c_{\mathrm{inv}}$ & 1680\tnote{a} $\pm$ 30\% & CHF/kW\\
$c_{\mathrm{maint}}$ & 84.0\tnote{a} $\pm$ 30\% & CHF/kW/y\\
$\tau$ & 20\tnote{a} & year(s)\\
$c_{\mathrm{p}}$ & 67.0\tnote{a} & \%\\
\bottomrule
\end{tabular}
\begin{tablenotes}
    \item Note: The conversion efficiencies are based on \cite{limpens_generating_2021} for the year 2050.
	\item[a] Based on the synthetic methanolation process from \cite{noauthor_input_nodate}.
\end{tablenotes}
\end{threeparttable}
}
\end{minipage}
\caption{Input and output flows and economic parameters of SYN\_METHANOLATION.}
\label{fig:SYN_METHANOLATION}
\end{figure}

\paragraph{High-value Chemicals}

High-value chemicals (HVC), such as ethylene, propylene, and aromatics, are essential for various industrial applications, particularly for plastic production.

Via oxidative coupling, methane can be converted into HVC (GAS\_TO\_HVC).

\begin{figure}[H]
\begin{minipage}[H]{.45\textwidth}
\centering
\includegraphics[width=0.65\linewidth]{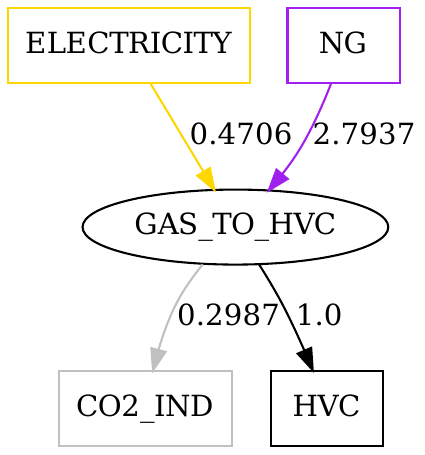}
\end{minipage}
\begin{minipage}{.5\textwidth}
\centering
\resizebox{\textwidth}{!}{%
\begin{threeparttable}
\begin{tabular}{l c c }
\toprule
\textbf{Parameter} & \textbf{Value} & \textbf{Unit}\\
\midrule
$c_{\mathrm{inv}}$ & 798\tnote{a} $\pm$ 30\% & CHF/kW\\
$c_{\mathrm{maint}}$ & 20.0\tnote{a} $\pm$ 30\% & CHF/kW/y\\
$\tau$ & 25\tnote{a} & year(s)\\
\bottomrule
\end{tabular}
\begin{tablenotes}
	\item Note: The conversion efficiencies are based on \cite{limpens_generating_2021} for the year 2050.
	\item[a] Taken from \cite{noauthor_input_nodate}.
\end{tablenotes}
\end{threeparttable}
}
\end{minipage}
\caption{Input and output flows and economic parameters of GAS\_TO\_HVC.}
\label{fig:GAS_TO_HVC}
\end{figure}

OIL\_TO\_HVC refers to refining processes that break down hydrocarbons in crude oil or its derivatives into lighter olefins (like ethylene and propylene) and aromatics.

\begin{figure}[H]
\begin{minipage}[H]{.45\textwidth}
\centering
\includegraphics[width=0.65\linewidth]{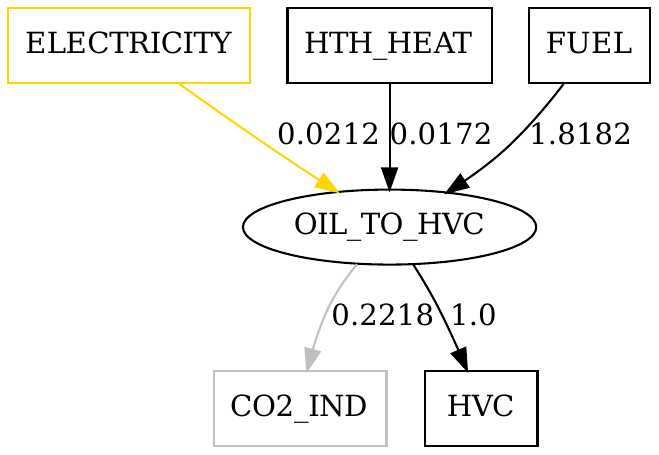}
\end{minipage}
\begin{minipage}{.5\textwidth}
\centering
\resizebox{\textwidth}{!}{%
\begin{threeparttable}
\begin{tabular}{l c c }
\toprule
\textbf{Parameter} & \textbf{Value} & \textbf{Unit}\\
\midrule
$c_{\mathrm{inv}}$ & 395\tnote{a} $\pm$ $30\%$ & CHF/kW\\
$c_{\mathrm{maint}}$ & 2.1\tnote{a} $\pm$ $30\%$ & CHF/kW/y\\
$\tau$ & 15\tnote{a} & year(s)\\
\bottomrule
\end{tabular}
\begin{tablenotes}
    \item Note: The conversion efficiencies are based on \cite{limpens_generating_2021} for the year 2050.
	\item[a] Taken from \cite{noauthor_input_nodate}.
\end{tablenotes}
\end{threeparttable}
}
\end{minipage}
\caption{Input and output flows and economic parameters of OIL\_TO\_HVC.}
\label{fig:OIL_TO_HVC}
\end{figure}

METHANOL\_TO\_HVC summarizes processes like methanol-to-olefins and methanol-to-aromatics, in which methanol is converted into high-value chemicals.

\begin{figure}[H]
\begin{minipage}[H]{.45\textwidth}
\centering
\includegraphics[width=0.8\linewidth]{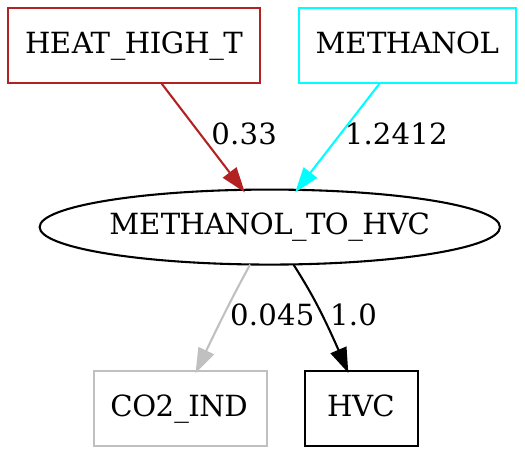}
\end{minipage}
\begin{minipage}{.5\textwidth}
\centering
\resizebox{\textwidth}{!}{%
\begin{threeparttable}
\begin{tabular}{l c c }
\toprule
\textbf{Parameter} & \textbf{Value} & \textbf{Unit}\\
\midrule
$c_{\mathrm{inv}}$ & 697\tnote{a} $\pm$ 30\% & CHF/kW\\
$c_{\mathrm{maint}}$ & 63.0\tnote{a} $\pm$ 30\% & CHF/kW/y\\
$\tau$ & 20\tnote{a} & year(s)\\
\bottomrule
\end{tabular}
\begin{tablenotes}
	\item Note: The conversion efficiencies are based on \cite{limpens_generating_2021} for the year 2050.
	\item[a] Taken from \cite{noauthor_input_nodate}.
\end{tablenotes}
\end{threeparttable}
}
\end{minipage}
\caption{Input and output flows and economic parameters of METHANOL\_TO\_HVC.}
\label{fig:METHANOL_TO_HVC}
\end{figure}

BIOMASS\_TO\_HVC first converts woody biomass into ethanol or dimethyl ether, which are then further processed into ethylene, a key high-value chemical.
\begin{figure}[H]
\begin{minipage}[H]{.45\textwidth}
\centering
\includegraphics[width=\linewidth]{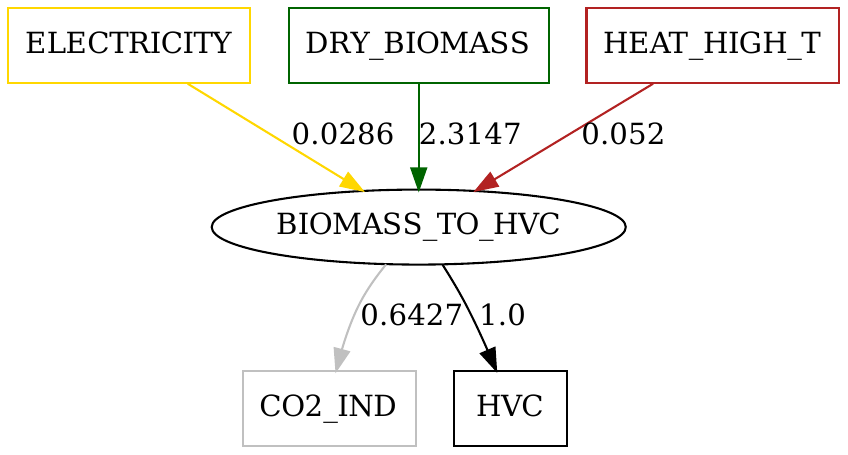}
\end{minipage}
\begin{minipage}{.5\textwidth}
\centering
\resizebox{\textwidth}{!}{%
\begin{threeparttable}
\begin{tabular}{l c c }
\toprule
\textbf{Parameter} & \textbf{Value} & \textbf{Unit}\\
\midrule
$c_{\mathrm{inv}}$ & 1743\tnote{a} $\pm$ 30\% & CHF/kW\\
$c_{\mathrm{maint}}$ & 52.0\tnote{a} $\pm$ 30\% & CHF/kW/y\\
$\tau$ & 20\tnote{a} & year(s)\\
\bottomrule
\end{tabular}
\begin{tablenotes}
	\item Note: The conversion efficiencies are based on \cite{limpens_generating_2021} for the year 2050.
	\item[a] Taken from \cite{noauthor_input_nodate}.
\end{tablenotes}
\end{threeparttable}
}
\end{minipage}
\caption{Input and output flows and economic parameters of BIOMASS\_TO\_HVC.}
\label{fig:BIOMASS_TO_HVC}
\end{figure}

In PLASTIC\_PYROLYSIS, plastic waste is thermally decomposed in the absence of oxygen to produce a naphtha-like product, which can be further processed. In NAPHTHA\_TO\_HVC, this naphtha is upgraded through catalytic reforming to produce high-value chemicals.

\begin{figure}[H]
\begin{minipage}[H]{.45\textwidth}
\centering
\includegraphics[width=\linewidth]{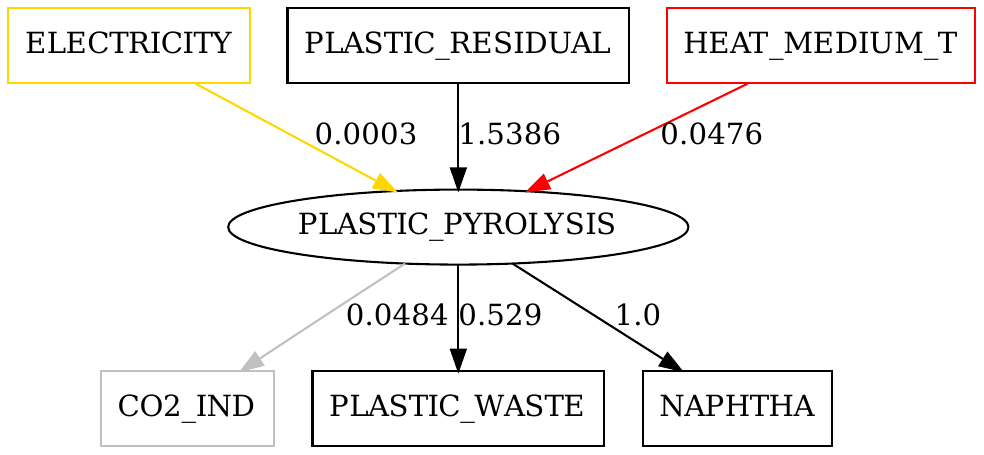}
\end{minipage}
\begin{minipage}{.5\textwidth}
\centering
\resizebox{\textwidth}{!}{%
\begin{threeparttable}
\begin{tabular}{l c c }
\toprule
\textbf{Parameter} & \textbf{Value} & \textbf{Unit}\\
\midrule
$c_{\mathrm{inv}}$ & 522.6\tnote{a} $\pm$ 30\% & CHF/kW\\
$c_{\mathrm{maint}}$ & 75.1\tnote{a} $\pm$ 30\% & CHF/kW/y\\
$\tau$ & 100\tnote{a} & year(s)\\
\bottomrule
\end{tabular}
\begin{tablenotes}
	\item[a] Own calculations based on \cite{meys_achieving_2021}, \cite{bachmann_towards_2023} and own estimates.
\end{tablenotes}
\end{threeparttable}
}
\end{minipage}
\caption{Input and output flows and economic parameters of PLASTIC\_PYROLYSIS.}
\label{fig:PLASTIC_PYROLYSIS}
\end{figure}

\begin{figure}[H]
\begin{minipage}[H]{.45\textwidth}
\centering
\includegraphics[width=\linewidth]{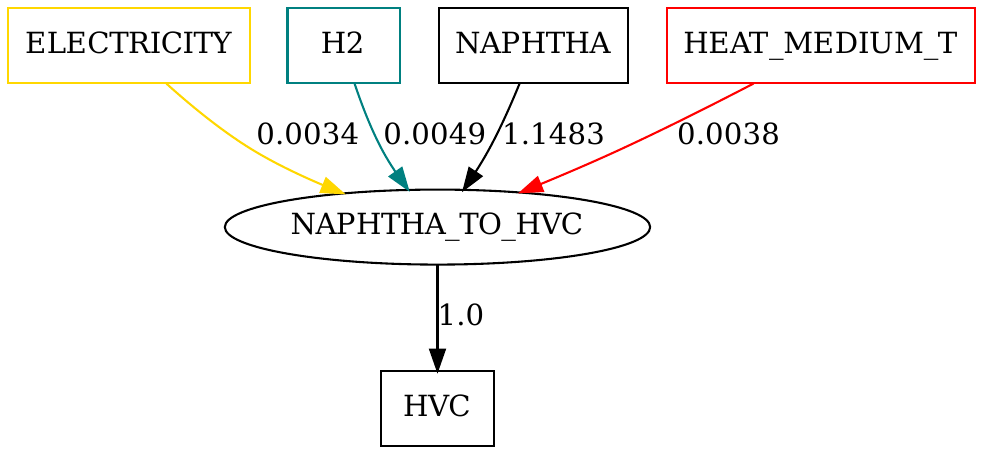}
\end{minipage}
\begin{minipage}{.5\textwidth}
\centering
\resizebox{\textwidth}{!}{%
\begin{threeparttable}
\begin{tabular}{l c c }
\toprule
\textbf{Parameter} & \textbf{Value} & \textbf{Unit}\\
\midrule
$c_{\mathrm{inv}}$ & 70.4\tnote{a} $\pm$ 30\% & CHF/kW\\
$c_{\mathrm{maint}}$ & 13.0\tnote{a} $\pm$ 30\% & CHF/kW/y\\
$\tau$ & 40\tnote{a} & year(s)\\
\bottomrule
\end{tabular}
\begin{tablenotes}
	\item[a] Own calculations based on \cite{meys_achieving_2021}, \cite{bachmann_towards_2023} and own estimates.
\end{tablenotes}
\end{threeparttable}
}
\end{minipage}
\caption{Input and output flows and economic parameters of NAPHTHA\_TO\_HVC.}
\label{fig:NAPHTHA_TO_HVC}
\end{figure}

In CHEMICAL\_RECYCLING, waste plastics are broken down into their monomers or other useful chemical intermediates, which subsequentially can be repurposed to produce new plastics or other valuable chemicals.

\begin{figure}[H]
\begin{minipage}[H]{.48\textwidth}
\centering
\includegraphics[width=\linewidth]{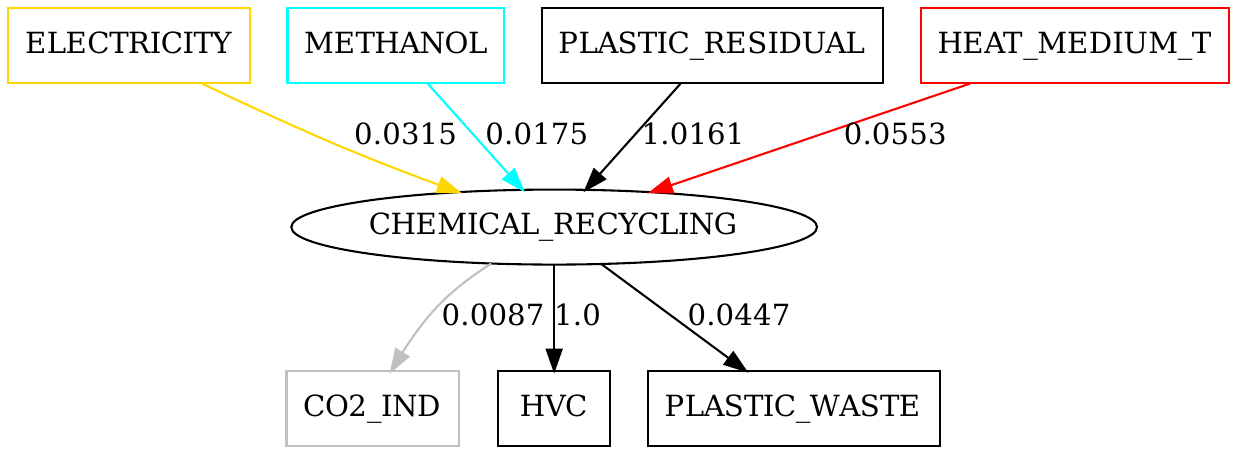}
\end{minipage}
\begin{minipage}{.5\textwidth}
\centering
\resizebox{\textwidth}{!}{%
\begin{threeparttable}
\begin{tabular}{l c c }
\toprule
\textbf{Parameter} & \textbf{Value} & \textbf{Unit}\\
\midrule
$c_{\mathrm{inv}}$ & 345.2\tnote{a} $\pm$ 30\% & CHF/kW\\
$c_{\mathrm{maint}}$ & 49.6\tnote{a} $\pm$ 30\% & CHF/kW/y\\
$\tau$ & 40\tnote{a} & year(s)\\
\bottomrule
\end{tabular}
\begin{tablenotes}
	\item[a] Own calculations based on \cite{meys_achieving_2021}, \cite{bachmann_towards_2023} and own estimates.
\end{tablenotes}
\end{threeparttable}
}
\end{minipage}
\caption{Input and output flows and economic parameters of CHEMICAL\_RECYCLING.}
\label{fig:CHEMICAL_RECYCLING}
\end{figure}

\paragraph{Plastic}
In our model, we group several materials (i.e., polyethylene (PE), polyethylenterephthalat (PET), polystyrene (PS), polypropylene (PP)) under the category of plastics. These materials are essential across industries, including packaging, construction, and automotive applications.

HVC\_TO\_PLASTIC summarizes the processes of polymerization of high-value chemicals to form long-chained molecules, which constitute various types of plastics.

\begin{figure}[H]
\begin{minipage}[H]{.48\textwidth}
\centering
\includegraphics[width=\linewidth]{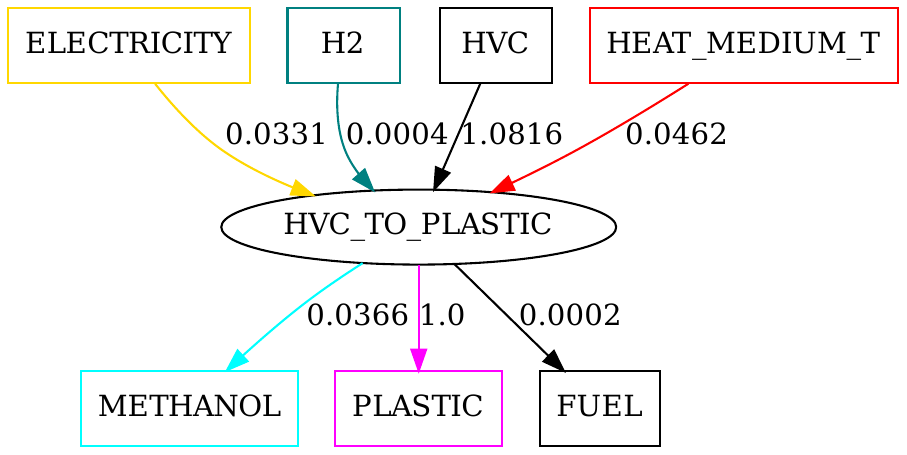}
\end{minipage}
\begin{minipage}{.5\textwidth}
\centering
\resizebox{\textwidth}{!}{%
\begin{threeparttable}
\begin{tabular}{l c c }
\toprule
\textbf{Parameter} & \textbf{Value} & \textbf{Unit}\\
\midrule
$c_{\mathrm{inv}}$ & 118.2\tnote{a} $\pm$ 30\% & CHF/kW\\
$c_{\mathrm{maint}}$ & 22.9\tnote{a} $\pm$ 30\% & CHF/kW/y\\
$\tau$ & 40\tnote{a} & year(s)\\
\bottomrule
\end{tabular}
\begin{tablenotes}
	\item[a] Own calculations based on \cite{meys_achieving_2021}, \cite{bachmann_towards_2023} and own estimates.
\end{tablenotes}
\end{threeparttable}
}
\end{minipage}
\caption{Input and output flows and economic parameters of HVC\_TO\_PLASTIC.}
\label{fig:HVC_TO_PLASTIC}
\end{figure}

In MECHANICAL\_RECYCLING, the plastic polymers are maintained. Instead, plastics are physically processed by shredding, melting, and reforming to produce new products. 

\begin{figure}[H]
\begin{minipage}[H]{.48\textwidth}
\centering
\includegraphics[width=\linewidth]{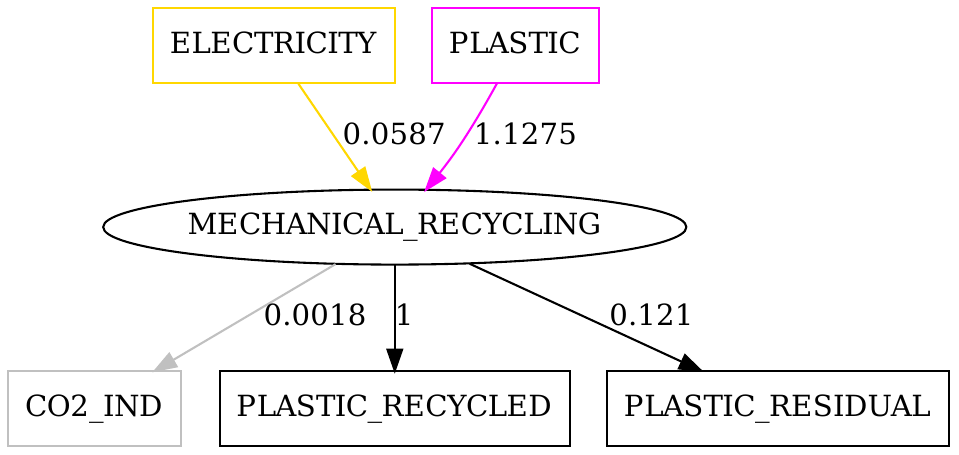}
\end{minipage}
\begin{minipage}{.5\textwidth}
\centering
\resizebox{\textwidth}{!}{%
\begin{threeparttable}
\begin{tabular}{l c c }
\toprule
\textbf{Parameter} & \textbf{Value} & \textbf{Unit}\\
\midrule
$c_{\mathrm{inv}}$ & 195.7\tnote{a} $\pm$ 30\% & CHF/kW\\
$c_{\mathrm{maint}}$ & 22.6\tnote{a} $\pm$ 30\% & CHF/kW/y\\
$\tau$ & 10\tnote{a} & year(s)\\
\bottomrule
\end{tabular}
\begin{tablenotes}
	\item[a] Own calculations based on \cite{meys_achieving_2021}, \cite{bachmann_towards_2023} and own estimates.
\end{tablenotes}
\end{threeparttable}
}
\end{minipage}
\caption{Input and output flows and economic parameters of MECHANICAL\_RECYCLING.}
\label{fig:MECHANICAL_RECYCLING}
\end{figure}

\subsubsection{CO$_2$ Capture}
\begin{figure}[H]
    \centering
    \includegraphics[width=\linewidth]{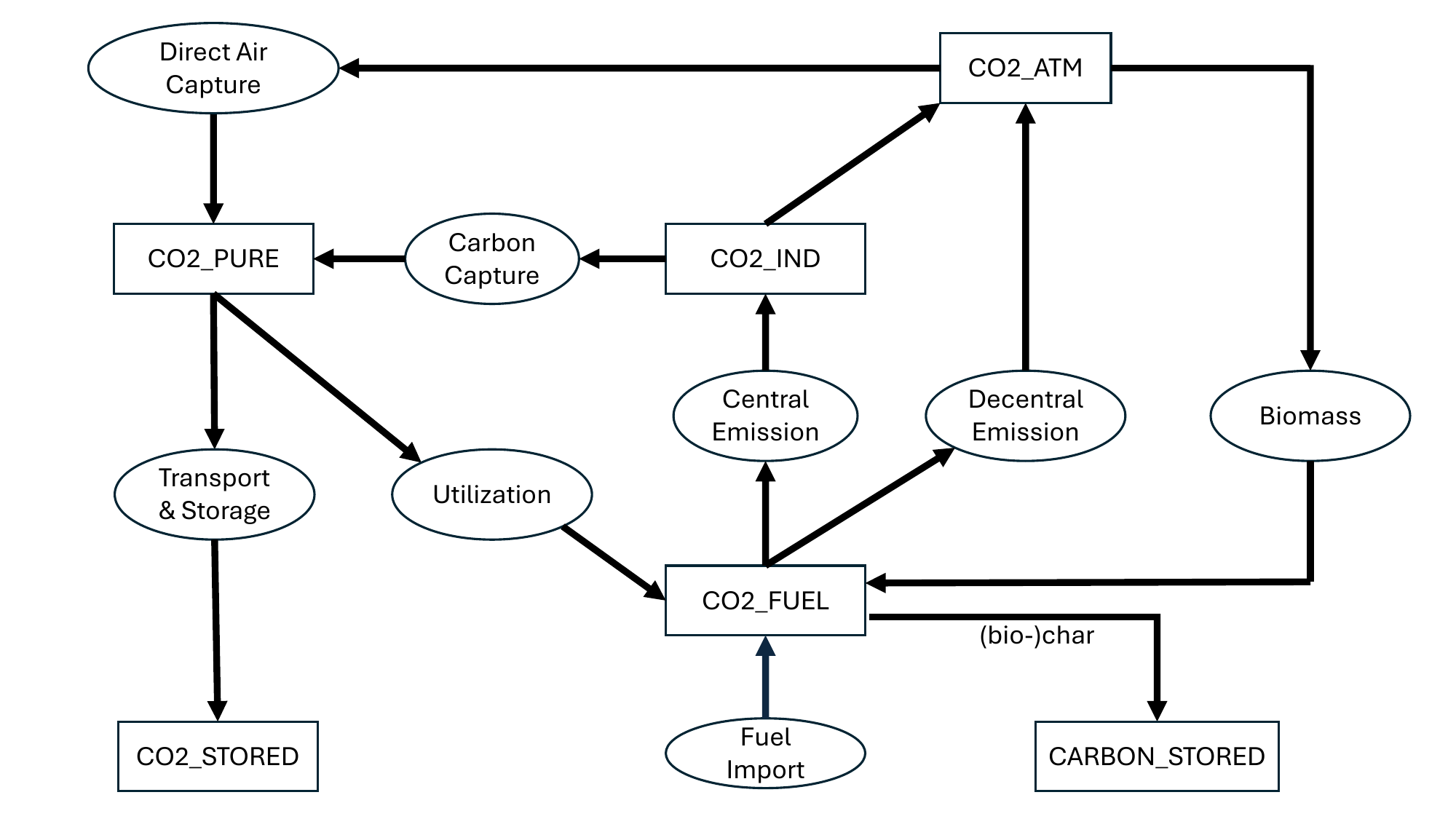}
    \caption{Modeling of the carbon flow in the energy system model.}
    \label{fig: carbon flow}
\end{figure}

The modeling of the carbon flow in our energy system model is schematically illustrated in Fig.~\ref{fig: carbon flow}.

Carbon enters the system via the import of fossil fuels (CO2\_FUEL). Burning the fuels releases the carbon as CO$_2$ either centrally (CO2\_IND), which allows for carbon capture, or decentrally, which releases the CO$_2$ to the atmosphere (CO2\_ATM).
From the atmosphere the CO$_2$ can either be removed by biomass or by direct air capture. Captured CO$_2$ (CO2\_PURE) can either be utilized as a carbon feedstock for sustainable chemicals and fuels or be stored permanently in geological sites (CO2\_STORED). Additionally, carbon can be removed from the system in the form of (bio-)char, which can be used in agriculture to enrich the soil (CARBON\_STORED).

Carbon capture gains importance for achieving net-zero energy systems by mitigating emissions from hard-to-abate sectors such as industry, aviation, and agriculture.

\begin{figure}[H]
\begin{minipage}[H]{.48\textwidth}
\centering
\includegraphics[width=\linewidth]{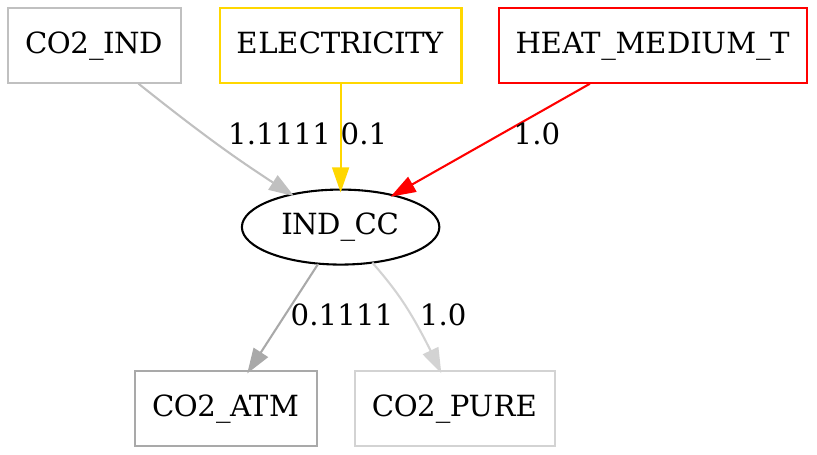}
\end{minipage}
\begin{minipage}{.5\textwidth}
\centering
\resizebox{\textwidth}{!}{%
\begin{threeparttable}
\begin{tabular}{l c c }
\toprule
\textbf{Parameter} & \textbf{Value} & \textbf{Unit}\\
\midrule
$c_{\mathrm{inv}}$ & 1800\tnote{a} $\pm$ 30\% & CHF/kg/h\\
$c_{\mathrm{maint}}$ & (2\% - 5\%)$\times c_{\mathrm{inv}}$ & CHF/kg/h/y\\
$\tau$ & 25\tnote{a} & year(s)\\
\bottomrule
\end{tabular}
\begin{tablenotes}
	\item[a] Based on the CO$_{2}$-separation from flue gas technology from \cite{guidati_gianfranco_value_2022}.
\end{tablenotes}
\end{threeparttable}
}
\end{minipage}
\caption{Input and output flows and economic parameters of IND\_CC.}
\label{fig:IND_CC}
\end{figure}

At industrial point-source emitters, carbon dioxide can be separated and captured from flue gas streams with high CO$_2$ concentrations (IND\_CC). We assume a typical CO$_2$ capture rate of 90\% (Fig.~\ref{fig:IND_CC}).

Carbon dioxide emissions from limestone calcination for cement production can be captured by CEM\_CC. We assume a capture rate of 90\%, with a lower heat demand compared to IND\_CC due to the potential for heat integration.

\begin{figure}[H]
\begin{minipage}[H]{.48\textwidth}
\centering
\includegraphics[width=\linewidth]{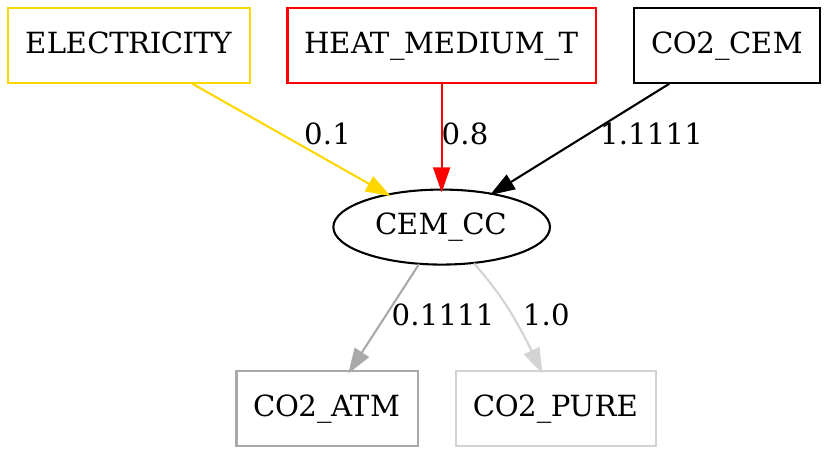}
\end{minipage}
\begin{minipage}{.5\textwidth}
\centering
\resizebox{\textwidth}{!}{%
\begin{threeparttable}
\begin{tabular}{l c c }
\toprule
\textbf{Parameter} & \textbf{Value} & \textbf{Unit}\\
\midrule
$c_{\mathrm{inv}}$ & 1800 $\pm$ 30\% & CHF/kg/h\\
$c_{\mathrm{maint}}$ & (2\% - 5\%)$\times c_{\mathrm{inv}}$ & CHF/kg/h/y\\
$\tau$ & 25 & year(s)\\
\bottomrule
\end{tabular}
\begin{tablenotes}
    \item Note: We estimate a lower heat demand for the CO$_{2}$ capture process due to the potential for heat integration in the cement plant.
	\item[a] Based on the CO$_{2}$-separation from flue gas technology from \cite{guidati_gianfranco_value_2022}.
\end{tablenotes}
\end{threeparttable}
}
\end{minipage}
\caption{Input and output flows and economic parameters of CEM\_CC.}
\label{fig:CEM_CC}
\end{figure}

Additionally, we consider that the heat required for carbon capture at a cement plant is provided by a heat pump (CEM\_CC\_HP). 

\begin{figure}[H]
\begin{minipage}[H]{.48\textwidth}
\centering
\includegraphics[width=0.65\linewidth]{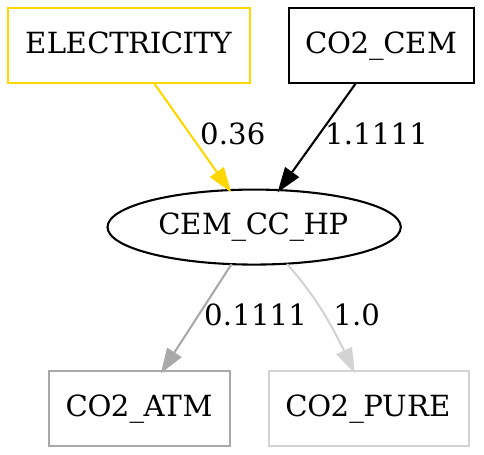}
\end{minipage}
\begin{minipage}{.5\textwidth}
\centering
\resizebox{\textwidth}{!}{%
\begin{threeparttable}
\begin{tabular}{l c c }
\toprule
\textbf{Parameter} & \textbf{Value} & \textbf{Unit}\\
\midrule
$c_{\mathrm{inv}}$ & 1800\tnote{a} $\pm$ 30\% & CHF/kg/h\\
$c_{\mathrm{maint}}$ & (2\% - 5\%)$\times c_{\mathrm{inv}}$ & CHF/kg/h/y\\
$\tau$ & 25\tnote{a} & year(s)\\
\bottomrule
\end{tabular}
\begin{tablenotes}
    \item Note: We assume that the heat can be provided by a heat pump with a COP of about 3.
	\item[a] Based on the CO$_{2}$-separation from flue gas technology from \cite{guidati_gianfranco_value_2022}.
\end{tablenotes}
\end{threeparttable}
}
\end{minipage}
\caption{Input and output flows and economic parameters of CEM\_CC\_HP.}
\label{fig:CEM_CC_HP}
\end{figure}

Direct air capture (DAC) removes carbon dioxide directly from ambient air (CO2\_ATM) using chemical sorbents or physical adsorption processes. The captured CO$_2$ can be stored permanently or utilized in industrial applications.

\begin{figure}[H]
\begin{minipage}[H]{.48\textwidth}
\centering
\includegraphics[width=\linewidth]{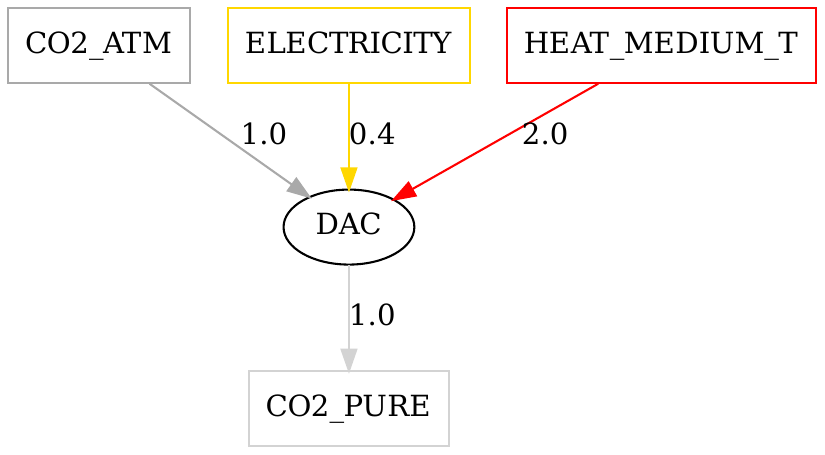}
\end{minipage}
\begin{minipage}{.5\textwidth}
\centering
\resizebox{\textwidth}{!}{%
\begin{threeparttable}
\begin{tabular}{l c c }
\toprule
\textbf{Parameter} & \textbf{Value} & \textbf{Unit}\\
\midrule
$c_{\mathrm{inv}}$ & 10000\tnote{a} $\pm$ 30\% & CHF/kg/h\\
$c_{\mathrm{maint}}$ & (2\% - 5\%)$\times c_{\mathrm{inv}}$ & CHF/kg/h/y\\
$\tau$ & 25\tnote{a} & year(s)\\
\bottomrule
\end{tabular}
\begin{tablenotes}
	\item[a] Based on the direct air capture technology from \cite{guidati_gianfranco_value_2022}.
\end{tablenotes}
\end{threeparttable}
}
\end{minipage}
\caption{Input and output flows and economic parameters of DAC.}
\label{fig:DAC}
\end{figure}

\subsubsection{Mobility}

In our model, we consider freight, aviation and personal transport for mobility. Personal transport is then further split into private and public transport with the share determined by $\%_{\mathrm{public}}$ given in Table \ref{tab: general system parameters}.

\paragraph{Freight Transport}
Freight transport is measured in units of tonne-kilometers [tkm]. In our model, the freight transport can either be provided by trains or trucks. The fraction is determined by $\%_{\mathrm{share\, train\,freight}}$.

\begin{figure}[H]
\begin{minipage}[H]{.48\textwidth}
\centering
\includegraphics[width=0.6\linewidth]{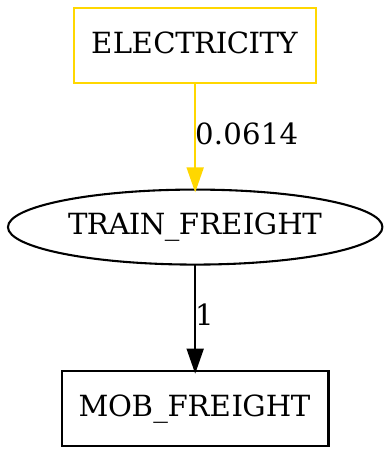}
\end{minipage}
\begin{minipage}{.5\textwidth}
\centering
\resizebox{\textwidth}{!}{%
\begin{threeparttable}
\begin{tabular}{l c c }
\toprule
\textbf{Parameter} & \textbf{Value} & \textbf{Unit}\\
\midrule
$c_{\mathrm{inv}}$ & 104.4\tnote{a} $\pm$ 30\% & CHF/tkm/h\\
$c_{\mathrm{maint}}$ & 2.1\tnote{a} $\pm$ 30\% & CHF/tkm/h/y\\
$\tau$ & 40\tnote{a} & year(s)\\
$c_{\mathrm{p}}$ & 34.2\tnote{a} & \%\\
\bottomrule
\end{tabular}
\begin{tablenotes}
	\item Note: The conversion efficiencies are based on \cite{limpens_generating_2021} for the year 2050.
	\item[a] Taken from \cite{noauthor_input_nodate}.
\end{tablenotes}
\end{threeparttable}
}
\end{minipage}
\caption{Input and output flows and economic parameters of TRAIN\_FREIGHT.}
\label{fig:TRAIN_FREIGHT}
\end{figure}

TRUCK\_EV represents trucks with electric motors that are entirely powered by electricity stored in batteries.

\begin{figure}[H]
\begin{minipage}[H]{.48\textwidth}
\centering
\includegraphics[width=0.45\linewidth]{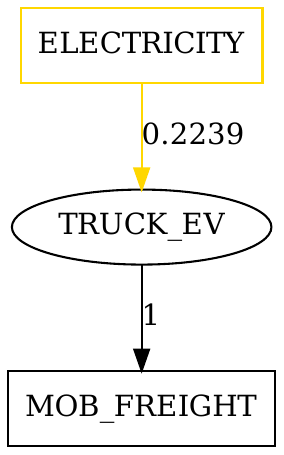}
\end{minipage}
\begin{minipage}{.5\textwidth}
\centering
\resizebox{\textwidth}{!}{%
\begin{threeparttable}
\begin{tabular}{l c c }
\toprule
\textbf{Parameter} & \textbf{Value} & \textbf{Unit}\\
\midrule
$c_{\mathrm{inv}}$ & 744.2\tnote{a} $\pm$ 30\% & CHF/tkm/h\\
$c_{\mathrm{maint}}$ & 23.1\tnote{a} $\pm$ 30\% & CHF/tkm/h/y\\
$\tau$ & 15\tnote{a} & year(s)\\
$c_{\mathrm{p}}$ & 9.3\tnote{a} & \%\\
\bottomrule
\end{tabular}
\begin{tablenotes}
	\item[a] Based on \cite{limpens_generating_2021} for the year 2050.
\end{tablenotes}
\end{threeparttable}
}
\end{minipage}
\caption{Input and output flows and economic parameters of TRUCK\_EV.}
\label{fig:TRUCK_EV}
\end{figure}

TRUCK\_FC represents a fuel-cell powered truck, which uses hydrogen and a fuel cell to generate electricity. This electricity is then used by electric motors to drive the truck.

\begin{figure}[H]
\begin{minipage}[H]{.48\textwidth}
\centering
\includegraphics[width=0.45\linewidth]{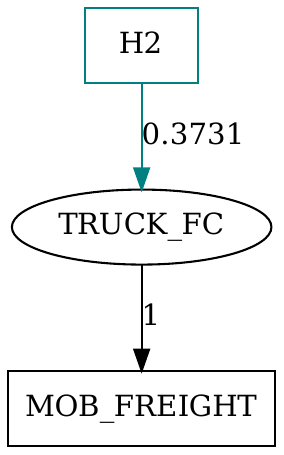}
\end{minipage}
\begin{minipage}{.5\textwidth}
\centering
\resizebox{\textwidth}{!}{%
\begin{threeparttable}
\begin{tabular}{l c c }
\toprule
\textbf{Parameter} & \textbf{Value} & \textbf{Unit}\\
\midrule
$c_{\mathrm{inv}}$ & 404.8\tnote{a} $\pm$ 30\% & CHF/tkm/h\\
$c_{\mathrm{maint}}$ & 12.1\tnote{a} $\pm$ 30\% & CHF/tkm/h/y\\
$\tau$ & 15\tnote{a} & year(s)\\
$c_{\mathrm{p}}$ & 9.3\tnote{a} & \%\\
\bottomrule
\end{tabular}
\begin{tablenotes}
	\item[a] Based on \cite{limpens_generating_2021} for the year 2050.
\end{tablenotes}
\end{threeparttable}
}
\end{minipage}
\caption{Input and output flows and economic parameters of TRUCK\_FC.}
\label{fig:TRUCK_FC}
\end{figure}

TRUCK\_NG represents a truck powered by natural gas, which is used as a cleaner alternative to gasoline or diesel. Compressed natural gas (CNG) or liquefied natural gas (LNG) is burned in an internal combustion engine to power the vehicle.

\begin{figure}[H]
\begin{minipage}[H]{.48\textwidth}
\centering
\includegraphics[width=.75\linewidth]{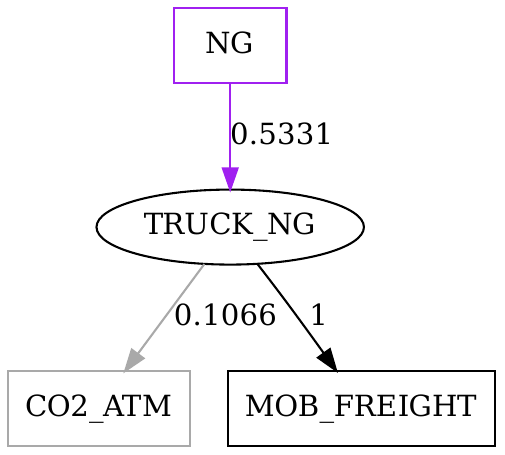}
\end{minipage}
\begin{minipage}{.5\textwidth}
\centering
\resizebox{\textwidth}{!}{%
\begin{threeparttable}
\begin{tabular}{l c c }
\toprule
\textbf{Parameter} & \textbf{Value} & \textbf{Unit}\\
\midrule
$c_{\mathrm{inv}}$ & 381.3\tnote{a} $\pm$ 30\% & CHF/tkm/h\\
$c_{\mathrm{maint}}$ & 18.6\tnote{a} $\pm$ 30\% & CHF/tkm/h/y\\
$\tau$ & 15\tnote{a} & year(s)\\
$c_{\mathrm{p}}$ & 9.3\tnote{a} & \%\\
\bottomrule
\end{tabular}
\begin{tablenotes}
	\item[a] Based on \cite{limpens_generating_2021} for the year 2050.
\end{tablenotes}
\end{threeparttable}
}
\end{minipage}
\caption{Input and output flows and economic parameters of TRUCK\_NG.}
\label{fig:TRUCK_NG}
\end{figure}

TRUCK\_DIESEL represents a truck powered by diesel fuel, which is burned in an internal combustion engine. 

\begin{figure}[H]
\begin{minipage}[H]{.48\textwidth}
\centering
\includegraphics[width=.75\linewidth]{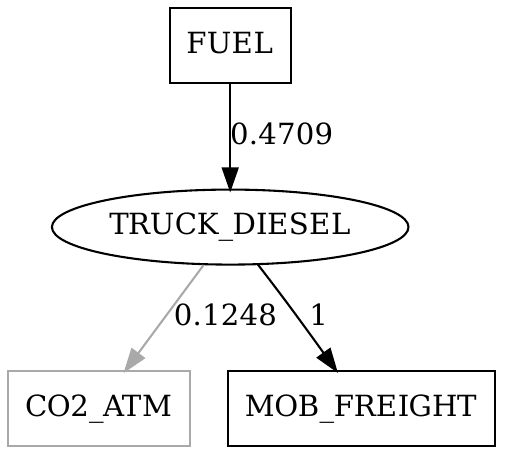}
\end{minipage}
\begin{minipage}{.5\textwidth}
\centering
\resizebox{\textwidth}{!}{%
\begin{threeparttable}
\begin{tabular}{l c c }
\toprule
\textbf{Parameter} & \textbf{Value} & \textbf{Unit}\\
\midrule
$c_{\mathrm{inv}}$ & 383.8\tnote{a} $\pm$ 30\% & CHF/tkm/h\\
$c_{\mathrm{maint}}$ & 18.6\tnote{a} $\pm$ 30\% & CHF/tkm/h/y\\
$\tau$ & 15\tnote{a} & year(s)\\
$c_{\mathrm{p}}$ & 9.3\tnote{a} & \%\\
\bottomrule
\end{tabular}
\begin{tablenotes}
	\item[a] Based on \cite{limpens_generating_2021} for the year 2050.
\end{tablenotes}
\end{threeparttable}
}
\end{minipage}
\caption{Input and output flows and economic parameters of TRUCK\_DIESEL.}
\label{fig:TRUCK_DIESEL}
\end{figure}

\paragraph{Private Mobility}
Private mobility is measured in units of passenger-kliometers [pkm] and can be provided by cars utilizing different drive technologies. Specifically, as for trucks, our model includes electric, hydrogen, natural gas and diesel-powered cars.

CAR\_EV represents cars equipped with electric motors that are fully powered by electricity stored in batteries.

\begin{figure}[H]
\begin{minipage}[H]{.48\textwidth}
\centering
\includegraphics[width=.45\linewidth]{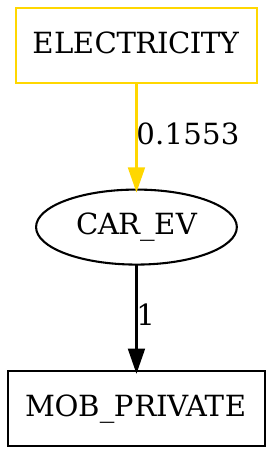}
\end{minipage}
\begin{minipage}{.5\textwidth}
\centering
\resizebox{\textwidth}{!}{%
\begin{threeparttable}
\begin{tabular}{l c c }
\toprule
\textbf{Parameter} & \textbf{Value} & \textbf{Unit}\\
\midrule
$c_{\mathrm{inv}}$ & 434.1\tnote{a} $\pm$ 30\% & CHF/pkm/h\\
$c_{\mathrm{maint}}$ & 10.0\tnote{a} $\pm$ 30\% & CHF/pkm/h/y\\
$\tau$ & 10\tnote{a} & year(s)\\
$c_{\mathrm{p}}$ & 5.1\tnote{a} & \%\\
\bottomrule
\end{tabular}
\begin{tablenotes}
	\item[a] Based on \cite{limpens_generating_2021} for the year 2050.
\end{tablenotes}
\end{threeparttable}
}
\end{minipage}
\caption{Input and output flows and economic parameters of CAR\_EV.}
\label{fig:CAR_EV}
\end{figure}

CAR\_FC represents cars powered by hydrogen fuel cells.

\begin{figure}[H]
\begin{minipage}[H]{.48\textwidth}
\centering
\includegraphics[width=.45\linewidth]{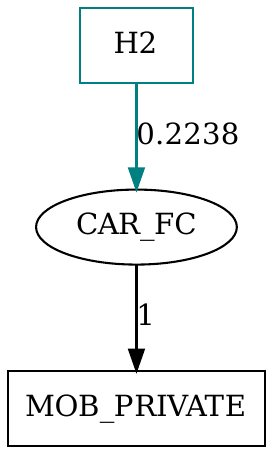}
\end{minipage}
\begin{minipage}{.5\textwidth}
\centering
\resizebox{\textwidth}{!}{%
\begin{threeparttable}
\begin{tabular}{l c c }
\toprule
\textbf{Parameter} & \textbf{Value} & \textbf{Unit}\\
\midrule
$c_{\mathrm{inv}}$ & 437.6\tnote{a} $\pm$ 30\% & CHF/pkm/h\\
$c_{\mathrm{maint}}$ & 10.0\tnote{a} $\pm$ 30\% & CHF/pkm/h/y\\
$\tau$ & 10\tnote{a} & year(s)\\
$c_{\mathrm{p}}$ & 5.1\tnote{a} & \%\\
\bottomrule
\end{tabular}
\begin{tablenotes}
	\item[a] Based on \cite{limpens_generating_2021} for the year 2050.
\end{tablenotes}
\end{threeparttable}
}
\end{minipage}
\caption{Input and output flows and economic parameters of CAR\_FC.}
\label{fig:CAR_FC}
\end{figure}

CAR\_NG represents cars powered by natural gas burned in internal combustion engines.

\begin{figure}[H]
\begin{minipage}[H]{.48\textwidth}
\centering
\includegraphics[width=.8\linewidth]{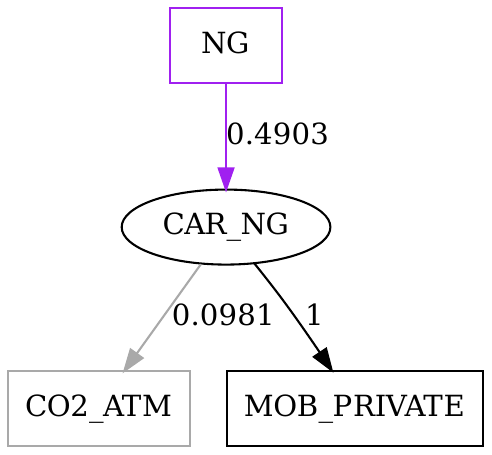}
\end{minipage}
\begin{minipage}{.5\textwidth}
\centering
\resizebox{\textwidth}{!}{%
\begin{threeparttable}
\begin{tabular}{l c c }
\toprule
\textbf{Parameter} & \textbf{Value} & \textbf{Unit}\\
\midrule
$c_{\mathrm{inv}}$ & 441.2\tnote{a} $\pm$ 30\% & CHF/pkm/h\\
$c_{\mathrm{maint}}$ & 24.0\tnote{a} $\pm$ 30\% & CHF/pkm/h/y\\
$\tau$ & 10\tnote{a} & year(s)\\
$c_{\mathrm{p}}$ & 5.1\tnote{a} & \%\\
\bottomrule
\end{tabular}
\begin{tablenotes}
	\item[a] Based on \cite{limpens_generating_2021} for the year 2050.
\end{tablenotes}
\end{threeparttable}
}
\end{minipage}
\caption{Input and output flows and economic parameters of CAR\_NG.}
\label{fig:CAR_NG}
\end{figure}

CAR\_DIESEL represents cars powered by liquid fuel burned in internal combustion engines.

\begin{figure}[H]
\begin{minipage}[H]{.48\textwidth}
\centering
\includegraphics[width=.8\linewidth]{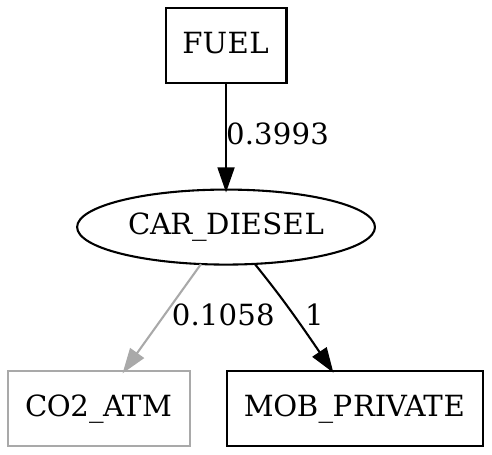}
\end{minipage}
\begin{minipage}{.5\textwidth}
\centering
\resizebox{\textwidth}{!}{%
\begin{threeparttable}
\begin{tabular}{l c c }
\toprule
\textbf{Parameter} & \textbf{Value} & \textbf{Unit}\\
\midrule
$c_{\mathrm{inv}}$ & 449.3\tnote{a} $\pm$ 30\% & CHF/pkm/h\\
$c_{\mathrm{maint}}$ & 24.0\tnote{a} $\pm$ 30\% & CHF/pkm/h/y\\
$\tau$ & 10\tnote{a} & year(s)\\
$c_{\mathrm{p}}$ & 5.1\tnote{a} & \%\\
\bottomrule
\end{tabular}
\begin{tablenotes}
	\item[a] Based on \cite{limpens_generating_2021} for the year 2050.
\end{tablenotes}
\end{threeparttable}
}
\end{minipage}
\caption{Input and output flows and economic parameters of CAR\_DIESEL.}
\label{fig:CAR_DIESEL}
\end{figure}

\paragraph{Public Mobility}
As private mobility, public mobility is measured in units of passener-kilometers [pkm]. It can be provided by trains, tramways or buses. The shares of trains and tramways are calculated based on \cite{marcucci_adriana_assumptions_2022} and given by $f_{\mathrm{perc}}$. The remaining share must be provided by buses.

Trains can provide inter-regional public transport. Our model considers only electricity-powered trains.

\begin{figure}[H]
\begin{minipage}[H]{.48\textwidth}
\centering
\includegraphics[width=.53\linewidth]{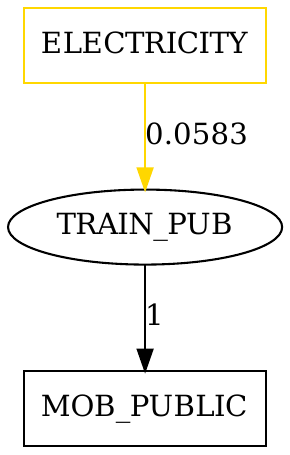}
\end{minipage}
\begin{minipage}{.5\textwidth}
\centering
\resizebox{\textwidth}{!}{%
\begin{threeparttable}
\begin{tabular}{l c c }
\toprule
\textbf{Parameter} & \textbf{Value} & \textbf{Unit}\\
\midrule
$c_{\mathrm{inv}}$ & 1506.0\tnote{a} $\pm$ 30\% & CHF/pkm/h\\
$c_{\mathrm{maint}}$ & 54.4\tnote{a} $\pm$ 30\% & CHF/pkm/h/y\\
$\tau$ & 40\tnote{a} & year(s)\\
$c_{\mathrm{p}}$ & 27.5\tnote{a} & \%\\
$f_{\mathrm{perc}}$ & 79.89 - 84.77\tnote{b} & \%\\
\bottomrule
\end{tabular}
\begin{tablenotes}
	\item[a] Based on \cite{limpens_generating_2021} for the year 2050.
    \item[b] Own estimation based on \cite{marcucci_cross_2022} \url{https://sweet-cross.ch/data/assumption-cross-scenarios/2022-09-30/}.
\end{tablenotes}
\end{threeparttable}
}
\end{minipage}
\caption{Input and output flows and economic parameters of TRAIN\_PUB.}
\label{fig:TRAIN_PUB}
\end{figure}

Tramways can provide public mobility in urban areas.

\begin{figure}[H]
\begin{minipage}[H]{.48\textwidth}
\centering
\includegraphics[width=.8\linewidth]{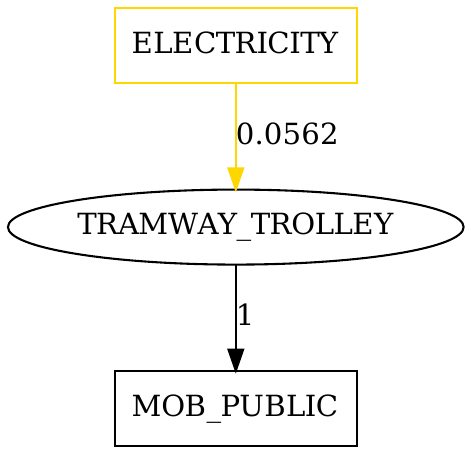}
\end{minipage}
\begin{minipage}{.5\textwidth}
\centering
\resizebox{\textwidth}{!}{%
\begin{threeparttable}
\begin{tabular}{l c c }
\toprule
\textbf{Parameter} & \textbf{Value} & \textbf{Unit}\\
\midrule
$c_{\mathrm{inv}}$ & 625\tnote{a} $\pm$ 30\% & CHF/pkm/h\\
$c_{\mathrm{maint}}$ & 12.5\tnote{a} $\pm$ 30\% & CHF/pkm/h/y\\
$\tau$ & 30\tnote{a} & year(s)\\
$c_{\mathrm{p}}$ & 34.2\tnote{a} & \%\\
$f_{\mathrm{perc}}$ & 5.40 - 7.41\tnote{b} & \%\\
\bottomrule
\end{tabular}
\begin{tablenotes}
	\item[a] Based on \cite{limpens_generating_2021} for the year 2050.
    \item[b] Own estimation based on \cite{marcucci_cross_2022} \url{https://sweet-cross.ch/data/assumption-cross-scenarios/2022-09-30/}.
\end{tablenotes}
\end{threeparttable}
}
\end{minipage}
\caption{Input and output flows and economic parameters of TRAMWAY\_TROLLEY.}
\label{fig:TRAMWAY_TROLLEY}
\end{figure}

As for trucks and cars, we include the four possible types of drives for buses: electric, hydrogen, natural gas, and diesel-powered.

BUS\_FC represents buses powered by hydrogen fuel cells.

\begin{figure}[H]
\begin{minipage}[H]{.48\textwidth}
\centering
\includegraphics[width=.42\linewidth]{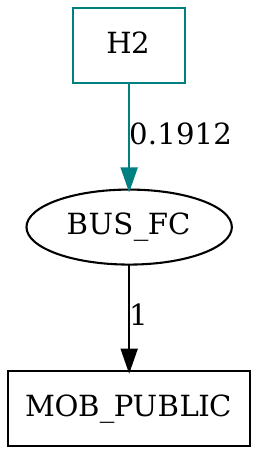}
\end{minipage}
\begin{minipage}{.5\textwidth}
\centering
\resizebox{\textwidth}{!}{%
\begin{threeparttable}
\begin{tabular}{l c c }
\toprule
\textbf{Parameter} & \textbf{Value} & \textbf{Unit}\\
\midrule
$c_{\mathrm{inv}}$ & 1048.4\tnote{a} $\pm$ 30\% & CHF/pkm/h\\
$c_{\mathrm{maint}}$ & 31.2\tnote{a} $\pm$ 30\% & CHF/pkm/h/y\\
$\tau$ & 15\tnote{a} & year(s)\\
$c_{\mathrm{p}}$ & 29.7\tnote{a} & \%\\
\bottomrule
\end{tabular}
\begin{tablenotes}
	\item[a] Based on \cite{limpens_generating_2021} for the year 2050.
\end{tablenotes}
\end{threeparttable}
}
\end{minipage}
\caption{Input and output flows and economic parameters of BUS\_FC.}
\label{fig:BUS_FC}
\end{figure}

BUS\_NG represents buses powered by natural gas burned in internal combustion engines.

\begin{figure}[H]
\begin{minipage}[H]{.48\textwidth}
\centering
\includegraphics[width=0.7\linewidth]{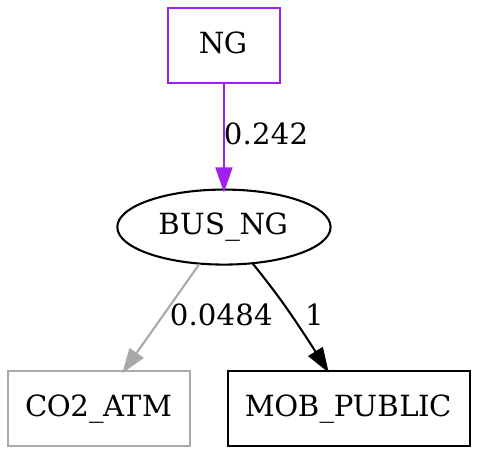}
\end{minipage}
\begin{minipage}{.5\textwidth}
\centering
\resizebox{\textwidth}{!}{%
\begin{threeparttable}
\begin{tabular}{l c c }
\toprule
\textbf{Parameter} & \textbf{Value} & \textbf{Unit}\\
\midrule
$c_{\mathrm{inv}}$ & 628.0\tnote{a} $\pm$ 30\% & CHF/pkm/h\\
$c_{\mathrm{maint}}$ & 30.6\tnote{a} $\pm$ 30\% & CHF/pkm/h/y\\
$\tau$ & 15\tnote{a} & year(s)\\
$c_{\mathrm{p}}$ & 29.7\tnote{a} & \%\\
\bottomrule
\end{tabular}
\begin{tablenotes}
	\item[a] Based on \cite{limpens_generating_2021} for the year 2050.
\end{tablenotes}
\end{threeparttable}
}
\end{minipage}
\caption{Input and output flows and economic parameters of BUS\_NG.}
\label{fig:BUS_NG}
\end{figure}

BUS\_DIESEL represents buses powered by diesel burned in internal combustion engines.

\begin{figure}[H]
\begin{minipage}[H]{.48\textwidth}
\centering
\includegraphics[width=0.7\linewidth]{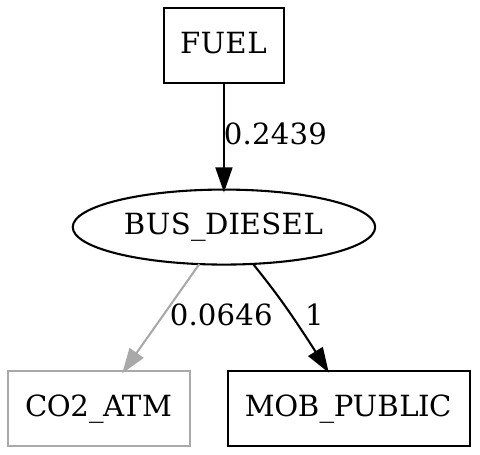}
\end{minipage}
\begin{minipage}{.5\textwidth}
\centering
\resizebox{\textwidth}{!}{%
\begin{threeparttable}
\begin{tabular}{l c c }
\toprule
\textbf{Parameter} & \textbf{Value} & \textbf{Unit}\\
\midrule
$c_{\mathrm{inv}}$ & 632.0\tnote{a} $\pm$ 30\% & CHF/pkm/h\\
$c_{\mathrm{maint}}$ & 30.6\tnote{a} $\pm$ 30\% & CHF/pkm/h/y\\
$\tau$ & 15\tnote{a} & year(s)\\
$c_{\mathrm{p}}$ & 29.7\tnote{a} & \%\\
\bottomrule
\end{tabular}
\begin{tablenotes}
	\item[a] Based on \cite{limpens_generating_2021} for the year 2050.
\end{tablenotes}
\end{threeparttable}
}
\end{minipage}
\caption{Input and output flows and economic parameters of BUS\_DIESEL.}
\label{fig:BUS_DIESEL}
\end{figure}

BUS\_EV represents buses equipped with electric motors that are fully powered by electricity stored in batteries.

\begin{figure}[H]
\begin{minipage}[H]{.48\textwidth}
\centering
\includegraphics[width=.42\linewidth]{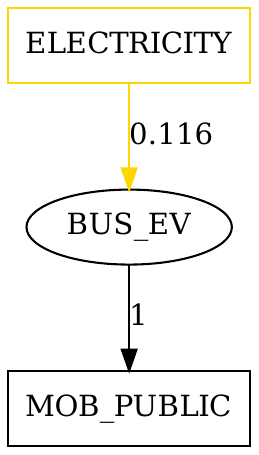}
\end{minipage}
\begin{minipage}{.5\textwidth}
\centering
\resizebox{\textwidth}{!}{%
\begin{threeparttable}
\begin{tabular}{l c c }
\toprule
\textbf{Parameter} & \textbf{Value} & \textbf{Unit}\\
\midrule
$c_{\mathrm{inv}}$ & 1225.4\tnote{a} $\pm$ 30\% & CHF/pkm/h\\
$c_{\mathrm{maint}}$ & 38.1\tnote{a} $\pm$ 30\% & CHF/pkm/h\\
$\tau$ & 15\tnote{b} & year(s)\\
$c_{\mathrm{p}}$ & 29.7\tnote{b} & \%\\
\bottomrule
\end{tabular}
\begin{tablenotes}
	\item[a] The efficiency and cost are estimated by taking the values of BUS\_DIESEL multiplied by the ratio of TRUCK\_EV to TRUCK\_DIESEL.
    \item[b] Assumed to be the same as for BUS\_DIESEL.
\end{tablenotes}
\end{threeparttable}
}
\end{minipage}
\caption{Input and output flows and economic parameters of BUS\_EV.}
\label{fig:BUS_EV}
\end{figure}

\paragraph{Aviation}
The demand for aviation mobility is measured in units of energy content in fuel [GWh$_{\mathrm{Fuel}}$], as we assume that aviation can only be provided from aircraft powered by liquid fuels, due to the high energy density required for the fuel.

\begin{figure}[H]
\begin{minipage}[H]{.48\textwidth}
\centering
\includegraphics[width=0.7\linewidth]{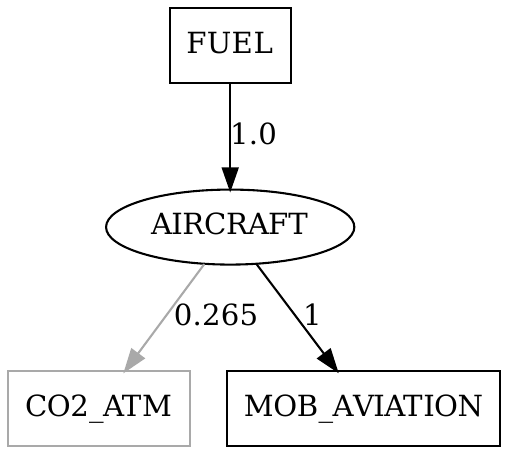}
\end{minipage}
\begin{minipage}{.5\textwidth}
\centering
\resizebox{\textwidth}{!}{%
\begin{threeparttable}
\begin{tabular}{l c c }
\toprule
\textbf{Parameter} & \textbf{Value} & \textbf{Unit}\\
\midrule
$c_{\mathrm{inv}}$ & 0\tnote{a} & CHF/kW\\
$c_{\mathrm{maint}}$ & 0\tnote{a} & CHF/kW/y\\
$\tau$ & 1\tnote{a} & year(s)\\
\bottomrule
\end{tabular}
\begin{tablenotes}
	\item[a] Due to lack of data, we do not assign costs to the aircraft. Since no competing technology exists, this assumption impacts only the objective value, not the solution.
\end{tablenotes}
\end{threeparttable}
}
\end{minipage}
\caption{Input and output flows and economic parameters of AIRCRAFT.}
\label{fig:AIRCRAFT}
\end{figure}

\subsection{Storage Technologies}
Storage technologies are essential for balancing supply and demand in energy systems. They store excess energy when supply exceeds demand and release it when needed, improving grid stability and enabling the integration of renewable energy sources. Each storage technology is characterized by its lifetime ($\tau$), investment ($c_{\mathrm{inv}}$) and maintenance costs ($c_{\mathrm{maint}}$), charge ($t_{\mathrm{charge}}$) and discharge times ($t_{\mathrm{discharge}}$), charging ($\eta_{\mathrm{in}}$) and discharging efficiencies ($\eta_{\mathrm{out}}$), as well as storage losses ($s_{\mathrm{loss}}$).
Storage solutions vary in capacity, duration, and application, ranging from short-term battery storage to long-term thermal and chemical storage options.

\subsubsection{Daily Storages}
Daily storages provide short-term storage. To avoid inter-day storage, they are modelled such that their storage level at the start and end of the day must be equal.
We categorize the daily storage into three types: electricity storage, i.e., batteries; thermal storage, e.g., water tanks; and chemical storage, e.g., natural gas and hydrogen buffer tanks.

\paragraph{Electricity Storage}

Electricity storage allows for the direct charging and discharging of electricity, primarily through batteries. We consider both stationary batteries and vehicle batteries as storage options.

For electric vehicles, bi-directional charging is included, meaning they can discharge their batteries to support grid stability. However, we assume that only 20\% of vehicles are plugged in and available for (bi-directional) charging at any given time. The total available vehicle battery capacity is determined by the number of vehicles multiplied by their storage capacity per unit.

\begin{table}[h]

\centering
\begin{threeparttable}
\caption{Economic parameters for electricity storage.}
\begin{tabular}{l c c c c }
\toprule
\textbf{Technology} & \boldmath$c_{\mathrm{inv}}$ & \boldmath$c_{\mathrm{maint}}$  & \boldmath$\tau$ & \textbf{capacity} \\
    & [CHF/kWh] & [CHF/kWh/y] & [years] & [kWh/vehicle] \\
\midrule
Battery                 & 156.66\tnote{a} $\pm$ 30\%       & 0.2422\tnote{a} $\pm$ 30\%     & 15\tnote{a} & - \\ 
Battery EV\tnote{b}     & -                       & -                     & -  & 80\tnote{c}\\
Battery BUS\tnote{b}    & -                       & -                     & -  & 400\tnote{d}\\
Battery TRUCK\tnote{b}  & -                       & -                     & -  & 400\tnote{e}\\
\bottomrule
\end{tabular}
\begin{tablenotes}
\item[a] Based on \cite{limpens_generating_2021}.
\item[b] The investment and maintenance costs of batteries for electric cars, buses and trucks are included in the vehicle costs.
\item[c] Rounded value for the year 2023 from figure ``Average Battery Capacity (kWh) of Available BEV Models per Year" \cite{noauthor_electric_nodate}.
\item[d] Based on MAN Lion's City E with 5 battery packs \cite{noauthor_man_nodate-1}.
\item[e] Based on MAN eTGX with 5 battery packs \cite{noauthor_man_nodate}.
\end{tablenotes}
\end{threeparttable}
\end{table}

\begin{table}[h]
\centering
\begin{threeparttable}
\caption{Technical parameters for batteries.}
\begin{tabular}{l c c c c c c}
\toprule
\textbf{Technology} & \boldmath$t_{\mathrm{charge}}$  & \boldmath$t_{\mathrm{discharge}}$  & \boldmath$s_{\mathrm{loss}}$ & \boldmath$s_{\mathrm{avail}}$  & \boldmath$\eta_{\mathrm{in}}$  & \boldmath$\eta_{\mathrm{out}}$  \\
    & [hours] & [hours] &  [\%]& [\%] & [\%] & [\%] \\
\midrule
Battery         & 4\tnote{a} & 4\tnote{a}  & 0.02\tnote{a} & 100\tnote{a} & 95\tnote{a} & 95\tnote{a} \\
Battery EV      & 4\tnote{a} & 10\tnote{a} & 0.02\tnote{a} & 20\tnote{b}  & 95\tnote{a} & 95\tnote{a} \\
Battery BUS     & 4\tnote{c} & 10\tnote{c} & 0.02\tnote{c} & 20\tnote{b}  & 95\tnote{c} & 95\tnote{c} \\
Battery TRUCK   & 4\tnote{c} & 10\tnote{c} & 0.02\tnote{c} & 20\tnote{b}  & 95\tnote{c} & 95\tnote{c} \\
\bottomrule
\end{tabular}
\begin{tablenotes}
\item[a] Based on \cite{limpens_generating_2021}.
\item[b] We assume that only 20\% of electric vehicles are plugged at a time and available for (bi-directional) charging \cite{limpens_energyscope_2019}.
\item[c] We assume that bus and truck batteries have the same characteristics as electric car (EV) batteries.
\end{tablenotes}
\end{threeparttable}
\end{table}

\clearpage
\paragraph{Thermal Storage}
Thermal storage systems store excess heat supply and release it when needed. They play an important role in district heating networks (DHN), industrial processes, and building heating, allowing for greater flexibility in energy systems.

\begin{table}[h]
\centering
\resizebox{\textwidth}{!}{%
\begin{threeparttable}
\caption{Economic parameters for thermal energy storage (TES).}
\begin{tabular}{l c c c c}
\toprule
\textbf{Technology} & \textbf{Resource} & \boldmath$c_{\mathrm{inv}}$ & \boldmath$c_{\mathrm{maint}}$ & \boldmath$\tau$ \\
    &  & [CHF/kWh] & [CHF/kWh/y] & [years] \\
\midrule
Hourly DHN TES      & DHN\_LTH & 3\tnote{a} $\pm$ 30\% & 0.0086\tnote{a} $\pm$ 30\% & 40\tnote{a} \\
Local TES (DEC)     & DEC\_LTH & 18.972\tnote{a} $\pm$ 30\% & 0.133\tnote{a} $\pm$ 30\% & 25\tnote{a} \\
Buffer Ground DEC\tnote{b}   & Anergy   & - & - & 20 \\
Hourly HTH TES      & HTH      & 28.03\tnote{a} $\pm$ 30\% & 0.3\tnote{a} $\pm$ 30\% & 25\tnote{a} \\
Hourly MTH TES      & MTH      & 28.03\tnote{c} $\pm$ 30\% & 0.3\tnote{c} $\pm$ 30\% & 25\tnote{c} \\
\bottomrule
\end{tabular}
\begin{tablenotes}
\item[a] Based on \cite{limpens_generating_2021}.
\item[b] The costs for the geothermal probes is already included in the costs of the ground source heat pump (DEC\_GHP).
\item[c] Assumed to be same as the hourly storage for High-temperature heat (Hourly HTH TES).
\end{tablenotes}
\end{threeparttable}}
\end{table}

\begin{table}[h]
\centering
\begin{threeparttable}
\caption{Technical parameters for thermal energy storage (TES).}
\begin{tabular}{l c c c c c}
\toprule
\textbf{Technology} & \boldmath$t_{\mathrm{charge}}$ & \boldmath$t_{\mathrm{discharge}}$ & \boldmath$s_{\mathrm{loss}}$ & \boldmath$\eta_{\mathrm{in}}$ & \boldmath$\eta_{\mathrm{out}}$ \\
    & [hours] & [hours] & [\%] & [\%] & [\%] \\
\midrule
Hourly DHN TES      & 60.3\tnote{a} & 60.3\tnote{a} & 0.00833\tnote{a} & 100\tnote{a} & 100\tnote{a} \\
Local TES (DEC)     & 4\tnote{a}    & 4\tnote{a}    & 0.00824\tnote{a} & 100\tnote{a} & 100\tnote{a} \\
Buffer Ground DEC   & 4\tnote{b}    & 4\tnote{b}    & 0.00824\tnote{b} & 100\tnote{b} & 100\tnote{b} \\
Hourly HTH TES      & 2\tnote{a}    & 2\tnote{a}    & 0.000355\tnote{a} & 100\tnote{a} & 100\tnote{a} \\
Hourly MTH TES      & 2\tnote{c}    & 2\tnote{c}    & 0.000355\tnote{c} & 100\tnote{c} & 100\tnote{c} \\
\bottomrule
\end{tabular}
\begin{tablenotes}
\item[a] Based on \cite{limpens_generating_2021}. 
\item[b] Assumed to be same as Local TES (DEC).
\item[c] Assumed to be same as Hourly HTH TES.
\end{tablenotes}
\end{threeparttable}
\end{table}

\clearpage
\paragraph{Chemical Storage}
Chemical storage enables energy to be stored in the form of chemical bonds and later released through conversion processes.

\begin{table}[h]
\centering
\resizebox{\textwidth}{!}{%
\begin{threeparttable}
\caption{Economic parameters for chemical storage.}
\begin{tabular}{l c c c c}
\toprule
\textbf{Technology} & \textbf{Resource} & \boldmath$c_{\mathrm{inv}}$ & \boldmath$c_{\mathrm{maint}}$ & \boldmath$\tau$ \\
    &  & [CHF/kWh] & [CHF/kWh/y] & [years] \\
\midrule
Buffer H2\tnote{a}          & H2       & 6.19\tnote{b} $\pm$ 30\% & 0.39\tnote{b} $\pm$ 30\% & 20\tnote{b} \\
Buffer NG\tnote{a}          & NG       & 2.0634\tnote{c} $\pm$ 30\% & 0.13\tnote{c} $\pm$ 30\% & 20\tnote{c} \\
Waste Storage      & Waste    & 0.00635\tnote{d} $\pm$ 30\% & 0.000397\tnote{d} $\pm$ 30\% & 20\tnote{d} \\
Plastic Storage    & Plastic  & 0.00635\tnote{d} $\pm$ 30\% & 0.000397\tnote{d} $\pm$ 30\% & 20\tnote{d} \\
Naphtha Storage    & Naphtha  & 0.00635\tnote{d} $\pm$ 30\% & 0.000397\tnote{d} $\pm$ 30\% & 20\tnote{d} \\
HVC Storage        & HVC      & 0.00635\tnote{d} $\pm$ 30\% & 0.000397\tnote{d} $\pm$ 30\% & 20\tnote{d} \\
Buffer $\mathrm{CO_2}$ & $\mathrm{CO_2}$ & 49.5\tnote{b}  $\pm$ 30\% & 0.495\tnote{b}  $\pm$ 30\% & 20\tnote{b} \\
\bottomrule
\end{tabular}
\begin{tablenotes}
\item[a] A buffer represents a short-term storage with fast charge/discharge times, e.g., a tank storage.
\item[b] Taken from \cite{limpens_generating_2021}.
\item[c] Based on Buffer H2. We estimate the costs for the natural gas buffer to be one-third that of the hydrogen buffer as natural gas has a three times higher energy density compared to hydrogen.
\item[d] Assumed to be the same as Fuel Storage (Tab. \ref{tab:seasonal storage}).

\end{tablenotes}
\end{threeparttable}}
\end{table}

\begin{table}[h]
\centering
\begin{threeparttable}
\caption{Technical parameters for chemical storages.}
\begin{tabular}{l c c c c c}
\toprule
\textbf{Technology} & \boldmath$t_{\mathrm{charge}}$ & \boldmath$t_{\mathrm{discharge}}$ & \boldmath$s_{\mathrm{loss}}$ & \boldmath$\eta_{\mathrm{in}}$ & \boldmath$\eta_{\mathrm{out}}$ \\
    & [hours] & [hours] & [\%] & [\%] & [\%] \\
\midrule
Buffer H2          & 4\tnote{a} & 4\tnote{a} & 0\tnote{a}   & 90\tnote{a}  & 98\tnote{a} \\
Buffer NG          & 4\tnote{b} & 4\tnote{b} & 0\tnote{b}   & 90\tnote{b}  & 98\tnote{b} \\
Waste Storage      & 4\tnote{c} & 4\tnote{c} & 0\tnote{c}   & 100\tnote{c} & 100\tnote{c} \\
Plastic Storage    & 4\tnote{c} & 4\tnote{c} & 0\tnote{c}   & 100\tnote{c} & 100\tnote{c} \\
Naphtha Storage    & 4\tnote{c} & 4\tnote{c} & 0\tnote{c}   & 100\tnote{c} & 100\tnote{c} \\
HVC Storage        & 4\tnote{c} & 4\tnote{c} & 0\tnote{c}   & 100\tnote{c} & 100\tnote{c} \\
Buffer $\mathrm{CO_2}$ & 1\tnote{a} & 1\tnote{a} & 0\tnote{a}   & 100\tnote{a} & 100\tnote{a} \\
\bottomrule
\end{tabular}
\begin{tablenotes}
\item[a] Based on \cite{limpens_generating_2021}. 
\item[b] Assumed to be the same as for Buffer H2.
\item[c] Own assumption.

\end{tablenotes}
\end{threeparttable}
\end{table}

\clearpage
\subsubsection{Seasonal Storage}
Seasonal storages help bridge seasonal gaps in supply and demand by storing excess energy produced in periods of high generation, such as summer, for use during periods of lower generation, such as winter.
\begin{table}[h]

\centering
\resizebox{\textwidth}{!}{%
\begin{threeparttable}
\caption{Economic parameters for seasonal storage technologies.}
\begin{tabular}{l c c c c c c}

\toprule
\textbf{Technology} & \boldmath$c_{\mathrm{inv}}$ & \boldmath$c_{\mathrm{maint}}$ & \boldmath$\tau$  & \boldmath$f_{\mathrm{min}}$ & \boldmath$f_{\mathrm{max}}$\\
    & [CHF/kWh] & [CHF/kWh] & [years] & [GWh]  & [GWh] \\
\midrule
NG cavern        & 0.051158\tnote{a} $\pm$ 30\% & 0.00130916\tnote{a} $\pm$ 30\% & 50\tnote{a} & 0 & $\infty$ \\
H2 cavern        & 0.153474\tnote{b} $\pm$ 30\% & 0.00392748\tnote{b} $\pm$ 30\% & 50\tnote{b} & 0 & $\infty$ \\
Hydro reservoir  & 0.464511\tnote{c} $\pm$ 30\% & (2\% - 5\%)$\times  c_{\mathrm{inv}}$ & 60\tnote{d} & 6500\tnote{e,f} & 8500\tnote{f} \\
Pumped hydro              & 4.98\tnote{g} $\pm$ 30\%     & 0.02\tnote{g} & 50\tnote{g} & 369\tnote{e,g} & 1700\tnote{g} \\
Seasonal TES     & 0.51718\tnote{a} $\pm$ 30\% & 0.002833\tnote{a} $\pm$ 30\% & 25\tnote{a} & 0 & $\infty$ \\
Fuel storage     & 0.00635\tnote{a} $\pm$ 30\%  & 0.000397\tnote{a} $\pm$ 30\% & 20\tnote{a} & 0 & $\infty$ \\
Coal storage     & 0.00635\tnote{h} $\pm$ 30\%  & 0.000397\tnote{h}  $\pm$ 30\% & 20\tnote{h}  & 0 & $\infty$ \\
Ammonia storage  & 0.00635\tnote{a} $\pm$ 30\%  & 0.000397\tnote{a} $\pm$ 30\% & 20\tnote{a} & 0 & $\infty$ \\
Methanol storage & 0.00635\tnote{a} $\pm$ 30\%  & 0.000397\tnote{a} $\pm$ 30\% & 20\tnote{a} & 0 & $\infty$ \\
Wood storage     & 0.00635\tnote{h}  $\pm$ 30\%  & 0.000397\tnote{h}  $\pm$ 30\% & 20\tnote{h}  & 0 & $\infty$ \\
\bottomrule
\label{tab:seasonal storage}
\end{tabular}
\begin{tablenotes}
    \item[a] Based on \cite{limpens_generating_2021}. 
    \item[b] Based on NG cavern. We estimate the costs for the hydrogen cavern to be three times higher than that for the hydrogen cavern as hydrogen has a three times lower energy density compared to natural gas.
    \item[c] Cost estimate for expanding an existing hydro reservoir based on the extension project at Lake Grimsel in Switzerland \cite{kwo_vergrosserung_nodate}.
    \item[d] Own assumptions.
    \item[e] We assume that the existing capacity will continue to be used.
    \item[f] Taken from Table 2 in \cite{guidati_gianfranco_value_2022}.
    \item[g] Taken from Table 27 in SI of \cite{limpens_energyscope_2019}.
    \item[h] Assumed to be equivalent to Fuel storage.
\end{tablenotes}
\end{threeparttable}
}
\end{table}

\begin{table}[h]
\centering
\begin{threeparttable}
\caption{Technical parameters for seasonal storage technologies.}
\begin{tabular}{l c c c c c}
\toprule
\textbf{Technology} & \boldmath$t_{\mathrm{charge}}$ & \boldmath$t_{\mathrm{discharge}}$ & \boldmath$s_{\mathrm{loss}}$ & \boldmath$\eta_{\mathrm{in}}$ & \boldmath$\eta_{\mathrm{out}}$ \\
    & [hours] & [hours] & [\%] & [\%] & [\%] \\
\midrule
NG cavern        & 2256\tnote{a} & 752\tnote{a}   & 0\tnote{a}     & 99\tnote{a}   & 99.5\tnote{a} \\
H2 cavern        & 2256\tnote{b} & 752\tnote{b}   & 0\tnote{b}    & 99\tnote{b}   & 99.5\tnote{b} \\
Hydro reservoir  & 1\tnote{c}     & 1\tnote{c}     & 0\tnote{c}     & 100\tnote{d}  & 100\tnote{d}  \\
PHS              & 203\tnote{e}   & 203\tnote{e}   & 0\tnote{e}     & 0.9\tnote{e} & 0.9\tnote{e} \\
Seasonal TES     & 150\tnote{a}   & 150\tnote{a}   & 0.00606\tnote{a} & 100\tnote{a} & 100\tnote{a}  \\
Fuel storage     & 168\tnote{a}   & 168\tnote{a}   & 0\tnote{a}     & 100\tnote{a}  & 100\tnote{a}  \\
Coal storage     & 168\tnote{f}   & 168\tnote{f}   & 0\tnote{f}     & 100\tnote{f}  & 100\tnote{f}  \\
Ammonia storage  & 168\tnote{a}   & 168\tnote{a}   & 0\tnote{a}     & 100\tnote{a}  & 100\tnote{a}  \\
Methanol storage & 168\tnote{a}   & 168\tnote{a}   & 0\tnote{a}     & 100\tnote{a}  & 100\tnote{a}  \\
Wood storage     & 168\tnote{f}   & 168\tnote{f}   & 0\tnote{f}     & 100\tnote{f}  & 100\tnote{f}  \\
\bottomrule
\label{tab:seasonal storage2}
\end{tabular}
\begin{tablenotes}
\item[a] Based on \cite{limpens_generating_2021}. 
\item[b] Assumed to be the same as for NG cavern.
\item[c] The charging of the Hydro reservoir is determined by the inflow timeseries (Fig. \ref{fig:hydro inflow}), while the discharging is limited by the installed capacity of the dam power plant (HYDRO\_DAM).
\item[d] Own assumptions.
\item[e] Taken from Table 28 in SI of \cite{limpens_energyscope_2019}.
\item[f] Assumed to be equivalent to Fuel storage.
\end{tablenotes}
\end{threeparttable}
\end{table}

\clearpage
\subsection{Grids}
Grids are a crucial part of the energy infrastructure, enabling the distribution of energy across different regions. We consider grids for electricity, district heating, natural gas, and hydrogen.

We model the required grid capacities to be capable of accommodating the flow of power/heat/gas at any time throughout the year.

\begin{table}[h]
\centering
\begin{threeparttable}
\caption{Grid properties.}
\begin{tabular}{l c c c c}
\toprule
 \textbf{Grid type} & $c_{\mathrm{inv}}$ & $c_{\mathrm{maint}}$ & $\tau$ & \textbf{losses} \\
        & [CHF/kW] &  [CHF/kW/y] & [years] & [\%] \\
\midrule
Electricity & 62.76\tnote{a} $\pm$ 30\% & (2\% - 5\%)$\times  c_{\mathrm{inv}}$ & 80\tnote{b} & 7\tnote{b} \\
DHN & 882\tnote{b} $\pm$ 30\% & (2\% - 5\%)$\times  c_{\mathrm{inv}}$ & 60\tnote{b} & 5\tnote{b} \\
Natural Gas & 8.67\tnote{a} $\pm$ 30\% & (2\% - 5\%)$\times  c_{\mathrm{inv}}$ & 25\tnote{c} & - \\
Hydrogen & 39.94\tnote{a} $\pm$ 30\% & (2\% - 5\%)$\times  c_{\mathrm{inv}}$ & 25\tnote{c} & - \\
\bottomrule
\end{tabular}
\begin{tablenotes}
    \item[a] Own calculations based on \cite{schnidrig_role_2023}.
    \item[b] Based on \cite{limpens_energyscope_2019}.
    \item[c] Own assumptions.
    
\end{tablenotes}
\end{threeparttable}
\end{table}

\subsection{Typical Days}

The model uses a typical days time clustering formulation as explained in \cite{limpens_energyscope_2019}. We selected 12 TDs for this study, corresponding to day 21, 49, 93, 112, 181, 192, 206, 258, 259, 313, 318, 351 of the year.

\newpage

\printbibliography[heading=bibintoc,title={Supplementary References}]